\documentclass{aa}  
\usepackage{graphicx}
\usepackage{multirow}
\usepackage{txfonts}
\usepackage{soul,ulem}
\def\cm#1{\ifmmode {\,{\rm cm^{-#1}}}                  
        \else \hbox{$\,${\rm cm$^{\rm -#1}$}}\fi}
\def\raw {\ifmmode\rightarrow\else$\rightarrow$\fi}
\def\ex#1{\ifmmode {\times 10^{#1}}         
        \else \hbox{{$\times 10^{\rm #1}$}}\fi}
\def\xmol#1{\ifmmode {\mbox{X({\rm #1})}}
            \else
            \hbox{\mbox{$X$(#1)}}
           \fi}

\newcommand{\ioe}{intermediate/outer envelope}

\newcommand{\mions}{\hcomas, \htrececomas, \somas, \ndoshmas, and \htresomas}
\newcommand{\iram}{IRAM 30\,m}
\newcommand{\hso}{{\it Herschel}}

\newcommand{\kms}{\mbox{km~s$^{-1}$}}
\newcommand{\s}{\mbox{$''$}}
\newcommand{\mloss}{\mbox{$\dot{M}$}}
\newcommand{\my}{\mbox{$M_{\odot}$~yr$^{-1}$}}
\newcommand{\ls}{\mbox{$L_{\odot}$}}
\newcommand{\msun}{\mbox{$M_{\odot}$}}

\newcommand{\rs}{\mbox{$R_{\star}$}}
\newcommand{\rin}{\mbox{$R_{\rm in}$}}
\newcommand{\rout}{\mbox{$R_{\rm out}$}}

\newcommand{\eu}{\mbox{E$_{\rm u}$}}

\newcommand{\vexp}{\mbox{$V_{\rm exp}$}}

\newcommand{\vsys}{\mbox{$V_{\rm sys}^{\rm LSR}$}} 
\newcommand{\vlsr}{\mbox{$V_{\rm LSR}$}} 
\newcommand{\h}{$^{\rm h}$}
\newcommand{\m}{$^{\rm m}$}

\newcommand{\ta}{\mbox{$T^*_{\rm A}$}}
\newcommand{\tmb}{\mbox{$T_{\rm MB}$}}
\newcommand{\texc}{\mbox{$T_{\rm ex}$}}
\newcommand{\trot}{\mbox{$T_{\rm rot}$}}
\newcommand{\tkin}{\mbox{$T_{\rm kin}$}}
\newcommand{\dens}{\mbox{$n_{\rm H_2}$}}
\newcommand{\nc}{\mbox{$n_{\rm crit}$}}
\newcommand{\ntot}{\mbox{$N_{\rm tot}$}}

\newcommand{\nhdos}{\mbox{$N_{\rm H_2}$}}

\newcommand{\trece}{$^{13}$CO\,($J$=1$-$0)}

\newcommand{\docem}{$^{12}$CO}
\newcommand{\trecem}{$^{13}$CO}

\newcommand{\water}{\mbox{H$_2$O}}

\newcommand{\waterdo}{\mbox{H$_2$$^{18}$O}}

\newcommand{\waterds}{\mbox{H$_2$$^{17}$O}}

\newcommand{\hcomas}{HCO$^+$}
\newcommand{\htrececomas}{H$^{13}$CO$^+$}
\newcommand{\hdocecomas}{H$^{12}$CO$^+$}
\newcommand{\somas}{SO$^+$}
\newcommand{\ndoshmas}{N$_2$H$^+$}
\newcommand{\htresomas}{H$_3$O$^+$}
\newcommand{\htresmas}{H$_3^+$}

\newcommand{\oh}{\mbox{OH\,231.8}}
\newcommand{\intta}{\mbox{$\int{T^*_{\rm A}d\upsilonup}$}}
\newcommand{\idosd}{\mbox{I2$_{\rm d}$}}

\begin{document} 

   \title{Molecular ions in the O-rich evolved star OH231.8+4.2: \hcomas, \htrececomas\ and first detection of SO$^+$, N$_2$H$^+$, and H$_3$O$^+$}


 \author{ C.~S\'anchez Contreras\inst{1}
          \and
          L.~Velilla Prieto\inst{1,2} 
          \and
          M.~Ag\'undez\inst{2,5}
          \and
          J.~Cernicharo\inst{2}
          \and
          G.~Quintana-Lacaci\inst{2}
          \and
          V.~Bujarrabal\inst{3}
          \and
          J.~Alcolea\inst{4} 
          \and
          J. R. Goicoechea\inst{2}
          \and 
          F.~Herpin\inst{5}
          \and
          K.~M.~Menten\inst{6}
          \and
          F.~Wyrowski\inst{6}
          }

\institute{Department of Astrophysics, Astrobiology Center (CSIC-INTA),  
Postal address: ESAC campus, P.O. Box 78, E-28691 Villanueva de la Ca\~nada, Madrid, Spain\\
              \email{csanchez@cab.inta-csic.es}
         \and
         Instituto de Ciencia de Materiales de Madrid, CSIC, c/ Sor Juana In\'es de la Cruz 3, 28049 Cantoblanco, Madrid, Spain
         \and
         Observatorio Astron\'omico Nacional (IGN), Ap 112, 28803 Alcal\'a de Henares, Madrid, Spain
         \and
         Observatorio Astron\'omico Nacional (IGN), Alfonso XII No 3, 28014 Madrid, Spain
         \and
         Universit\'e de Bordeaux, LAB, UMR 5804, F-33270, Floirac, France
         \and
         Max-Planck-Institut f\"{u}r Radioastronomie, Auf dem H\"{u}gel 69, 53121 Bonn, Germany
             }


   \date{}

 
  \abstract {OH\,231.8+4.2, a bipolar outflow around 
a Mira-type variable star, 
    displays a unique molecular richness amongst circumstellar
    envelopes (CSEs) around O-rich AGB and post-AGB stars.
We report line observations of the \hcomas\ and
\htrececomas\ molecular ions and the first detection of SO$^+$,
N$_2$H$^+$, and (tentatively) H$_3$O$^+$ in this source.  \somas\ and
\htresomas\ have not been detected before in CSEs around evolved
stars. These data have been obtained as part of a full mm-wave and
far-IR spectral line survey carried out with the \iram\ 
radio telescope and with \hso/HIFI.
Except for \htresomas, all the molecular ions detected in this work
display emission lines with broad profiles (FWHM$\sim$\,50-90\,\kms),
which indicates that these ions are abundant in the fast bipolar
outflow of \oh. The narrow profile (FWHM$\sim$14\,\kms) and high
critical densities ($>$10$^6$\,\cm3) of the \htresomas\ transitions
observed are consistent with this ion arising from denser, inner (and
presumably warmer) layers of the fossil remnant of the slow AGB CSE at
the core of the nebula. From rotational diagram analysis, we
deduce excitation temperatures of \texc$\sim$10-20\,K for all ions
except for \htresomas, which is most consistent with
\texc$\approx$100\,K. Although uncertain, the higher excitation
temperature suspected for \htresomas\ is similar to that recently
found for \water\ and a few other molecules, which
selectively trace a previously
unidentified, warm nebular component. The column densities of the molecular ions reported here
are in the range \ntot$\approx$[1-8]$\times$10$^{13}$\,\cm2, leading
to beam-averaged fractional abundances relative to H$_2$ of
$X$(\hcomas)$\approx$10$^{-8}$,
$X$(\htrececomas)$\approx$2$\times$10$^{-9}$,
$X$(\somas)$\approx$4$\times$10$^{-9}$,
$X$(\ndoshmas)$\approx$2$\times$10$^{-9}$, and
$X$(\htresomas)$\approx$7$\times$10$^{-9}$\,\cm2. We have performed
chemical kinetics models to investigate the formation of these ions in
\oh\ as the result of standard gas phase reactions initiated by
cosmic-ray and UV-photon ionization. The model predicts that \hcomas,
\somas, and \htresomas\ can form with abundances comparable to the
observed average values in the external layers of the slow central
core (at $\sim$[3-8]$\times$10$^{16}$\,cm); \htresomas\ would also
form quite abundantly in regions closer to the center
($X$(\htresomas)$\sim$10$^{-9}$ at $\sim$10$^{16}$\,cm). For
\ndoshmas,\ the model abundance is lower than the observed
value by more than two orders of magnitude. The model fails to
reproduce the abundance enrichment of \hcomas, \somas, and
\ndoshmas\ in the lobes, which is directly inferred from the broad
emission profiles of these ions. Also, in disagreement with the narrow
\htresomas\ spectra, the model predicts that this ion should form
in relatively large, detectable amounts ($\approx$10$^{-9}$) in the
external layers of the slow central core and in the high-velocity lobes. 
Some of the model-data discrepancies are reduced, but not suppressed,
by lowering the water content and enhancing the elemental nitrogen
abundance in the envelope.
The remarkable chemistry of \oh\ probably reflects the molecular
regeneration process within its envelope after the passage of fast shocks
that accelerated and dissociated molecules in the AGB wind
$\sim$800\,yr ago.}


\titlerunning{Molecular ions in \oh}

   \maketitle
%

\section{Introduction}
\label{intro}

%


OH\,231.8+4.2 (hereafter, \oh) is a well-studied bipolar nebula around
an OH/IR source\footnote{OH/IR objects are infrared-bright evolved
  stellar objects with a dense envelope showing prominent OH maser
  emission.}. Although its evolutionary stage is not clear because of its
many unusual properties, it is believed to be a planetary nebula (PN)
precursor probably caught in a short-lived transitional phase.  The
obscured central star, named QX\,Pup, is classified as M9-10\,III and
has a Mira-like variability consistent with an evolved asymptotic
giant branch (AGB) star \citep{coh81,fea83,kas92,san04}. The evolution
of this object may have been complex since it has a binary companion
star (of type A0\,V), which has been indirectly identified from the
analysis of the stellar spectrum reflected by the nebular dust
\citep{coh85,san04}. The system, located at $\sim$1500 pc
\citep{choi12}, has a total luminosity of $\sim$10$^4$\ls\ and its
systemic velocity relative to the Local Standard of Rest (LSR) is
\vsys$\sim$34\,\kms. \oh\ is likely a member of the open cluster M\,46
with a progenitor mass of $\sim$3\,\msun\ \citep{jur85}.

\oh\ is surrounded by a massive ($\sim$\,1\,\msun) and predominantly
cold ($\sim$10-40\,K)
molecular envelope well characterized by mm-wavelength emission from
CO and other molecules (Fig.\,\ref{mapa}).  The molecular gas is
located in a very elongated and clumpy structure with two major
components: ($i$) a central core (referred to as {\sl clump I3}) with
an angular diameter of $\sim$6-8\arcsec, a total mass of
$\sim$0.64\,\msun\ , and low expansion velocity of
\vexp$\sim$6-35\,\kms; and ($ii$) a highly collimated
$\la$6\s$\times$57\s\ bipolar outflow, with a mass of
$\sim$0.3\msun\ and expansion velocities that increase linearly with
the distance from the center, reaching values of up to $\sim$200 and
430\,\kms\ at the tip of the north and south lobe, respectively. The
temperature in the lobes is notably low, $\sim$10-20\,K
\citep{san97,alc01}.

The molecular envelope of \oh\ is markedly different from the slow,
roughly round expanding CSEs of most AGB stars; its pronounced axial
symmetry, large expansion velocities, and the presence of shocks are
common in objects that have left the AGB phase and are evolving to the
PN stage, the so-called pre-PNe \citep{neri98,buj01,cc10,san12}. It is
believed that the nebula of \oh\ was created as the result of a huge
mass-loss event that occurred during the late-AGB evolution of the primary at a
rate of \mloss$\approx$10$^{-4}$\,\my. With a total linear momentum
of $\sim$27\,\msun\,\kms, the bipolar flow is interpreted as the
result of a sudden axial acceleration of the envelope. The linear
distance-velocity relation observed in the CO-outflow (with a
projected velocity gradient of
$\nabla$$\upsilonup$$\sim$6.5\,\kms\,arcsec$^{-1}$) suggests that such
an acceleration took place $\sim$\,800\,yr ago in less than
$\sim$150\,yr. The low-velocity, low-latitude central core of the
outflow is thought to be the fossil remnant of the AGB star's CSE.

It is probable that the acceleration of the bipolar lobes resulted
from the violent collision of underlying jets (emanating from the
stellar companion) and the slowly expanding AGB envelope
\citep{san00,alc01,buj02,san04}; this is one plausible scenario that
has been proposed to explain the shaping and acceleration of bipolar
pre-PNe and PNe \citep[e.g.,][]{sah98,bal02}.  Recently, \cite{sab14}
have found indications of a well-organized magnetic field parallel to
the major axis of the CO-outflow of \oh\ that could point to a
magnetic outflow launching mechanism. Alternatively, as noticed by
these authors, the magnetic field could have been dragged by the fast
outflow, which may have been driven by a different mechanism. For
example, the underlying jets could have been launched by the main-sequence companion powered by mass accretion from the mass-losing AGB
star through an FU Ori type outburst \citep{san04} or a much more violent
($\la$100\,days) Intermediate-luminosity Optical Transient (ILOT) like
event \citep{sok12}.

\oh\ has a remarkably rich and unusual chemistry amongst CSEs
around O-rich low-to-intermediate-mass evolved stars.  In addition to
the typical oxygen-rich content, with molecules such as H$_2$O, OH or
SiO \citep{bow84,morr87,zij01,san02,des07}, \oh\ displays strong lines
of many different molecular species, including many containing carbon.
High-angular-resolution mapping  of the HCO$^+$\,($J$=1--0) emission
(Fig.\,\ref{mapa}) indicates that this ion is comparatively more
abundant in the fast lobes that in the slow central core
\citep{san00}. Single-dish maps of the SiO\,($J$=5--4)
emission show that the abundance of this molecule could also be enhanced in the
lobes \citep{san97}. The spectrum of \oh\ is unusually rich
in lines from S- and N-bearing molecules. For example, it was the
first O-rich CSE in which H$_2$S, NS, CS, and OCS were detected, and
we have recently reported the first detection of HNCO, HNCS, HC$_3$N,
and NO \citep{vel14}.  Some of these S- and N-compounds are present in
the envelope at levels not expected in O-rich CSEs around
low-to-intermediate mass stars. In the case of \oh, it has been
proposed that extra Si and S are released into the gas phase from the
sputtering of dust grains by shocks. Shocks might also initiate
(endothermic) reactions that trigger the N and S chemistry and could
also be additional suppliers of free atoms and ions
\citep[][]{morr87,san00,vel14}.

We have recently completed a sensitive molecular line survey of this
object in the mm-wavelength and far-IR range with the IRAM\,30m
telescope and \hso\ \citep[][full survey data to be published by
  Velilla Prieto et al., Sanchez Contreras et al., in
  prep.]{vel14,san14}. We  have detected hundreds of molecular
transitions, discovered $>$30 new species (including isotopologues),
and extended the rotational ladders for many others.  This has led to
very detailed information on the global physicochemical structure of
this envelope. From a preliminary analysis of the survey data \citep{san14},
we find
two main temperature components: $a)$ a predominantly cold
($\sim$10-40\,K) component mainly traced by CO (but also by, e.g.,
HCN, HNC, HNCO, and HCO$^+$) and $b)$ a warm ($\approx$100\,K)
molecular component that is selectively traced by certain molecules,
such as H$_2$O, amongst a few others (e.g., CS, H$_2$S, and SiS). 

In this paper, we report the first detection of the molecular ions
\somas, \ndoshmas, and (tentatively) \htresomas\ in \oh\ as part of
our surveys. We also present and analyse several lines of \hcomas\ and
\htrececomas, which are known to be present in \oh\ from earlier works
\citep{morr87,san97,san00}.  Molecular ions are believed to be
significant contributors to the molecule formation process in
circumstellar environments, which are active sites of molecular
synthesis.
In the interstellar medium (including molecular clouds, star-forming
regions, PDRs, etc.) more than 30 molecular ions have been detected,
however, in the envelopes around low-to-intermediate mass evolved stars
detections still remain scarce and limited to
\hcomas\ in most cases \citep[see, e.g.,][]{cer11}. 
In only a few objects in the post-AGB or PN phase, 
other positive ions have been detected, namely, CO$^+$, \ndoshmas,
CH$^+$, OH$^+$, and (tentatively) HCS$^+$. Except for OH$^+$, so far
these ions have been exclusively identified in C-rich objects, which
show in general a richer chemistry than their O-rich analogues. In
particular, toward the young PN NGC7027, where \hcomas, \htrececomas,
CO$^+$, \ndoshmas, CH$^+$, and, tentatively, HCS$^+$, are observed
\citep[][and references therein]{cer97,has01,zha08} and the pre-PN
CRL618, 
where \hcomas\ and \ndoshmas\ are detected \citep[][]{buj88,par07}.
Recently, the ion OH$^+$ has been detected with \hso\ in five PNe
\citep{etx14,ale14}. To our knowledge, in circumstellar envelopes
negatively charged molecular anions have been found to date only in
the C-rich AGB star IRC+10216 \citep[e.g.,][]{mcc06,cer07,agu10a}.


\section{Observations} 
\label{obs}

These data are part of a spectral line survey in
the mm/far-IR wavelength range carried out with the \iram\ radio telescope
\citep[$\sim$79-355\,GHz;][]{san11,vel13} and the \hso\ space observatory
\citep[$\sim$479-1244\,GHz;][]{san14}.
The spectra are presented in units of antenna-temperature (\ta), which
can be converted to a main-beam temperature (\tmb) scale via
\tmb=\ta/$\eta_{\rm eff}$, where $\eta_{\rm eff}$ is the frequency
dependent ratio between the main-beam efficiency ($\eta_{\rm mb}$) and the forward
efficiency ($\eta_{\rm l}$) of the telescope (values are provided in \S\,\ref{obs-30m}
and \S\,\ref{obs-hso} for \iram\ and \hso, respectively).

\begin{figure}[!htbp]
\centering 
\includegraphics[width=0.95\hsize]{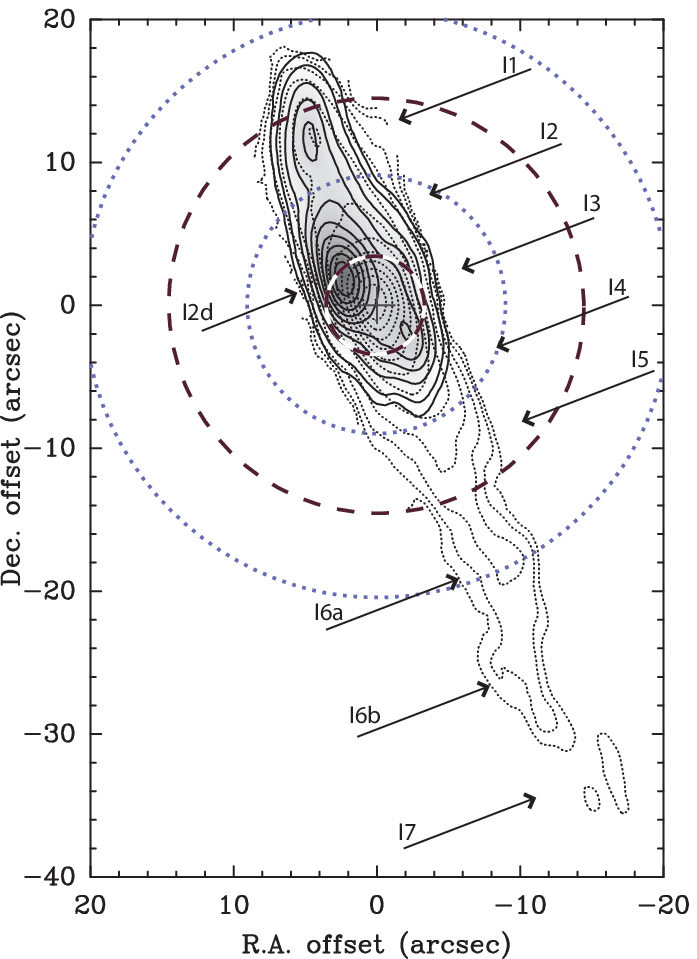}
\caption{Distribution of the \hcomas\,(1--0) (gray scale and solid
  contours) and \docem\,(1--0) (dotted contours) velocity-integrated
  emission in \oh\ from previous interferometric observations; adapted from Fig.\,2 of \cite{san00}. The smallest and largest HPBW
  of the \iram\ (long dashed) and \hso\ (dotted) telescopes in these
  observations are represented by the circumferences. We labeled the different
  regions/clumps in the molecular outflow as I1 to
  I7,  as in \cite{alc01}. Clump I2d
  corresponds to the region where the \hcomas\ emission peaks. The
  \vlsr\ range (in \kms) of each clump is I1) [$-$80:$-$30], I2)
               [$-$30:$+$10], I2d) [$-$20:$+$60], I3) [$+$10:$+$55],
               I4) [$+$55:$+$80], I5) [$+$80:$+$150], I6a)
               [$+$150:$+$205], I6b) [$+$205:$+$230], and I7)
               [$+$230:$+$285]; see Table\,2 in \cite{alc01}, and also
               \cite{san97,san00}, for more detail on the physical
               properties of the clumps.}
\label{mapa}
\end{figure}

In both surveys, we have observed one single position toward the
center of \oh\ (with J2000 coordinates R.A.\=7\h42\m16\fs830;
Dec.\,=$-$14\degr42\arcmin52\farcs10). Considering
the half power beam width (HPBW) of the \iram\ and \hso\ telescopes at
the observed frequencies (HPBW$_{\rm
  30m}$\,$\sim$7\arcsec-29\arcsec\ and HPBW$_{\rm
  HSO}$\,$\sim$18\arcsec-41\arcsec), this pointing fully covers the
slow central core at all frequencies and the fast bipolar lobes up to
some extent depending on the frequency (Fig.\,\ref{mapa}).

\subsection{Observations with \iram/EMIR}
\label{obs-30m}

Our mm/submm-wavelength survey was performed with the \iram\,
radio telescope (Pico Veleta, Granada, Spain) using the new generation
heterodyne Eight MIxer Receiver (EMIR).  Spectra were taken in several
observational campaigns between years 2009 and 2013.
We covered the whole accessible frequency range $\sim$79-355\,GHz,
using the four EMIR bands E090=\,3\,mm, E150=\,2\,mm, E230=\,1\,mm, and
E330=\,0.8\,mm \citep[][]{car12}. The EMIR was operated in single sideband
(SSB) mode for band E150, and in dual sideband (2SB) mode for bands
E090, E230 and E330. In all cases, the image sideband was rejected
with an average sideband rejection of $\sim$14\,dB.  Each receiver
band was connected to different backends; here we report data observed
with the WILMA autocorrelator, which provides a spectral resolution of
2\,MHz,
and the Fast Fourier Transform Spectrometer (FTS) in its 195\,kHz
spectral resolution mode.
Observations were done  
in wobbler switching mode with a wobbler throw of 120\arcsec.  Pointing
and focus were checked regularly (every $\sim$1.5 and $\sim$4\,h,
respectively) on strong nearby sources. 
On-source integration times per tuning step were typically
$\sim$1\,h. Calibration scans on the standard two load system were
taken every $\sim$18\,min. Errors in the absolute flux calibration are
expected to be $\la$25\%. 
The parameters of the beam of the \iram\ telescope used in this work
are described to a good accuracy as a function of the frequency ($\nu$) 
by 


\begin{equation}
\left \{ 
\begin{array}{l l}
\mathrm{HPBW}(\arcsec)=\frac{2460}{\nu[\mathrm{GHz}]}  & \\
\eta_{\rm eff}=0.9\exp{-(\frac{\nu[\mathrm{GHz}]}{399.7})^2} & \\ 
\end{array} 
\right . 
\end{equation}




\noindent 
according to measurement updates performed in August 2013\footnote{\tt http://www.iram.es/IRAMES/mainWiki/EmirforAstronomers}.

We reduced the data  using CLASS\footnote{CLASS is a worldwide
  software to process, reduce, and analyze heterodyne line observations
  maintained by the Institut de Radioastronomie Millimetrique (IRAM)
  and distributed with the GILDAS software, see {\tt
    http://www.iram.fr/IRAMFR/GILDAS}.} following the standard
procedure, which includes killing bad channels, subtracting baseline,
and averaging individual, good quality scans to produce final spectra.


\subsection{Observations with \hso/HIFI} 
\label{obs-hso}

Observations were carried out with the \hso\ Space Observatory
\citep{pil10} and its HIFI's wideband spectrometer
\citep[WBS][]{graa10} in several runs in November 2011 and April-May
2012. Our survey, executed in spectral scan mode, covers the frequency
range $\sim$479-1244\,GHz (bands 1a-5a) with an average spectral resolution of
$\Delta$$\nu$=1.1\,MHz.  Observations were performed in the dual beam
switching (DBS) mode with a 3\arcmin\ chop throw.
The two orthogonal receivers of HIFI (horizontal H, and vertical V)
were used simultaneously. 

We reprocessed raw \hso/HIFI data  from level 0 to level 2
running the {\tt hifiPipeline} task of HIPE\footnote{HIPE is a joint
  development by the Herschel Science Ground Segment Consortium,
  consisting of ESA, the NASA Herschel Science Center, and the HIFI,
  PACS, and SPIRE consortia.} \citep[versions 9.0-11.0; ][]{ott10}.
Afterward,  level 2 spectra were saved to FITS format and then
imported by CLASS where standard data reduction routines were
applied: blanking bad-quality data, removing spurs, fitting and
subtracting baselines, stitching of the spectrometer subbands, and
combining individual spectra.  We averaged the spectra from both V and
H polarisations (with equal weights), reducing the noise in the final
product.

HIFI is a double-sideband (DSB) heterodyne
instrument and, therefore, every spectrum contains the
upper and lower sideband data folded together.
We performed the deconvolution of the DSB data into single-sideband (SSB) format in CLASS using the task {\tt deconv}.
We  assumed a side-band gain ratio of one. We performed the DSB-convolution  using all bands simultaneously, which takes advantage
of redundant observations in the frequency overlap regions between
bands and optimizes the rms of the final SSB spectra. The absolute
flux calibration uncertainty of the full, deconvolved SSB spectrum is
estimated to be $\sim$10\% (HIFI Observers’ Manual, version 2.4,
\S\, 5.7). Here we adopt a more conservative value of $\sim$20\%.

We adopted the values of HPBW and $\eta_{\rm eff}$ most recently
updated and reported by Mueller et al.\,(2014)\footnote{{\sl The HIFI
    Beam: Release \#1. Release Note for Astronomers} at {\tt
    http://herschel.esac.esa.int/twiki/bin/view/Public/Hifi\-CalibrationWeb\#HIFI\_performance\_and\_calibration}},
which supersede previous estimates in \cite{roe12}. HIFI has 14 mixers
(7 frequency bands in 2 polarizations), each of which has a
frequency-dependent antenna pattern (beam). Detailed models of the
beam have been obtained independently for each mixer and separately for
polarizations H and V (i.e., a total of 7$\times$2$\times$2=28 beam
models). The HPBW of HIFI for one spot frequency per mixer and the
adopted values for the Ruze-like scaling of the main-beam efficiency
($\eta_{\rm mb}$) within the frequency range of each mixer are
provided in Table 3 and Table 2, respectively, of Mueller et
al.\,(2014). For convenience, we reproduce the values of these
beam-model parameters in Table\,\ref{t-neff} of our
Appendix\,\ref{app-neff}. Taking these values into account, the HPBW
and $\eta_{\rm eff}$ can be described by 



\begin{equation} \label{eq-neff}
\left \{ 
\begin{array}{l}
\mathrm{HPBW}(\arcsec)=\frac{20950}{\nu[\mathrm{GHz}]}  \\
\eta_{\rm eff}=\frac{\eta_{\rm mb}}{\eta_{\rm l}}=\frac{\eta_{\rm mb,0}}{\eta_{\rm l}}\exp{[-(\frac{4\pi\sigma_{\rm mb}}{\lambda})^2]} \\ 
\end{array} 
\right . 
\end{equation}

\noindent 
with the forward efficiency, $\eta_{\rm l}$, being 0.96 at all
frequencies (Mueller et al.\, 2014, Roelfsema et al.\, 2012), and
where $\lambda$ and $\sigma_{\rm mb}$ are both expressed in $\mu$m.
As shown in Table\,\ref{t-neff}, $\eta_{\rm mb,0}$ and $\sigma_{\rm
  mb}$ are different for the different mixers of HIFI.


\section{Results}
\label{s-res}

%
\begin{center}
\begin{table}
\caption{\label{t_res} Parameters of the transitions used in this
  research: 1) rest frequency (MHz); 2) rotational quantum
  numbers; 3) upper level energy (\eu); 4) spontaneous emission
  Einstein's coefficient (A$_{\rm ul}$); and 5) integrated intensity
  or line flux ($\int{T^*_{A}d\upsilonup}$) and its error (within
  parenthesis); for nondetections, we provide 3$\sigma$-upper limits
  to the line flux adopting a spectral window of 30\,\kms, except for
  \htresomas\ for which we use 12\,\kms\ (see \S\,\ref{r-h3o+}).
The permanent dipole moment ($\mu$) of the ions are also indicated.
}

\small
\begin{tabular}{r l r c r}      
\hline\hline                  
Rest Freq.\ & Transition  & E$_{\rm u}$     &   A$_{\rm ul}$  &  $\int{T^*_{A}d\upsilonup}$\\    
(MHz)       & QNs         & (K)            &     (s$^{-1}$)  &  (K\,\kms) \\     
\hline                        
\multicolumn{4}{c}{{\bf H$^{12}$CO$^+$}  \hspace{0.5cm} $\mu$= 3.888\,Debyes} &\\ 
89188.5 & 1   - 0   & 4.3 & 4.161E-05 & 4.56 (0.05) \\ 
267557.6 & 3   - 2   & 25.7 & 1.444E-03 & 9.1 (0.30)\tablefootmark{*} \\ 
535061.6 & 6   - 5   & 89.9 & 1.244E-02 & 0.15\tablefootmark{\dag}(0.05) \\
624208.4 & 7   - 6   & 119.8 & 1.997E-02 & $<$0.11 \\
\multicolumn{4}{c}{{\bf H$^{13}$CO$^+$}  \hspace{0.5cm} $\mu$= 3.888\,Debyes} &\\ 
86754.3 & 1   - 0   & 4.2 & 3.829E-05 & 0.50 (0.05) \\
260255.3 & 3   - 2   & 25.0 & 1.329E-03 & 1.85 (0.11) \\
346998.3 & 4   - 3   & 41.6 & 3.267E-03 & 0.76 (0.10) \\
520459.9 & 6   - 5   & 87.4 & 1.145E-02 & $<$0.11 \\
\multicolumn{4}{c}{{\bf SO$^+$}  \hspace{0.5cm} $\mu$= 2.3\,Debyes} &\\ 
162198.6 & 7/2-5/2(e) & 16.7 & 1.108E-04 & 0.64 (0.04) \\
208965.4 & 9/2-7/2(f) & 26.8 & 2.457E-04 & 0.50 (0.08) \\
254977.9 & 11/2-9/2(e) & 38.9 & 4.566E-04 & 0.55 (0.15)\\
301361.5 & 13/2-11/2(e) & 53.4 & 7.655E-04 & $<$0.15 \\ 
301736.8 & 13/2-11/2(f) & 53.5 & 7.684E-04 & $<$0.15 \\ 
347740.0 & 15/2-13/2(e) & 70.1 & 1.189E-03 & 0.25\tablefootmark{\dag}(0.15)\\ 
348115.2 & 15/2-13/2(f) & 70.2 & 1.193E-03 & $<$0.17\\ 
486837.2 & 20/2-18/2(f) & 133.5 & 3.330E-03 &  $<$0.11 \\ 
487212.1 & 20/2-18/2(e) & 133.7 & 3.338E-03 & $<$0.11 \\
\multicolumn{4}{c}{{\bf N$_2$H$^+$}  \hspace{0.5cm} $\mu$= 3.4\,Debyes} &\\ 
93173.1 & 1   - 0   & 4.5 & 3.628E-05 & 0.54 (0.04) \\
279510.8 & 3   - 2   & 26.8 & 1.259E-03 & 1.17 (0.10) \\
558965.1 & 6   - 5   & 93.9 & 1.085E-02  & $<$0.12 \\
652094.2 & 7   - 6   & 125.2 & 1.741E-02 & $<$0.14 \\
\multicolumn{4}{c}{{\bf H$_3$O$^+$}  \hspace{0.5cm} $\mu$= 1.44\,Debyes} &\\ 
307192.4 & 1$^-_1$ - 2$^+_{1}$ & 79.5 & 3.498E-04 & 0.28 (0.09) \\
984708.7 & 0$^-_0$ - 1$^+_{0}$ & 54.6 & 2.305E-02 & 0.40\tablefootmark{\dag}(0.20) \\
1031299.5 & 4$^+_3$ - 3$^-_3$ & 232.2 & 5.148E-03 & $<$0.31 \\
1069827.6 & 4$^+_2$ - 3$^-_2$ & 268.8 & 9.850E-03 & $<$0.23 \\
\hline 
\end{tabular}
\tablefoottext{*}{Flux after removing the contribution of the SO$_2$ line ($\sim$50\%, see \S\,\ref{r-hco+}).} \\
\tablefoottext{\dag}{Tentative detection.}
\end{table}
\end{center}
%


We  detected emission from the molecular ions \hcomas,
\htrececomas, \somas, \ndoshmas, and (tentatively) \htresomas\ toward
\oh\ (Table\,\ref{t_res} and Figs.\,\ref{hco+}-\ref{h3o+}). Except for
\hcomas\,$J$=1--0 \citep{morr87,san97}, \hcomas\,$J$=3--2, and
\htrececomas\,$J$=1--0 \citep{san97}, all transitions reported here
are first detections in this source. \somas\ and \htresomas\ have not
been detected before in CSEs around evolved stars (either oxygen- or
carbon-rich). \cite{qui13} recently reported the discovery of \ndoshmas\ emission toward the massive
yellow hypergiant IRC+10420, but no detection of this molecule in O-rich
low-to-intermediate mass evolved stars (AGB or post-AGB) has been
published to date. Detection of \ndoshmas\ has been reported in two
C-rich objects, the young PN NGC7027 and the pre-PN CRL618 (see
\S\,\ref{intro}).

In Table\,\ref{t_res} we list the main parameters of the molecular
transitions observed, which are all in the ground vibrational state.
The flux is computed by integrating
the area under the line profile. The
corresponding formal errors of the flux are 
given in parenthesis in the last column of the table and have been
calculated according to the error propagation theory as

\[
\label{errorbar}
\sigma = {\rm rms}\sqrt{{\rm FWZI}\times\Delta \upsilonup}, 
\]

where FWZI
is the spectral window (in \kms) over which the line is integrated
and rms is the root mean square statistical noise (in kelvins) of the
spectrum near the line for a spectral resolution of $\Delta
\upsilonup$ (in \kms). This estimate of the flux error does not
include additional uncertainties that may result from baseline
subtraction and absolute flux calibration.  Typically, we binned our original
spectra  to match a common velocity resolution of
$\Delta \upsilonup$\,$\sim$2\,\kms, except for the 2\,MHz-resolution
WILMA data at frequencies lower than $\sim$300\,GHz, for which
$\Delta \upsilonup$ is already larger than 2\,\kms\ (\S\,\ref{obs}). 

We obtained the line frequencies and other spectroscopic parameters discussed here   from 
the Cologne Database for Molecular Spectroscopy
\citep[CDMS;][]{mul05}, the Jet Propulsion Laboratory (JPL) molecular
spectroscopy database \citep{pick98}, and the MADEX catalogue
\citep{madex}.

\subsection{\hdocecomas\ and \htrececomas} 
\label{r-hco+} 

\begin{figure*}[!htbp]
   \centering
\includegraphics*[bb=135 130 556 738,angle=-90,width=0.475\hsize]{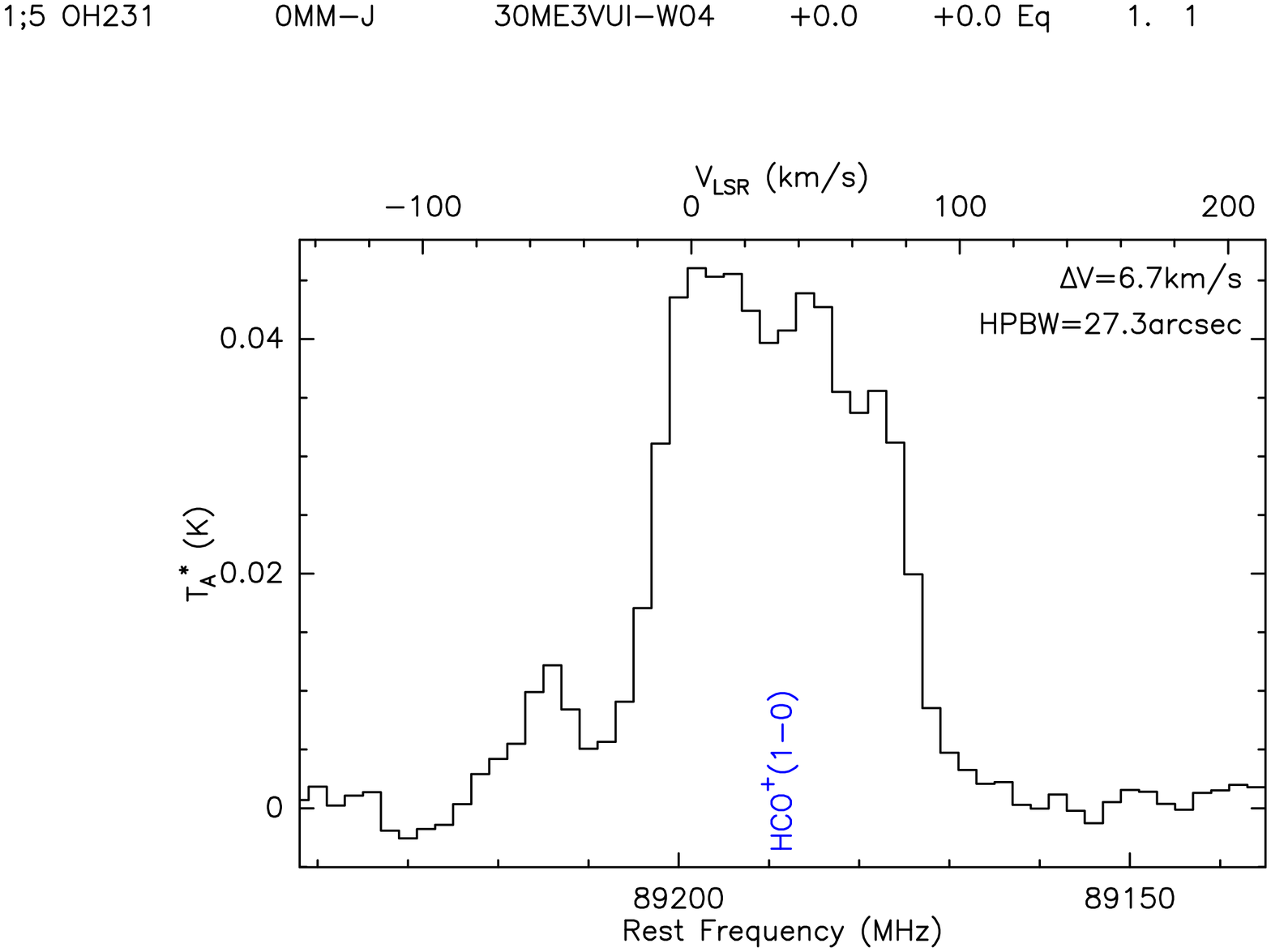} 
\includegraphics*[bb=135 130 556 738,angle=-90,width=0.475\hsize]{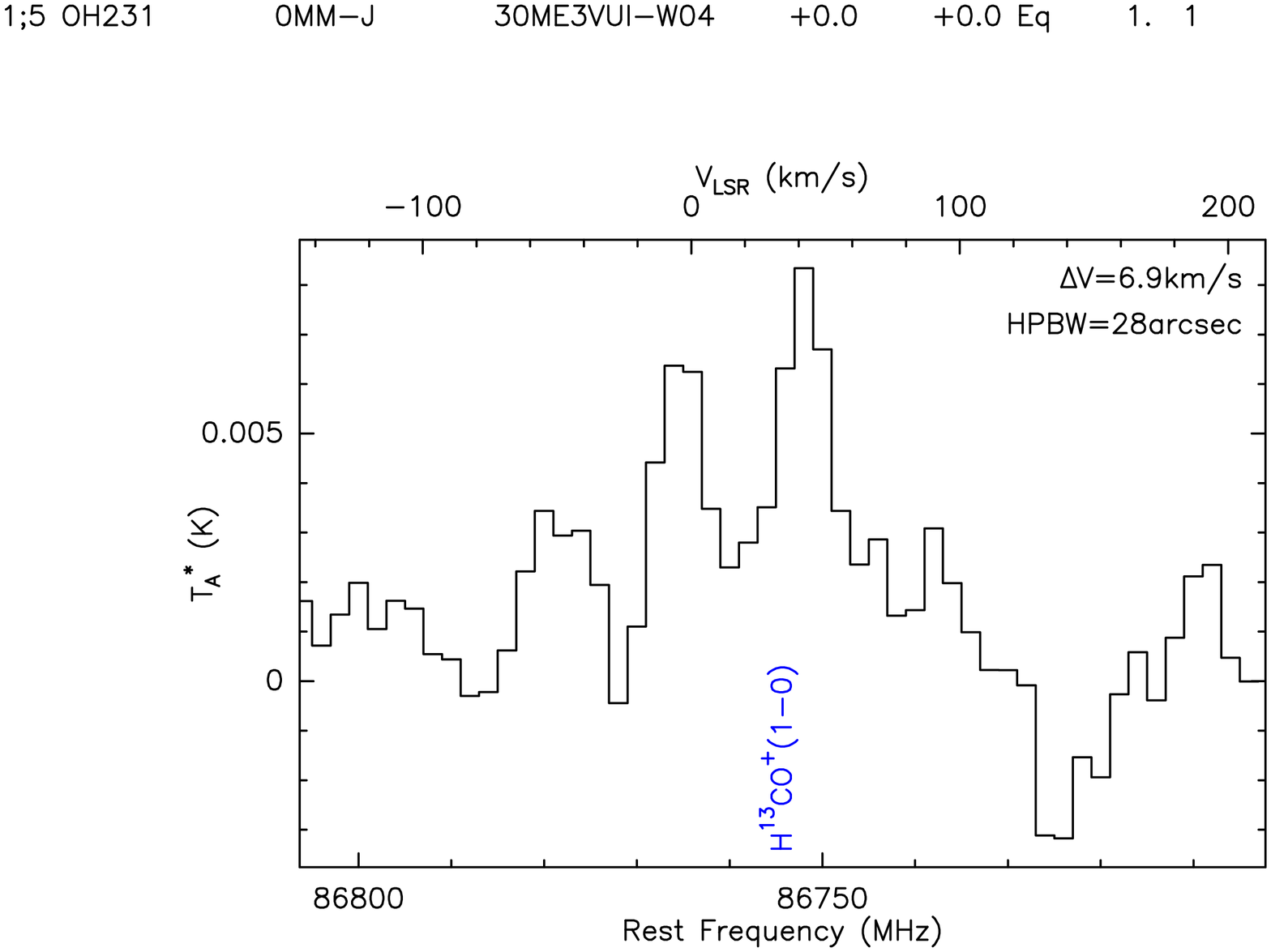} 

\includegraphics*[bb=135 130 556 738,angle=-90,width=0.475\hsize]{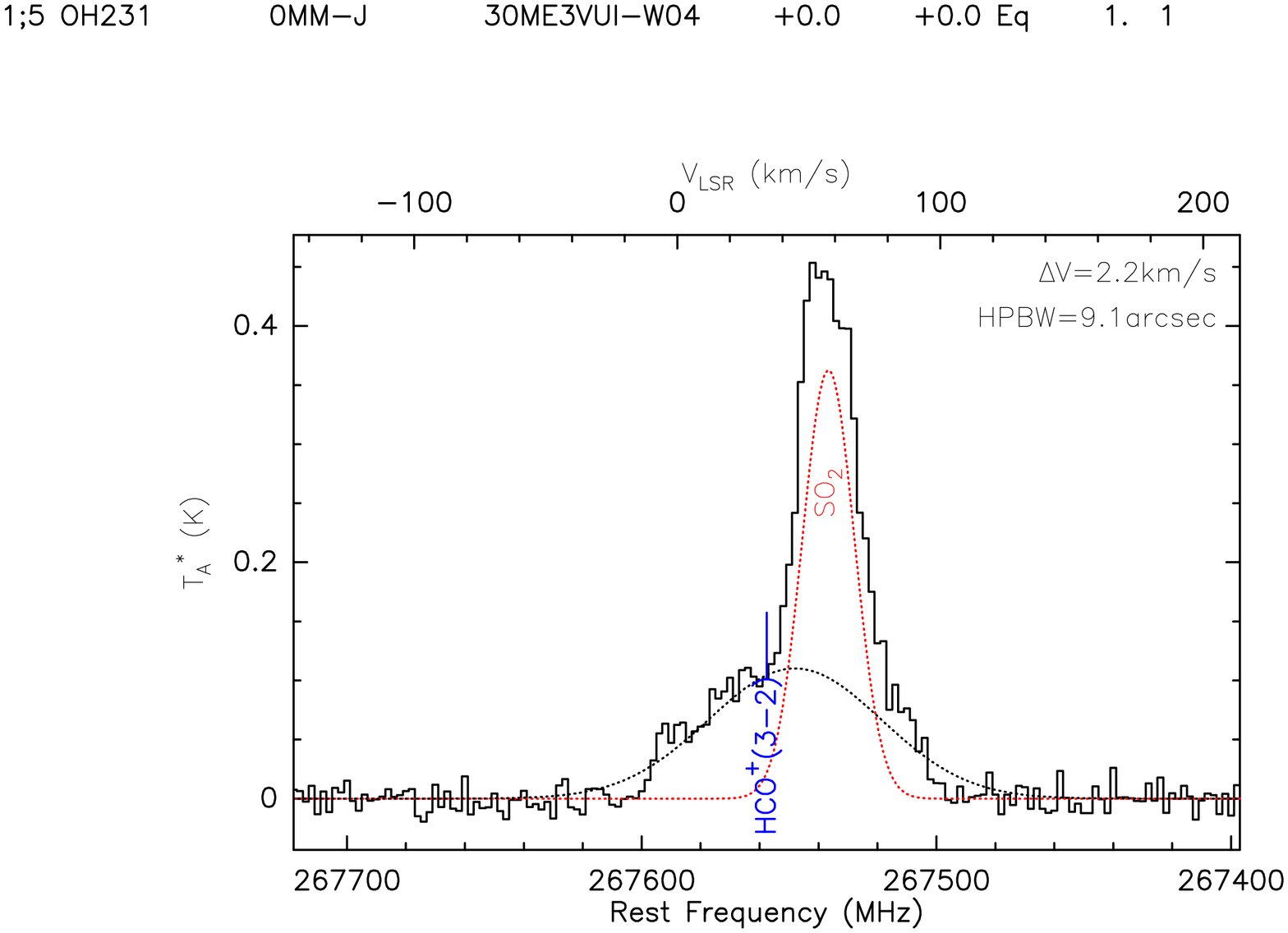}
\includegraphics*[bb=135 130 556 738,angle=-90,width=0.475\hsize]{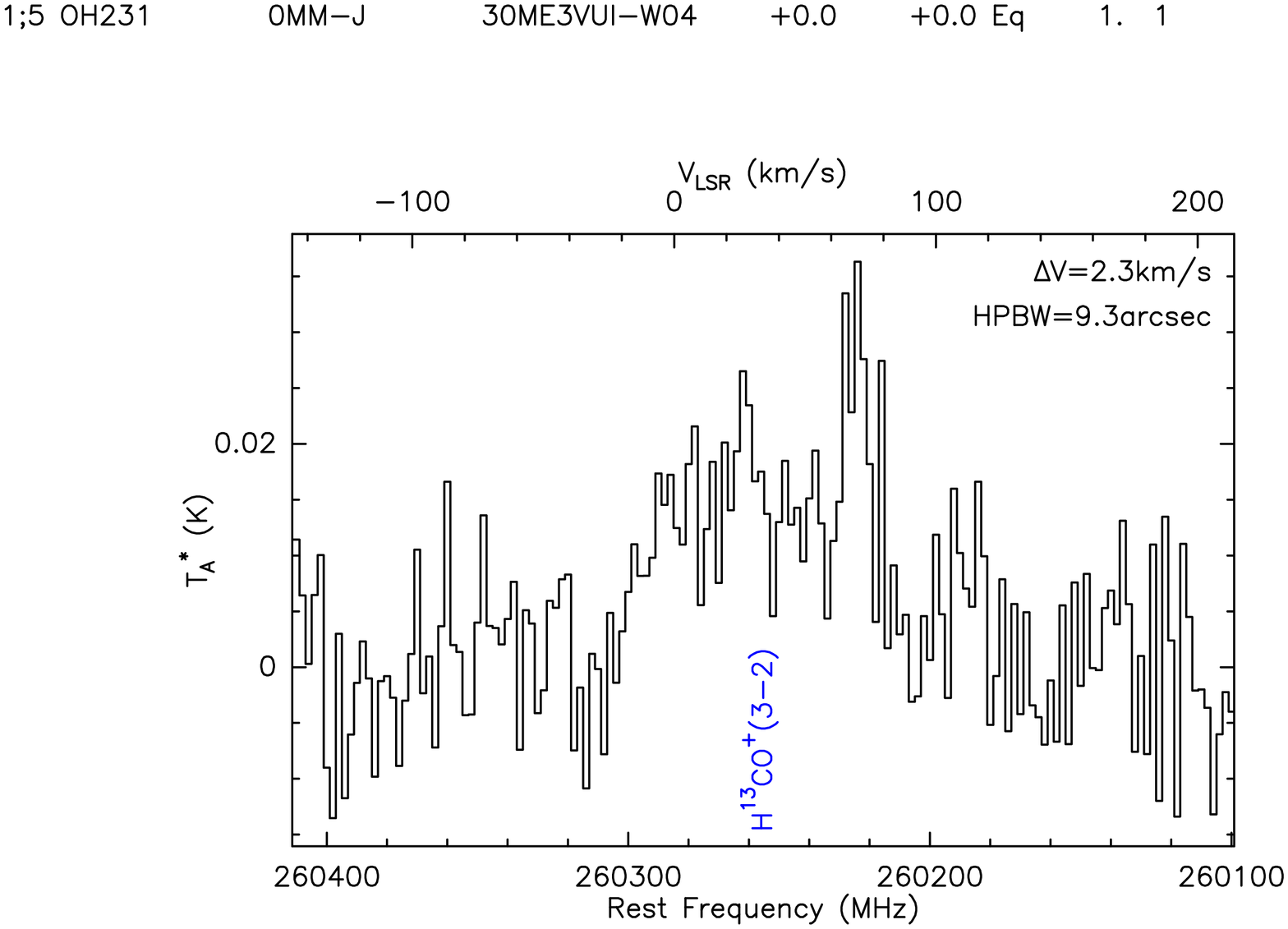} 

\includegraphics*[bb=135 130 556 738,angle=-90,width=0.475\hsize]{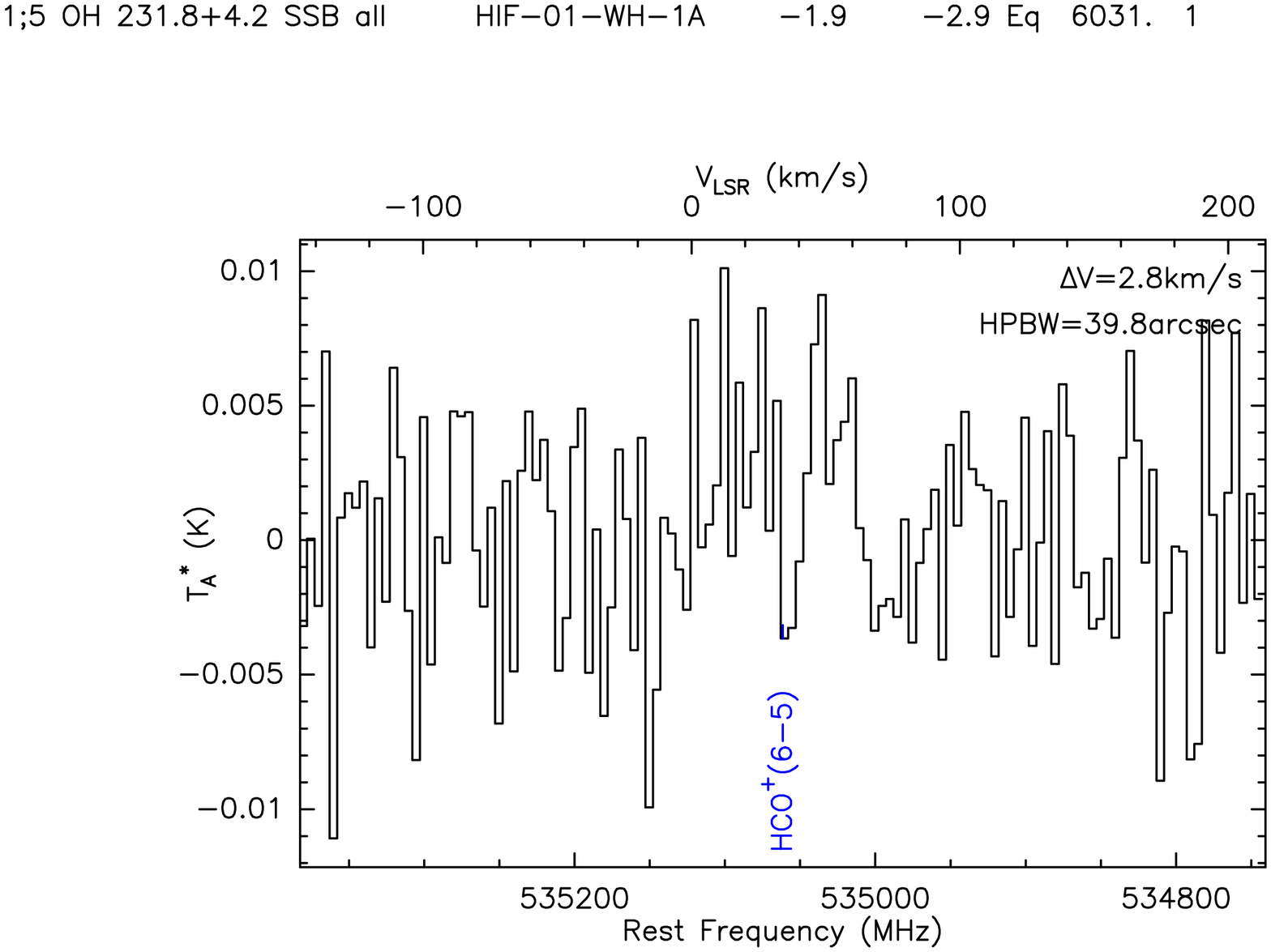} 
\includegraphics*[bb=135 130 556 738,angle=-90,width=0.475\hsize]{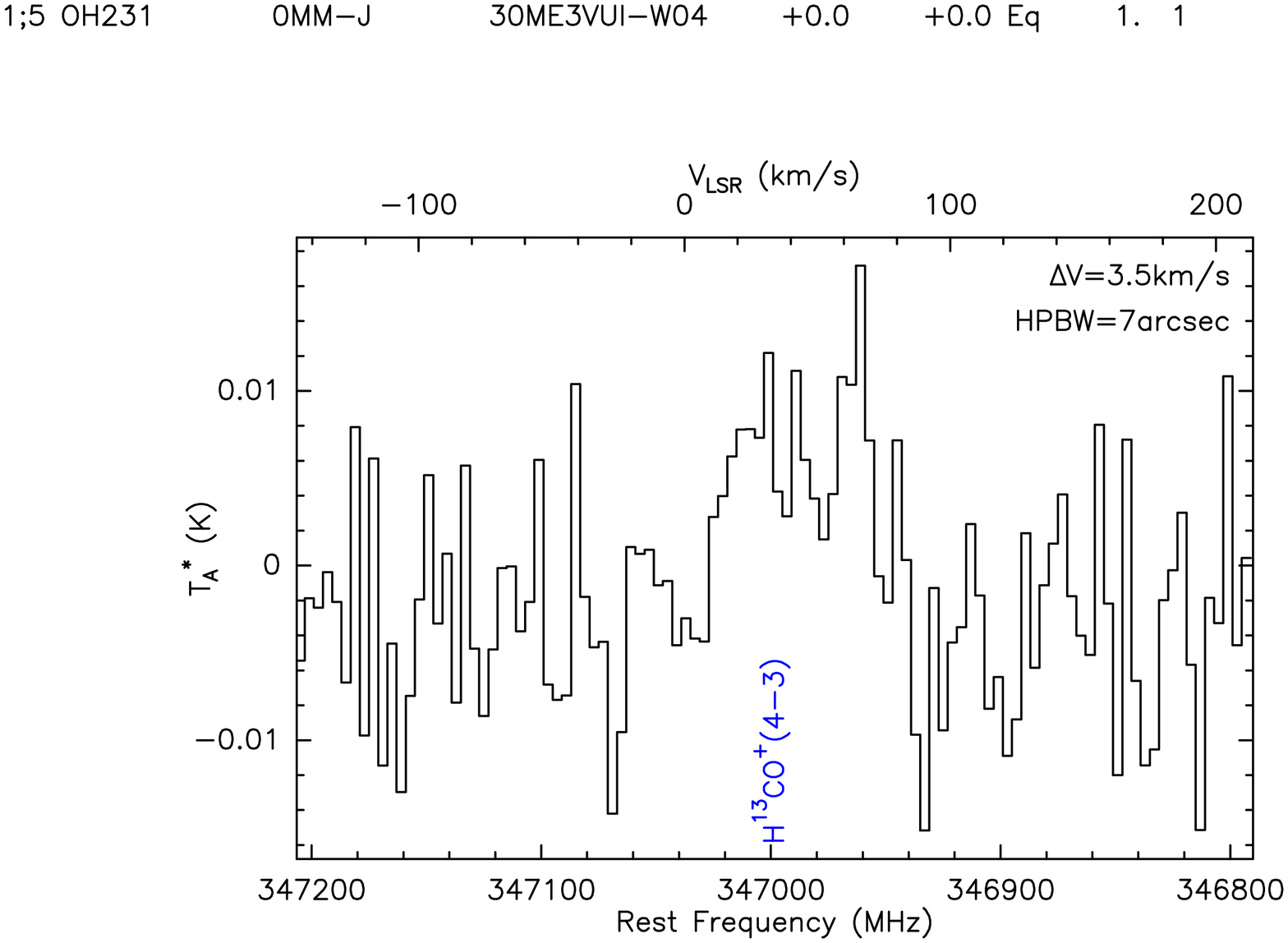}
   \caption{Spectra of \hcomas\ (left) and \htrececomas\ (right)
     emission lines in \oh. 
The velocity resolution ($\Delta$V) and half power beam width (HPBW)
are indicated inside the boxes at the top-right corner.  A two
Gaussian fit to the blend of the \hcomas\,$J$=3--2 and SO$_2$ lines is
shown (dotted lines).}
   \label{hco+}
   \end{figure*}

The three \hcomas\ lines we observed  toward \oh\ are shown in
Fig.\,\ref{hco+}.  The single-dish profiles of the \mbox{$J$=1--0} and
\mbox{$J$=3--2} transitions were already known from previous
observations \citep{san97,san00}. As expected, we find similar
shapes and absolute intensities (within the calibration
uncertainties).
The \hcomas\,$J$=1--0 transition, observed over a full velocity range
of $\sim$\,200\,\kms\ (\vlsr=[$-$80:+120]\kms), is characterized by a
broad flattened 
profile centered around \vlsr$\sim$\,30\,\kms\ with a full width
at half maximum of FWHM$\sim$\,90\,\kms\ and 
%
%
a weaker, blue-shifted component centered at
\vlsr$\sim$$-$53\,\kms\ with FWHM$\sim$25\,\kms.


The \hcomas\,$J$=3--2 line is also broad, covering the velocity range
\vlsr=[$-$20:+105]\,\kms. There is a prominent feature centered at
\vlsr$\sim$55\,\kms\ with FWHM$\sim$24\,\kms, which represents roughly
half of the total flux measured in the profile.  This feature,
observed but not discussed in previous works, is a blend of
\hcomas\,$J$=3--2 with the SO$_2$\,$J_{K,k}$=13$_{3,11}$--13$_{2,12}$
line at 267.5\,GHz (\eu=105\,K).  SO$_2$ shows a wealth of intense
transitions that dominate the mm/far-IR spectrum of \oh. We have used
these other SO$_2$ transitions to estimate the total column density and
excitation temperature of this molecule with the classic population
diagram method (to be published in S\'anchez Contreras et al., in
prep).
Using this technique, we confirm the identification of the SO$_2$ line
as the main contributor to the \vlsr$\sim$\,55\,\kms\ feature and
remove its emission from the total flux measured.
We find that the expected intensity of this SO$_2$ transition accounts
for 50\% of the total flux measured in the blend at 267\,GHz
(\intta$\sim$18\,K\kms) and, thus, the remaining $\sim$50\% is due to
\hcomas\,$J$=3--2 (Table\,\ref{t_res}).  A two Gaussian fit to the
blended profile is consistent with the \hcomas\ line being centered at
\vlsr$\sim$40\,\kms\ and having FWHM$\sim$80\,\kms\ (although these
figures are rather uncertain).

Regarding \hcomas\,$J$=6--5, although the low S/N ratio prevents an
accurate characterization of the profile, the shape of this weak
line is also consistent  with a broad profile (FWZI$\sim$60\,\kms) 
with two peaks around \vlsr$\sim$20 and 50\,\kms, similar to the profile 
of the \htrececomas\,$J$=4--3 (also in Fig.\,\ref{hco+}, right column).


As already noted and discussed by, e.g., \cite{morr87} and
\cite{san97}, the broad \hcomas\ profiles are notably different from
those of most molecules, including \docem\ and \trecem. The latter
display an intense narrow emission core (in the range
\vlsr=[$+$10:$+$55]\kms) that arises at the low-velocity, low-latitude parts
of the nebula (clump I3), which contain $\sim$2/3 of the total nebular
mass, plus much weaker broad wings produced by the less massive but faster
bipolar lobes. (Profiles of \trecem\ transitions from our survey are
included as additional material in Appendix\,\ref{app-1},
Fig.\,\ref{13co}).

The spatio-kinematic distribution of \hcomas\,$J$=1--0 is known with
$\sim$4\arcsec$\times$2\arcsec-resolution from interferometric maps by
\cite{san00} -- see Fig.\,\ref{mapa}. These data show that, indeed,
the different line profiles of \hcomas\ and CO come from
significant differences between the spatial distribution of these
molecules. In contrast to CO, the \hcomas\,\mbox{$J$=1--0} emission is
enhanced in the bipolar lobes and it shows a deficit of emission
from the massive slow central region; this explains the broad
flattened \hcomas\ profiles. The \hcomas\ emission peaks at a compact
region (referred to as clump \idosd) at $\sim$3-4\arcsec\ along
PA$\sim$50\degr\ from the center.
There is a large velocity dispersion, of $\sim$80\,\kms, within this
clump that, indeed, produces the bulk of the \hcomas\,$J$=1--0
emission in the velocity range \vlsr=[$-$20:+60]\,\kms.
The extent of the \hcomas\ emission in the outflow is smaller
than that of CO 
(in the southern lobe, the \hcomas\ emission abruptly falls beyond clump
I5) although the axial velocity gradient is similar in both cases.

The single-dish profile of the $J$=1--0 transition (Fig.\,\ref{hco+})
is consistent with the whole \hcomas\ emission completely filling the
telescope beam at 89\,GHz, HPBW=27\farcs3: the bluest
\hcomas\,$J$=1--0 spectral feature at \vlsr=$-$53\,\kms\ traces the
tip of the approaching North lobe (clump I1), while the reddest wing
emission (up to \vlsr=$+$120\,\kms) originates in the South lobe
(mainly from clump I4 and partially from clump I5).

The smaller full width of the \hcomas\,$J$=3--2 transition
($\sim$120\,\kms), compared with the $J$=1--0 line, is consistent with
the smaller beam size of the \iram\ telescope at 267\,GHz
(HPWB=9\farcs1) and, thus, the smaller fraction of the axial outflow
lying within the beam.  Given the velocity range over which the
\hcomas\,$J$=3--2 emission is observed in our single-dish data, most
of the emission must arise at the nebular center, the bright clump
\idosd\ and, partially, at the base of the South lobe.

The full width of the \hcomas\,$J$=6--5 line is $\sim$60\,\kms. At this
frequency, the beam of \hso\ is sufficiently large (HPBW=39\farcs8,
i.e.,\,larger than the \iram\ beam at 89\,GHz) as to include the
bipolar \hcomas-flow in its full extent. Therefore, one would
expect to observe \hcomas\,$J$=6--5 emission over a velocity range
comparable to that of the $J$=1--0 line ($\sim$200\,\kms). The
narrower $J$=6--5 profile could indicate that the emission from this
higher excitation transition, which has \eu=89\,K, is restricted to
regions closer to the nebula center, with smaller expansion
velocities. In particular, the emission could mainly be associated with 
the bright emission clump \idosd.
 However,  these conclusions are very
tentative 
because of the low S/N of this transition.





The \htrececomas\ line profiles are shown in Fig.\,\ref{hco+} (right
column). As expected, these transitions are weaker than their main
isotope analogues. The spectra are broad, with a full width that
ranges from $\sim$190\,\kms\ in the $J$=1--0 transition to
$\sim$70\,\kms\ in the $J$=4--3. They lack a dominant narrow core
component but, instead, show broad, multiple-peaked shapes.
For the $J$=1--0 transition, we clearly detect emisson at
\vlsr=$-$50\,\kms\ from the tip of the North lobe (clump I1).
In all transitions, we detect emission from a velocity range that is
consistent with emission produced at the bright clump \idosd\ and at
the base of the South lobe (clump I4). The emission contribution of
the latter, clump I4, concentrates in the range
\vlsr$\sim$[60:80]\,\kms, this spectral feature being rather
prominent and sharp in the $J$=3--2 and 4--3 lines. 

\subsection{\somas}
\label{s-so+}

\somas\ is a reactive radical with a $^2 \Pi$-electronic state.  The
energy levels of a $^2 \Pi$-molecule exhibit the additional splittings
due to the electron spin and orbital angular momentum interactions
($\Omega$- and $\Lambda$-doublets). The rotational levels are defined
with three quantum numbers, $\Omega$, $J$, $l$, which represent the
absolute value of the projection of the total electronic angular
momentum on the molecular axis, the total angular momentum from
rotation and electronic motion, and the parity, respectively \citep{ama91}.

\begin{figure*}[!htbp]
   \centering
\includegraphics*[bb=  110 87 786 580,width=0.475\hsize]{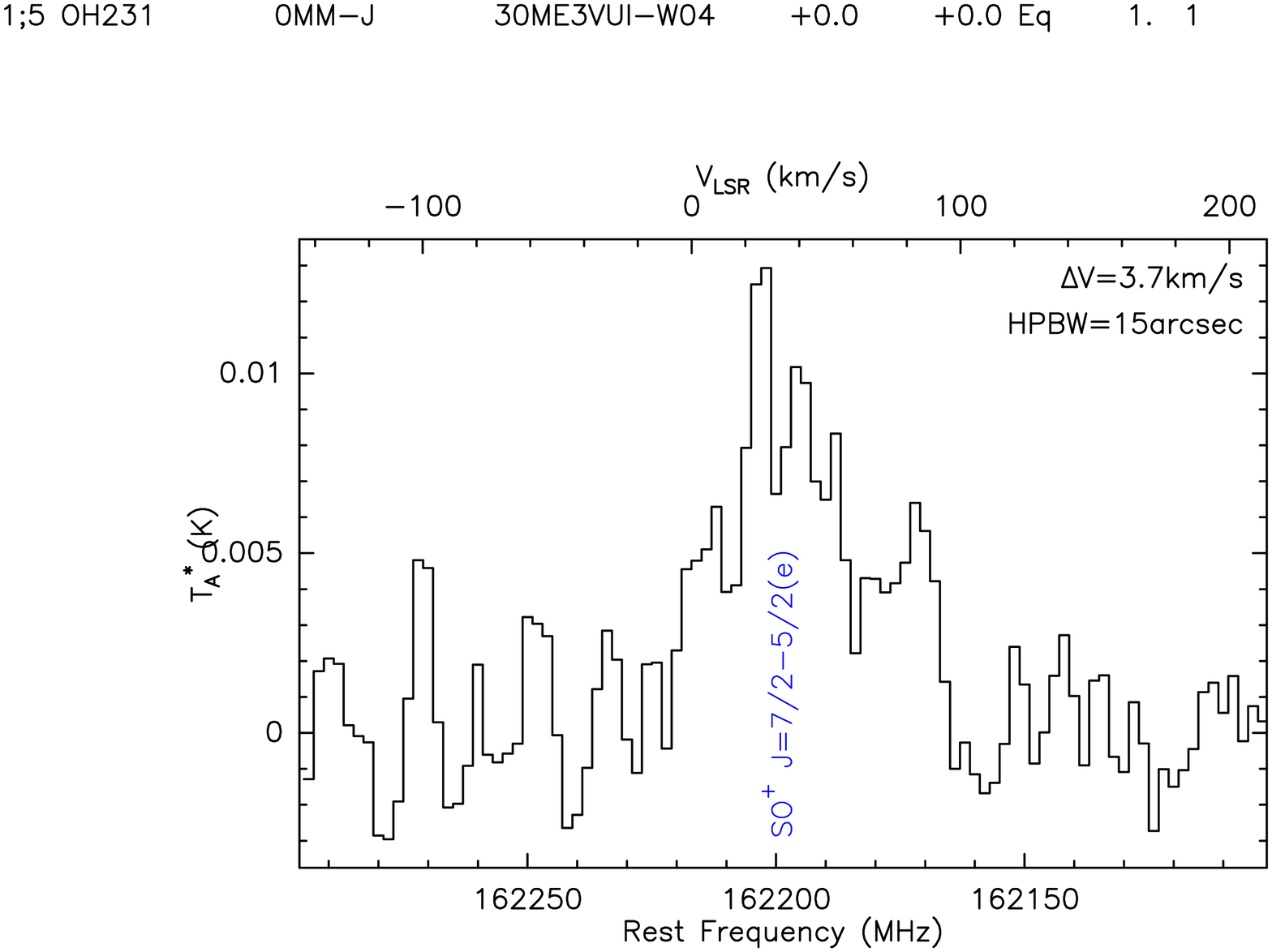}
\includegraphics*[bb=  110 87 786 580,width=0.475\hsize]{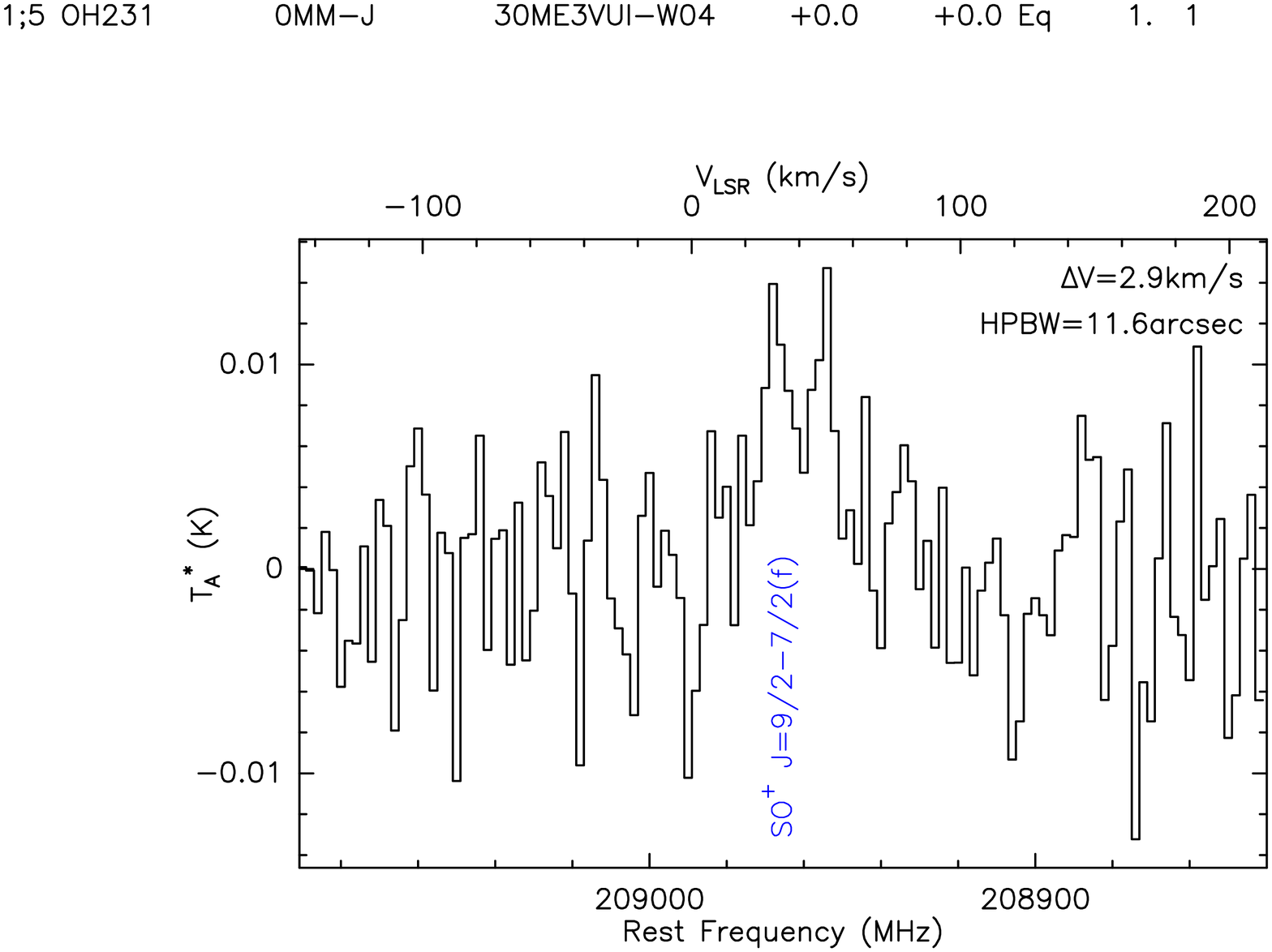}
\includegraphics*[bb=  110 87 786 580,width=0.475\hsize]{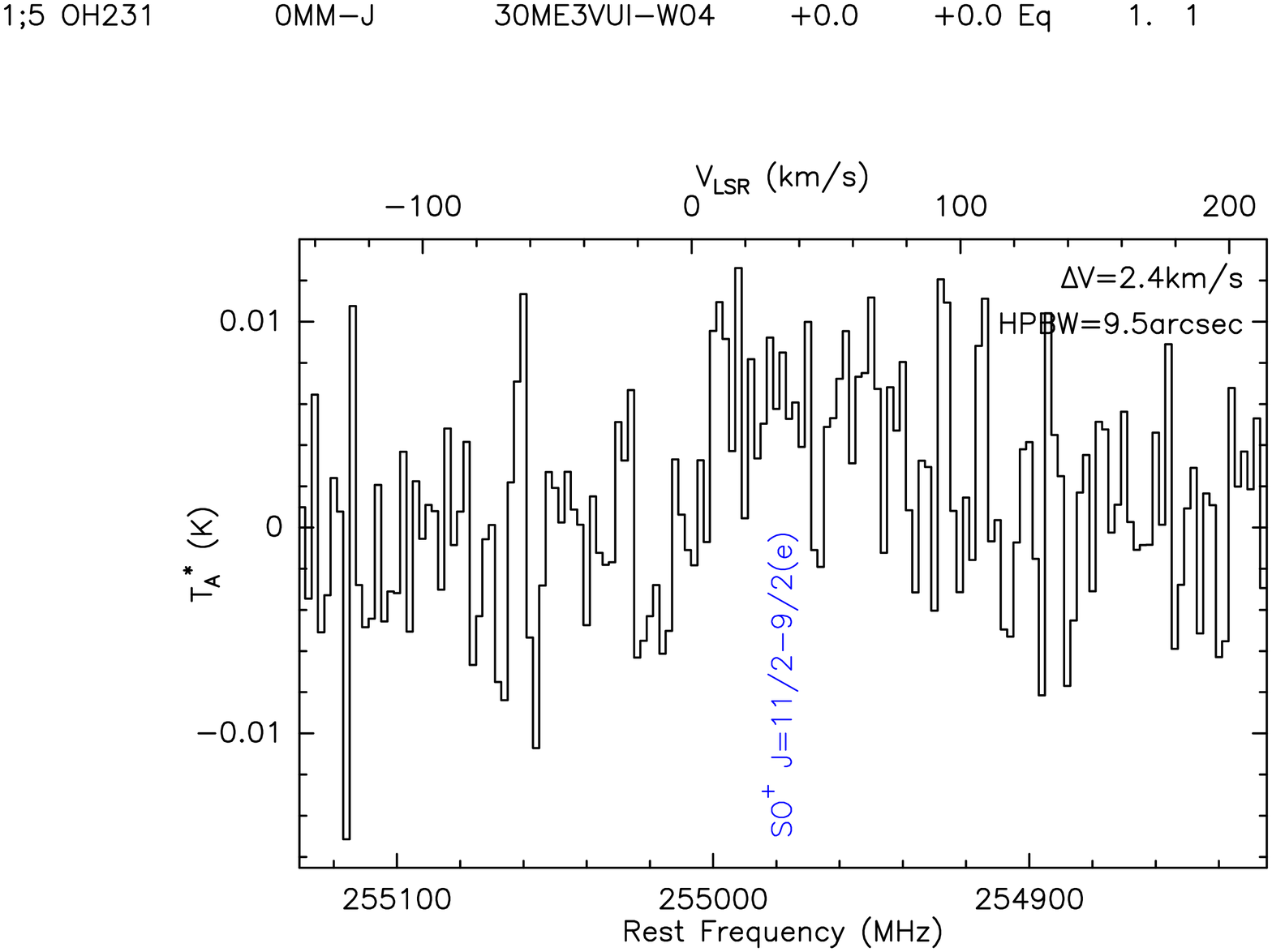}
\includegraphics*[bb=  110 87 786 580,width=0.475\hsize]{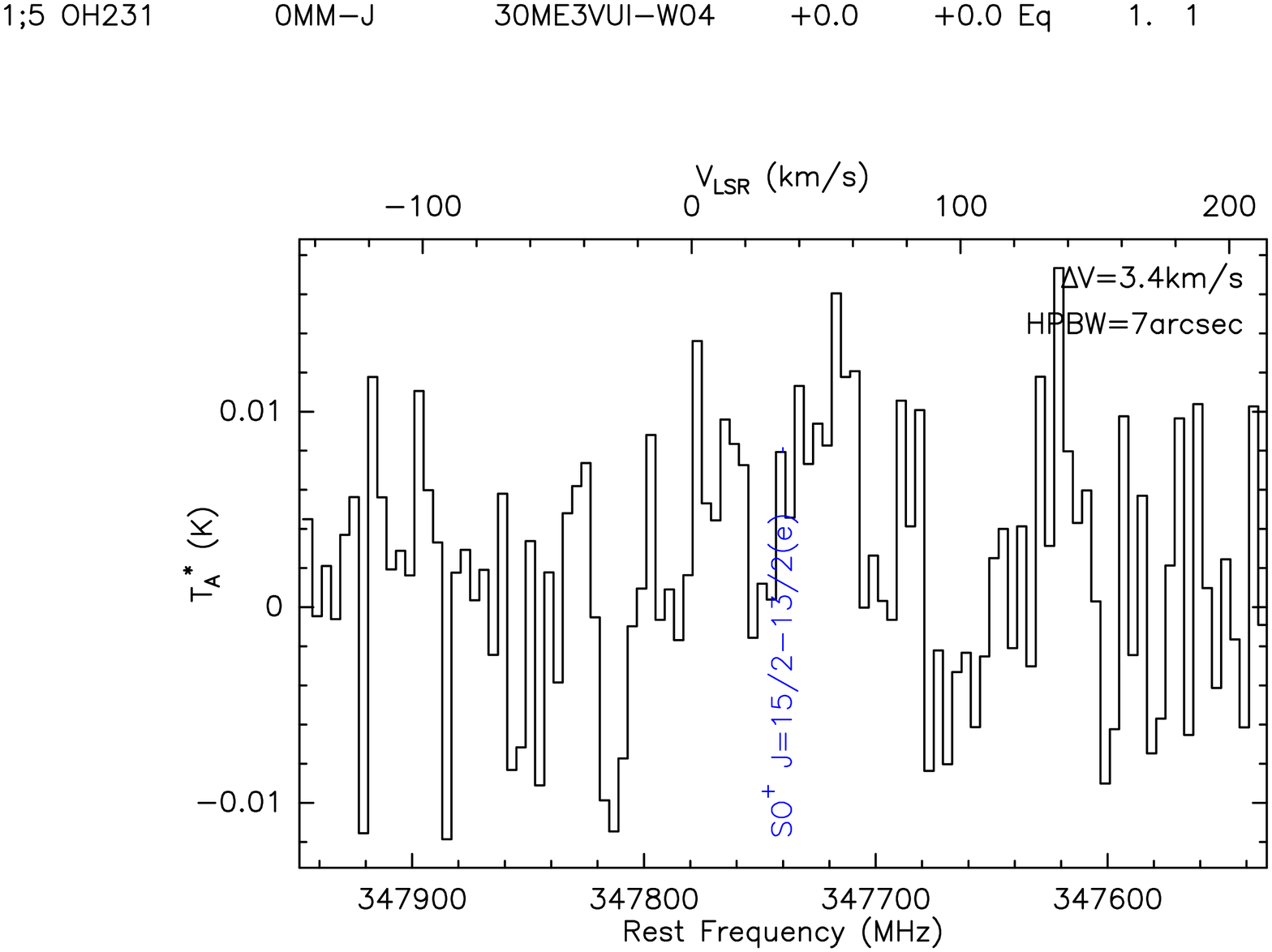}
      \caption{Same as in Fig.\,\ref{hco+} except for \somas. See also figure \ref{II-so+} in the appendix.}
         \label{I-so+}
   \end{figure*}

A total of twenty \somas\ transitions (five $\Lambda$-doublets in each
of the $\Omega$=1/2 and $\Omega$=3/2 states) lie within the frequency
range covered in our mm-wave survey with \iram.  The
\somas\ transitions detected toward \oh\ are listed in
Table\,\ref{t_res} and their profiles are shown in
Fig.\,\ref{I-so+}. The lines detected correspond to the lowest
rotational levels of the $^2 \Pi_{1/2}$ ladder of \somas, which
require low- excitation temperatures, between \eu=17 and 70\,K.  The
nondetection of the $^2 \Pi_{3/2}$ transitions is in very good
agreement with the expectations given that the lowest levels are at
\eu$\ga$500\,K above the ground, much higher than the average
temperature (\tkin$\sim$10-40\,K) of the molecular outflow of
\oh\ (\S\,\ref{intro} and \S\,\ref{rotdiagr}). In Table\,\ref{t_res} we
have not included additional \somas\ transitions that may be tentatively
detected but that appear partially or fully blended with other stronger lines
or in particularly noisy spectral regions  (the spectra over the full
range of frequencies covering the five \somas\ doublets observed 
are shown in Fig.\,\ref{II-so+} of the Appendix).


The \somas\ line profiles are broad, with FWHM$\sim$60$\pm$10\,\kms\ and
centered around \vlsr$\sim$40\,\kms\ (Fig.\,\ref{I-so+}).
The moderate S/N of the spectra does not justify an accurate analysis
of the profiles, however, it is sufficient to establish that, as for
\hcomas, the \somas\ lines lack  a prominent, narrow central core, but
instead are dominated by a broad spectral component. This suggests a
similar distribution for \somas\ and \hcomas\ and, therefore, it is
likely that the emission from both molecules arises mainly at the fast 
outflow. 


\subsection{\ndoshmas}  
\label{r-n2h+}

\ndoshmas\ is a linear, triatomic molecule.  Its spectral lines show
hyperfine structure due to the electric quadrupole moment of the
nitrogen nucleus. The line widths of the individual hyperfine
components, which are separated by a few \kms, are expected to 
blend because of 
the large expansion velocities of \oh's molecular
outflow.  In Table\,\ref{t_res} we list the \ndoshmas\ $J$=1--0 and
3--2 transitions detected in the mm-wavelength range and also provide
flux upper limits for nondetections from our far-IR survey with \hso\ up
to upper level energies of \eu$\sim$160\,K.



The spectra of the two \ndoshmas\ transitions detected are shown in
Fig.\,\ref{n2h+}. Both lines have rather broad, somewhat structured
profiles: the $J$=1--0 line is centered at 45$\pm$4\,\kms\ and has
FWHM=85$\pm$10\,\kms\ and the $J$=3--2 line is centered at
39$\pm$4\,\kms\ and has FWHM=70$\pm$10\,\kms.  
The smaller line width of the
\ndoshmas\,$J$=3--2 transition is consistent with the smaller beam size
at this frequency (HPBW$\sim$8\farcs7), which implies that a smaller
fraction of the outflow is probed. 


The deconvolved width of the lines at half maximum, taking  the hyperfine structure and the spectral resolution of the data into
account,
is $\sim$70\,\kms, indicating that most of the \ndoshmas\ emission
arises in the fast outflow.  As for \hcomas\ and \somas,
\ndoshmas\ lines present broad profiles, lacking of a prominent,
narrow core component.

\subsection{\htresomas}
\label{r-h3o+}




The molecular ion \htresomas, hydronium, is isoelectronic with ammonia
(NH$_3$) and has a similar physical structure: \htresomas\ is a
trigonal pyramid with the oxygen atom at the apex and the hydrogen
atoms at the base.
Like ammonia, \htresomas\ undergoes inversions as the oxygen atom
tunnels through the plane of the hydrogen atoms leading to the
characteristic inversion splitting of its rotational levels. 
An energy level diagram of this molecule for the lowest four
rotational levels is shown by, e.g., \cite{mam87}. The
rotation-inversions transitions are described by three quantum numbers
$J^p_K$, where the parity number $p$ takes a $+$ or $-$ value \cite[see, e.g.,][]{koz11}.
\htresomas\ has both ortho- and para- modifications: transitions with
$K$=3$n$, where $n$ is an integer $\geq$0, are from the {\it ortho} and the
rest are {\it para} variety.

\begin{figure*}[!htbp]
  \centering
 \includegraphics*[bb=110 87 786 580,width=0.475\hsize]{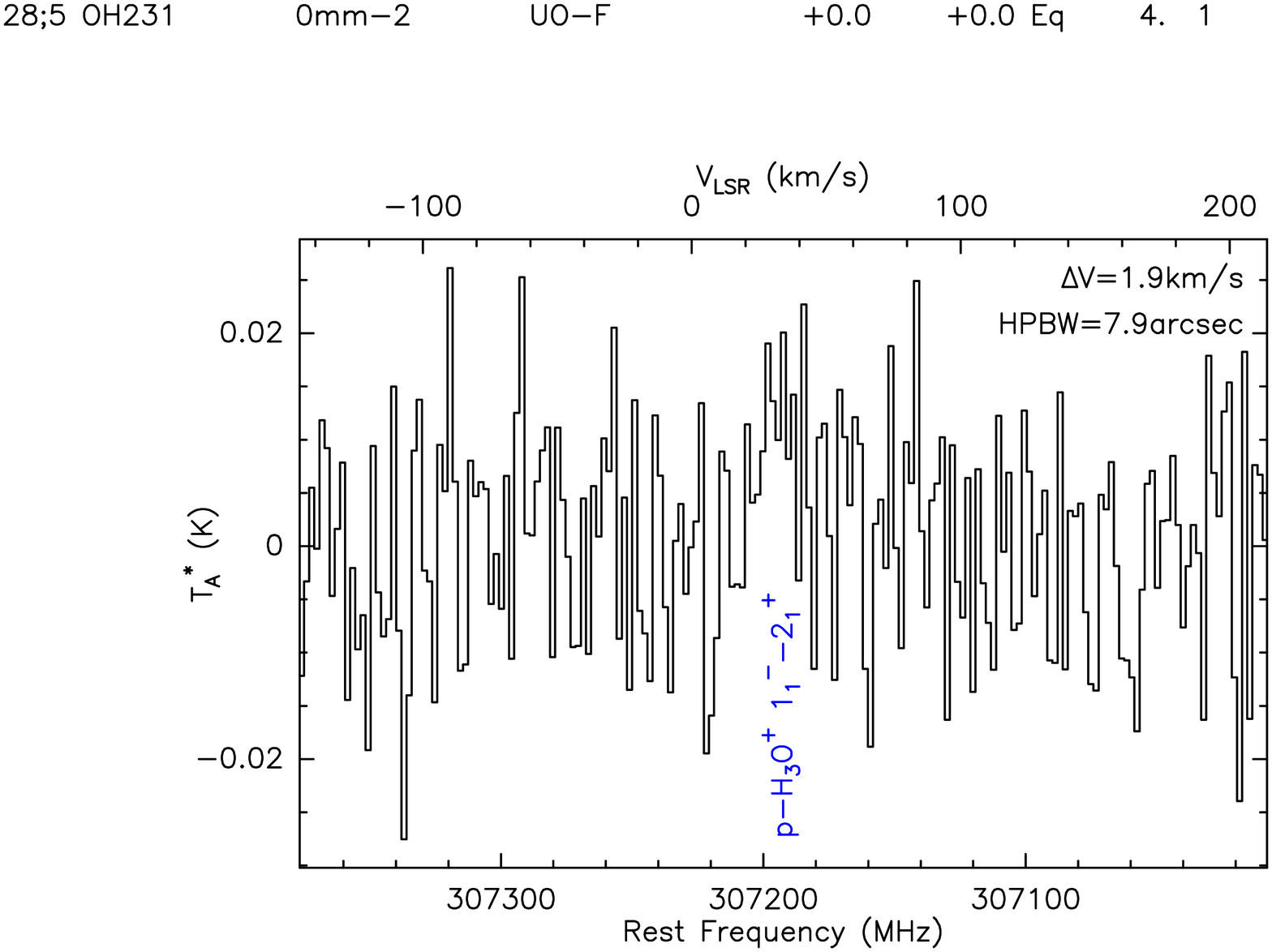}
 \includegraphics*[bb=110 87 786 580,width=0.475\hsize]{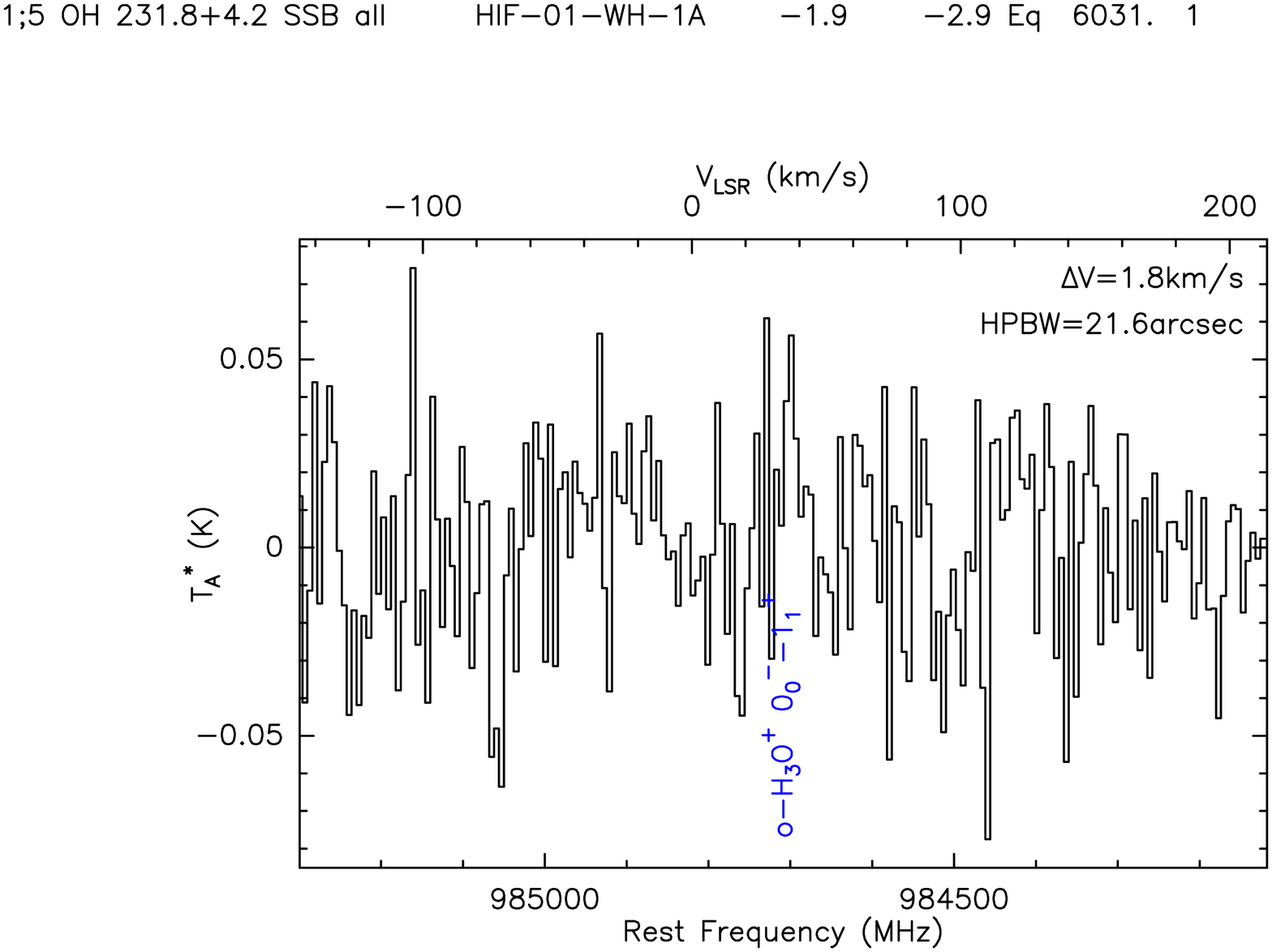}
  \caption{As in Fig.\,\ref{hco+} except for H$_3$O$^+$. Detection of the o-\htresomas\,$J^p_K$=0$^-_0$-1$^+_1$ line at 985\,GHz is tentative.}
  \label{h3o+}
  \end{figure*}

The spectrum of the p-\htresomas\,$J^p_K$=1$^-_1$-2$^+_1$ line at
307\,GHz is shown in Fig.\,\ref{h3o+}. A weak emission line is
observed centered at \vlsr=34$\pm$3\,\kms\ and with a width of
FWHM$\sim$14$\pm$6\,\kms. This profile is significantly narrower that
those of the other molecular ions described here, but
consistent with the low expansion velocities found in the central
regions of \oh\ (\S\,\ref{intro}). Although the \htresomas\ emission
is weak, it satisfies
a simple objective detection criterion: 
more than three adjacent channels of width 3.9\,\kms, corresponding to an
anticipated line-FWHM of $\ga$12\kms, 
are above 1.5$\sigma$ (this is adopting a gaussian profile with a
$\ga$3$\sigma$ level at the peak). 
Moreover, the line centroid coincides with the systemic velocity of
the source within errors, which 
reinforces the detection case. 



We observed the o-\htresomas\,$J^p_K$=0$^-_0$-1$^+_1$ line at 985\,GHz and other
transitions of this molecular ion at higher frequencies  in our survey with \hso\ toward \oh. The spectrum at
985\,GHz (shown in Fig.\,\ref{h3o+}) is consistent with weak emission
centered at \vlsr=37$\pm$3\,\kms\ and FWHM=8$\pm$5\,\kms. The line
flux measured over a 14\,\kms-wide window is 0.4($\pm$0.2)\,K\,\kms, so
we take a conservative position and consider this as a tentative
detection. Nevertheless, it is worth noting that very weak emission
from this transition is indeed expected if \htresomas\ arises in the
warm ($\approx$100\,K) envelope component that we recently discovered 
in our \hso\  survey  (see \S\,\ref{intro} and also
\S\,\ref{rotdiagr}). This warm component is found to be selectively
traced by certain molecules, in particular, by \water\ (as well as  by CS,
H$_2$S, and SiS). Since the chemistry of \htresomas\ and \water\ are
expected to be intimately related to each other, we consider the
relatively high temperature suspected from the weakness of the
\htresomas\ 985\,GHz transition (compared to the \htresomas\ line at 307\,GHz)
consistent with both species  being produced
in similar regions.

\section{Analysis}

\subsection{Column densities and excitation temperatures}
\label{rotdiagr}

We have obtained beam-averaged column densities (\ntot) and excitation
temperatures (\texc) for the different ions observed using the
standard population diagram technique  \citep[e.g.,][]{gol99}. In this
method,
the natural logarithm of the column density per statistical weight
(N$_u$/g$_u$) is plotted against the energy of the upper level above
the ground state (E$_u$) for a number of transitions of the same
molecule. Assuming that the lines are optically thin and thermalized,
i.e.,\,all levels are under local thermodynamic equilibrium (LTE)
conditions at a given unique temperature,
N$_u$/g$_u$ and \eu\ are related by the following formula:

\begin{equation}
\label{equ-rd}
\ln \frac{N_u}{g_u} = 
\ln \frac{3kW_{\rm ul}}{8\pi^3\nu_{\rm ul}\,S_{\rm ul}\mu^2} = 
\ln \frac{\ntot}{Z(\texc)} - \frac{\eu}{k\texc} 
,\end{equation}

\noindent 
where $k$ is the Boltzmann constant; W$_{\rm ul}$ is the source
brightness temperature integrated over velocity; $\nu_{\rm ul}$ and
S$_{\rm ul}$ are the frequency and line strength of the transition,
respectively; $\mu$ is the appropriate component of the permanent
dipole moment of the molecule; Z(\texc) its partition function;
and $u$ and $l$ refer to the upper and lower levels involved in the
transitions. According to eq.\,\ref{equ-rd}, for a given molecule a
straight-line fit to the points in the population diagram provides
\ntot/Z(\texc) from the $y$-axis intercept and \texc\ from the slope
of the fit. The line flux W is given by

\begin{equation} 
\label{equ-W}
W = f_{\rm d}^{-1} \eta_{\rm eff}^{-1} \intta,  
\end{equation}

\noindent 
where $\eta_{\rm eff}$ is the ratio between the main-beam efficiency
and the forward efficiency of the telescope at a given frequency
(values are given in \S\,\ref{obs}) and $f_{\rm d}$ is the dilution
correction factor. The latter is estimated as
\[ 
f_{\rm d} = 1 - \exp \{-\frac{\Omega_{\rm src}}{\Omega_{\rm MB}}\}, \hfill \left\{
\begin{array}{l l}
\Omega_{\rm src} = \frac{\pi}{4}\theta_{\rm a}\theta_{\rm b} \\
\Omega_{\rm MB} = \frac{\pi}{4\ln2} \mathrm{HPBW}^2 
\end{array} \right.\]

\noindent 
where $\Omega_{\rm src}$ and $\Omega_{\rm MB}$ are the solid angular
extent of the source and the telescope main beam, respectively. These
solid angles are expressed above as a function of the telescope beam
(HPBW) and the angular major ($\theta_{\rm a}$) and minor
($\theta_{\rm b}$) axes of a uniform elliptical source (see,
e.g.,\,Kramer 1997), which has been taken to roughly represent the
elongated molecular outflow of \oh.

Based on the angular extent of the \hcomas$J$=1--0 emission in
\oh\ deduced from interferometric mapping 
\citep{san00}; see discussion in \S\,\ref{r-hco+}, we adopt a
characteristic angular size of $\theta_{\rm a} \times \theta_{\rm b}$
$\sim$\,4\arcsec$\times$12\arcsec\ for all the species discussed here.  Although the spatial distribution of \htrececomas,
\somas, and \ndoshmas\ is unknown, their relatively broad emission
profiles and similar excitation requirements justifies our assumption.
In the case of \htresomas, the narrower line profile and possibly
higher \texc\ inferred (see below) may suggest a more compact
distribution. 
We have also considered a smaller size of the
\htresomas-emitting area of $\theta_{\rm a}\times\theta_{\rm
  b}$\,$\sim$\,4\arcsec$\times$4\arcsec, which leads to a lower \texc\
but a comparable value for \ntot. We note that the adopted angular
size affects both N$_{\rm u}$/g$_{\rm u}$ and \texc: on the one hand,
the smaller the size, the larger the value of N$_{\rm u}$/g$_{\rm u}$
obtained from the fit; on the other hand, $f_{\rm d}$ changes by a
different factor for different transitions, which translates into a
modification of the slope of the straight-line fit to the points in
the population diagram and, thus, into a different \texc. In the
particular case of \htresomas, both effects compensate to yield a
similar value of \ntot\ when adopting a smaller size of
$\sim$\,4\arcsec$\times$4\arcsec.

\begin{figure*}[!htbp]
  \centering
\includegraphics*[bb=170 20 515 690,angle=270,width=0.51\hsize]{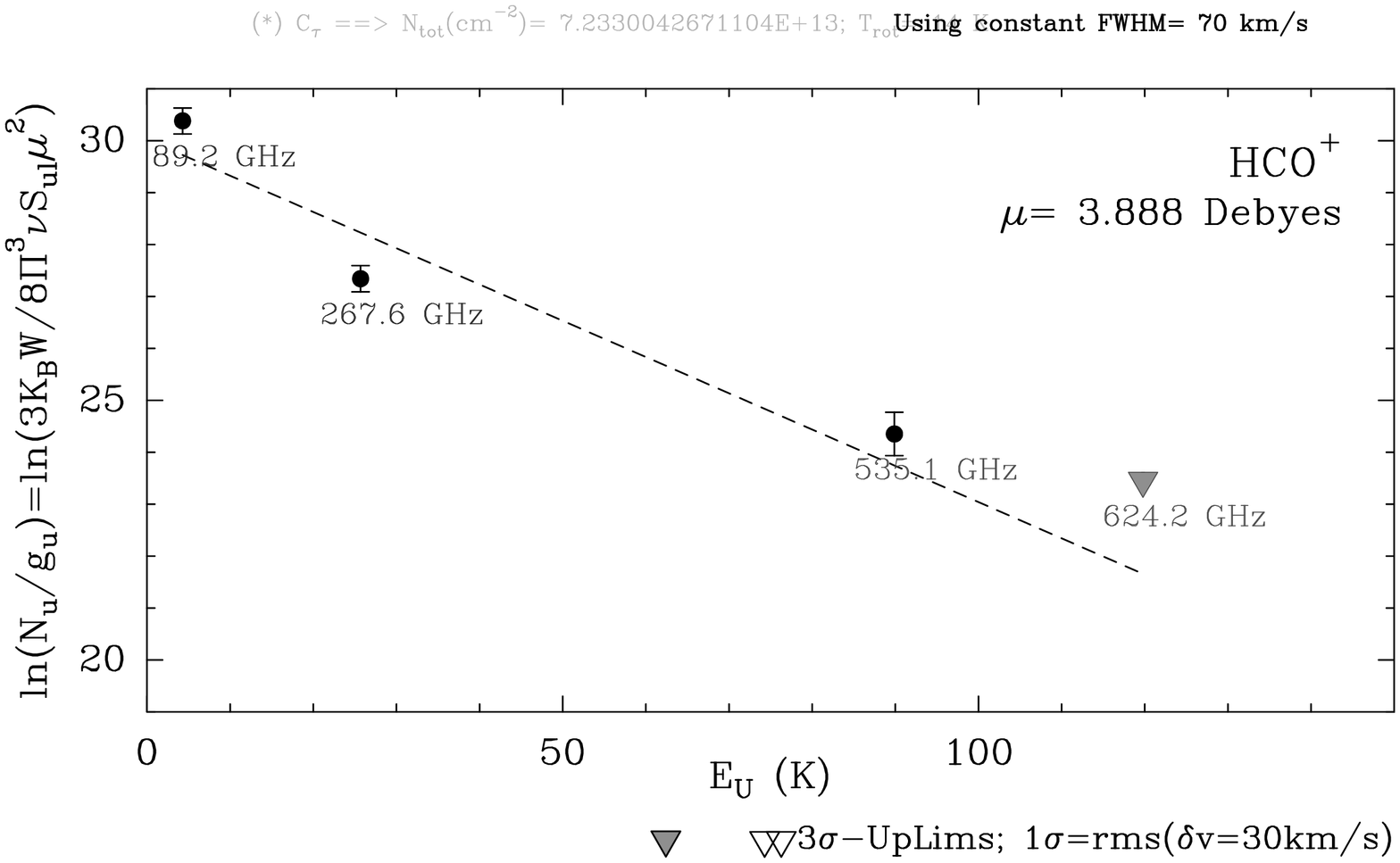}

\includegraphics*[bb=170 20 515 690,angle=270,width=0.51\hsize]{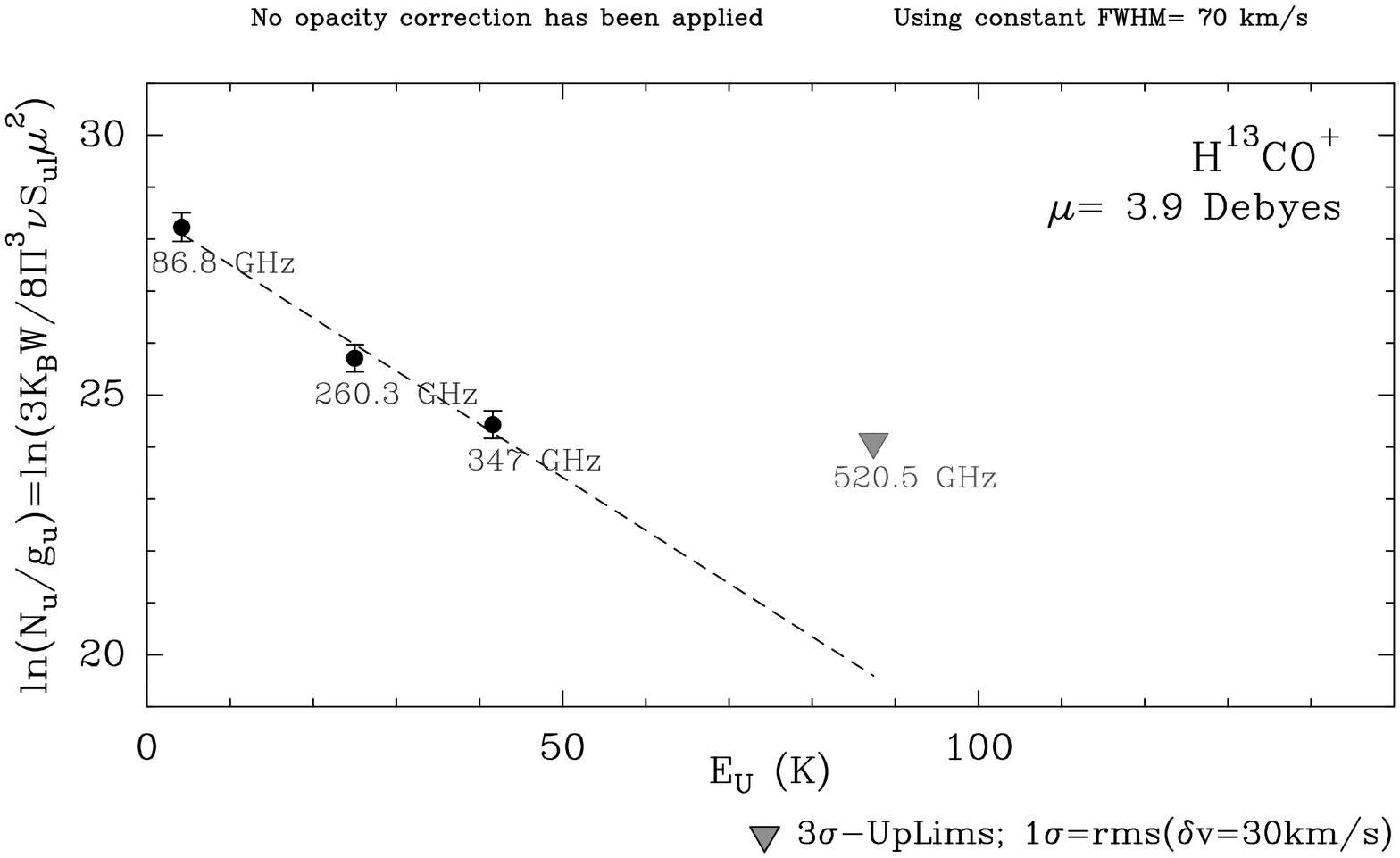}

\includegraphics*[bb=170 20 515 690,angle=270,width=0.51\hsize]{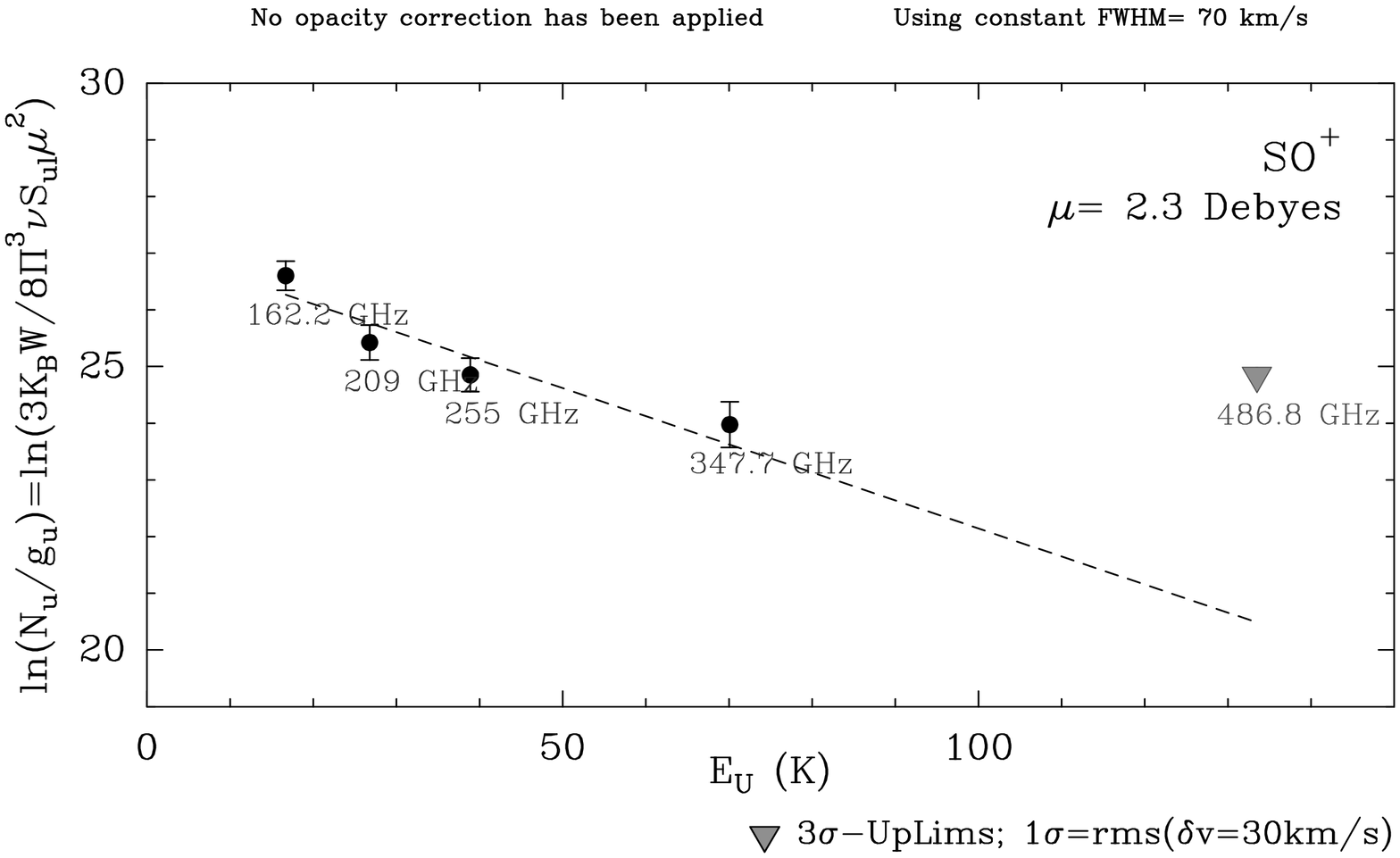}

\includegraphics*[bb=170 20 515 690,angle=270,width=0.51\hsize]{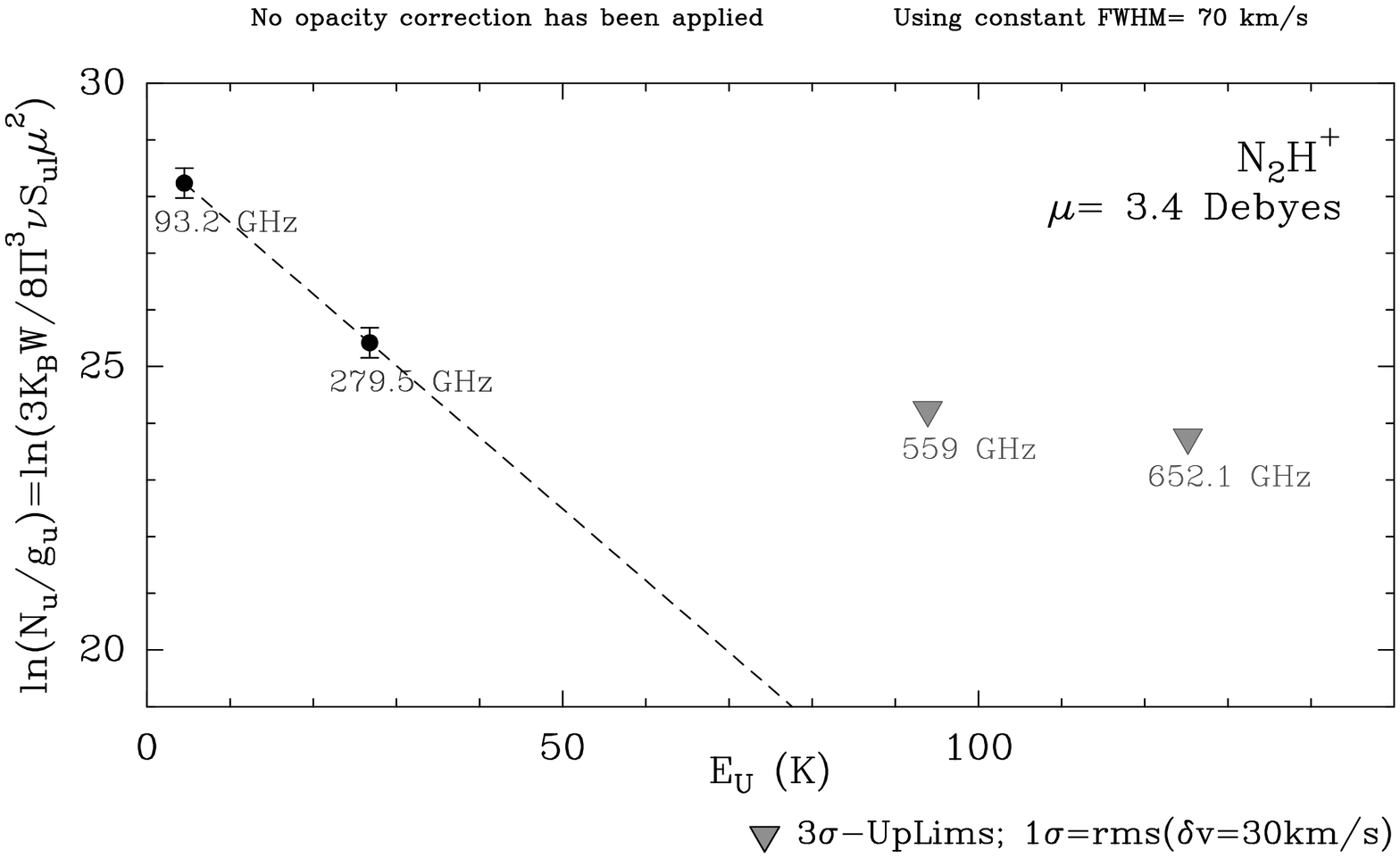}

\includegraphics*[bb=170 20 515 690,angle=270,width=0.51\hsize]{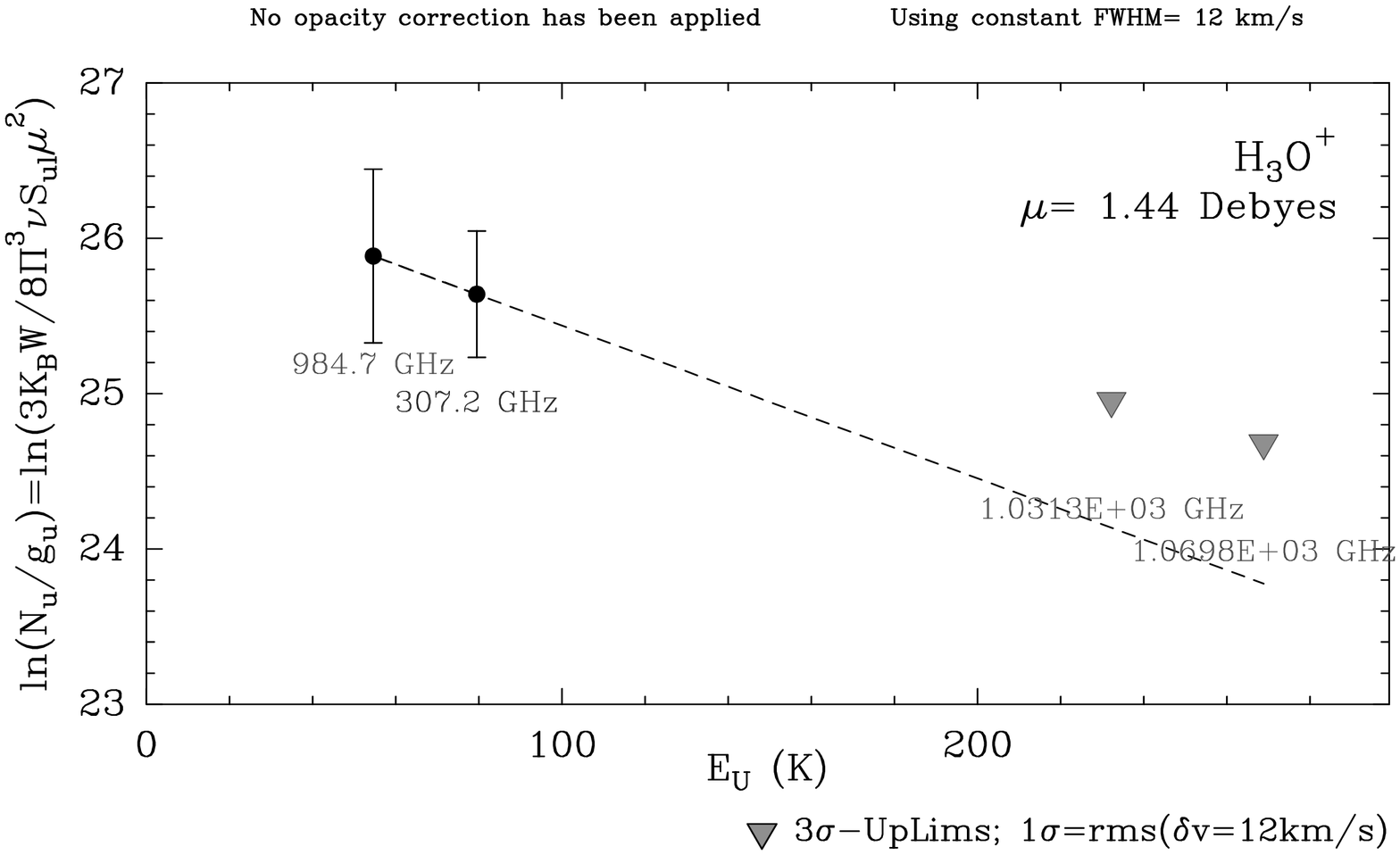}
     \caption{Population diagrams. Line fluxes (circles) and upper
       limits (triangles) are from Table\,\ref{t_res}. The error bars
       include the rms of the spectra and absolute flux
       calibration uncertainties, up to $\sim$25\%. The frequencies of
       the transitions used are indicated in GHz. The column density
       and excitation temperature obtained from the linear fit to the
       data (dashed line) are given in Table\,\ref{tabun}.
}
  \label{f-rd}
  \end{figure*}

In Fig.\,\ref{f-rd} we plot the population diagrams for the molecular ions
detected in \oh. The partition function, Z(\texc), 
has been computed for each molecule by explicit summation of a finite
number of rotational energy levels from the ground vibrational state
using MADEX \citep{madex}.  For \htresomas\ we have adopted an
ortho-to-para ratio of 1, which is the expected value in statistical
equilibrium conditions at $T$$\geq$50\,K \citep{phi92}. 
The values of \ntot\ and \texc\ obtained for the different species are
given in Table\,\ref{tabun}.  In this table, we also include results
for \trecem\ obtained in \cite{vel14} using the same technique and
assumptions. As discussed by these and other authors,
\trecem\ transitions are optically thin, or only moderately opaque at
the line peak toward the densest central parts, and are expected to
be thermalized over the bulk of the nebula, with high average
densities of $\sim$10$^5$-10$^6$\,\cm3, and even larger in
low-latitude regions closer to the nebular nucleus.

For \hcomas, \htrececomas, \somas, and \ndoshmas, the values of
\texc\ obtained are low, $\sim$8-20\,K, in agreement with previous
estimates of the kinetic temperature in the outflow, which is $\sim$10-20\,K in
the fast lobes and, somewhat larger, $\sim$10-40\,K in the
low-velocity, low-latitude component\footnote{These temperatures are
  derived from low-$J$ transitions and, thus, correspond to the external
  layers of the central component, deeper regions are expected to be
  progressively warmer closer to the central star.} (see
\S\,\ref{intro}).
The excitation temperature inferred for \htresomas, $\approx$100\,K,
is significantly larger than for the rest of the molecular ions
(Fig.\,\ref{f-rd}). This value of \texc\ is very uncertain, however,
the upper limit to the flux ratio between the 985\,GHz and the 307\,GHz 
transitions indicates temperatures certainy larger than 40\,K for the
emitting region; otherwise, the 985\,GHz line should have been much
stronger, and thus well detected,
given the relatively low energy of the upper state level of this
transition (\eu$\sim$55\,K) compared to that of the 307\,GHz line
(\eu$\sim$80\,K). Moreover, it is suggestive
that the excitation temperature guessed for \htresomas\ matches very
well that deduced for \water\ and its main isotopologues, \waterdo\ and
\waterds, from our recent survey with \hso\ 
\citep{san14}.
Considering that the chemistry of \htresomas\ is expected to be
closely linked to that of \water, the obtained result is probably more
than an unlucky coincidence. 

On top of that, the narrow profile of the
\htresomas$J^p_K$=1$^-_1$-2$^+_{1}$ line at 307\,GHz suggests that the
emission arises in the dense central parts of the nebula where the
expansion velocities are smaller and the temperatures are presumably
higher.
Support for  this comes from the fact that the narrow line width
and lowish LSR velocity we find for H$_3$O$^+$ is, remarkably, also
observed for the ($J,K)$ = (2,2) and (3,3) lines of NH$_3$ by
\cite{men95}. According to these authors, these lines' profiles are
``characterized by a narrow line width of $\leq$\,15\,\kms\ (FWHM)
and a centroid velocity around 30\,\kms", which is in excellent
agreement with the values we find for H$_3$O$^+$ (see Section
\ref{r-h3o+}). This gives, first, credence to the reality of our
H$_3$O$^+$ detection. Secondly, the NH$_3$ (2,2) and (3,3) lines
arise from levels that are at energies 64 and 123 K
 above the ground state, respectively, and from their ratio Menten \&\ Alcolea
calculate a rotation temperature of $\approx 50$~K from which they
conclude that ``the narrow component is emitted from a relatively warm
region''. By association, this supports our conjecture that the
H$_3$O$^+$ emission arises from a warm environment. In contrast, the
lower excitation NH$_3$ (1,1) line, which arises from the molecule's
inversion-split para-ground state, shows broader emission than the
(2,2) and (3,3) lines over a 73~km~s$^{-1}$ (FWZI) velocity range with
a (higher) centroid of 41~km~s$^{-1}$.

For the values of \texc\ and \ntot\ implied by the population
diagrams, all the transitions detected are optically thin
($\tau$$<$\,0.1). The largest optical depth (at the line peak)
is found for the \hcomas\,$J$=3--2 transition, for which
$\tau$$\sim$\,0.20. In this case, we applied a small opactity correction factor
of $\tau$/(1$-$$e^{-\tau}$)$\sim$1.1, resulting in an insignificant effect on
the \texc\ and \ntot\ deduced from the fit. 
The optical depth of the
\hcomas\,$J$=1--0 and 3--2 lines is only expected to be larger than 1
for column densities $\ga$4$\times$10$^{14}$\,\cm2; for
the rest of the molecular ions under discussion even larger column
densities, $\ga$10$^{15}$\,\cm2, would be necessary to produce optically
thick emission.

In the previous work by \cite{san00}, a large opacity for the
\hcomas\,$J$=1--0 transition is also ruled out based on the
\hcomas/\htrececomas ($J$=1--0) intensity ratio, which agrees well
with the low $^{12}$C/$^{13}$C isotopic ratio estimated from CO and
other molecules in the fast outflow, where the emission is known to be
optically thin (see also \S\,\ref{s-abund}).

\subsubsection{Non-LTE excitation effects}
\label{nonLTE}

When the local density of molecular hydrogen (\dens) is insufficient
to thermalize the transitions of a given molecule, departures from a
linear relation in the population diagram are expected. We 
have performed a non-LTE excitation analysis of the \hcomas\ and
\ndoshmas\ emission.
We have considered radiative and collisional processes and solved the
statistical equilibrium 
equations under the
\textup{\textup{{\sl large velocity gradient} }}(LVG) approximation using the code MADEX
\citep[][and references therein]{madex}. We have adopted collisional coefficients for
\hcomas\ and \ndoshmas\  from \cite{flo99} and
\cite{dan05}, respectively. For \somas\ and \htresomas\ there are no
collisional coefficients available in the literature, and thus non-LTE
calculations cannot be done.

\begin{figure*}[!htbp]
   \centering
   \includegraphics*[bb= 110 87 786 580,width=0.475\hsize]{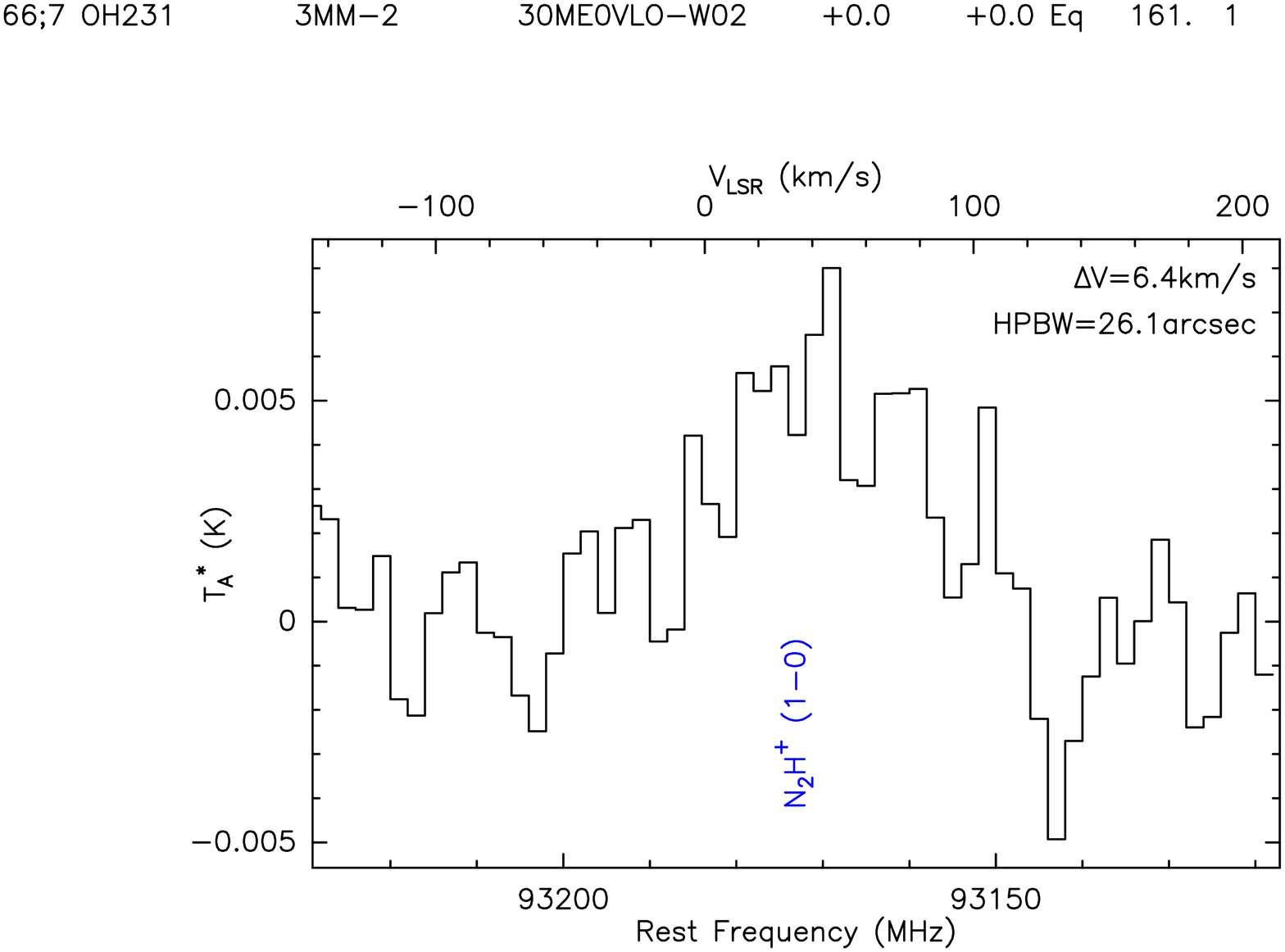}
   \includegraphics*[bb= 110 87 786 580,width=0.475\hsize]{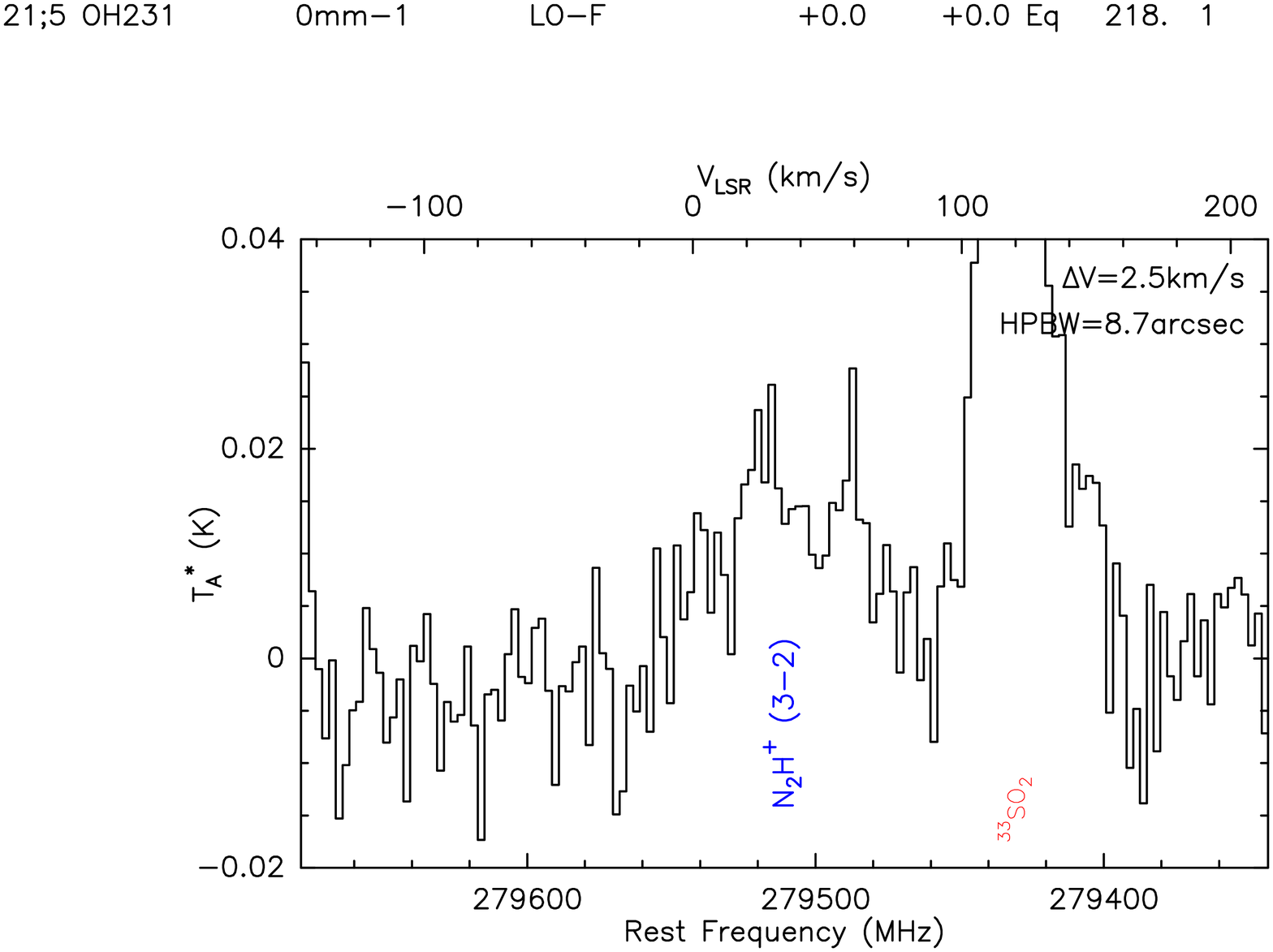}
      \caption{Same as in Fig.\,\ref{hco+} except for \ndoshmas.}
         \label{n2h+}
   \end{figure*}

The critical densities of the mm-wave transitions of \hcomas,
\htrececomas, and \ndoshmas\ observed by us are in the range
\nc$\approx$10$^5$-10$^6$\,\cm3. These values are comparable to the
mean densities in the central core component ($\ga$10$^6$-10$^7$\,\cm3) and
moderately larger that the typical densities of the high-velocity
clumps where the bulk of the wing emission arises
(\dens$\sim$10$^5$\,\cm3). Note that the lowest H$_2$ densities in
  the molecular outflow of \oh, \dens$\sim$10$^4$-10$^3$\,\cm3 are
  only found at very large distances from the star in the south lobe
  (clumps I5 and beyond; see Alcolea et al.\ 2001), but these distant,
  most tenuous and rapidly expanding regions (at \vlsr$\ga$120\,\kms)
  do not contribute to the emission profiles of the
  molecular ions under discussion, as shown in
  Figs.\,\ref{hco+}-\ref{h3o+}. 

For our non-LTE excitation calculations, we have adopted a
representative density of \dens=10$^5$\,\cm3. We find that for both
\hcomas\ and \ndoshmas, only mild LTE deviations are expected at this
density, which result in values of \texc\ and \ntot\ (as deduced from
the population diagram) slightly lower than the real \tkin\ and
\ntot. This means that the temperatures and abundances quoted
in Table\,\ref{tabun} could be lower limits. We have explored the range of input \tkin\ and \ntot\ that would still be consistent
with the observations if the characteristic density of the emitting
regions is \dens$\sim$10$^5$\,\cm3. For both \hcomas\ and \ndoshmas,
we find that the population diagram could also be reproduced with a
temperature of \tkin$\sim$30-40\,K; for these temperatures, the total
column densities would be
\ntot(\hcomas)$\sim$9$\times$10$^{13}$\,\cm2
and
\ntot(\ndoshmas)$\sim$1.5$\times$10$^{13}$\,\cm2. 
Therefore, the column densities deduced under non-LTE and LTE
conditions do not differ significantly (less than about 15\%). 



\subsection{Fractional abundances}
\label{s-abund}

We computed the fractional abundances of the ions relative to H$_2$,
$X$=\ntot/\nhdos\  as
$X$=$X$(\trecem)$\times$\ntot/\ntot(\trecem) and adopted
$X$(\trecem)=5$\times$10$^{-5}$, as estimated by \cite{morr87}. This
value of the \trecem\ abundance is in the high end of the typical
range for O-rich stars; in the case of \oh, this is consistent with the 
particularly low $^{12}$C/$^{13}$C isotopic ratio, $\sim$5-10,
measured in this object \citep[and other O-rich CSEs;][and references
  therein]{san00,mil09,ram14}. We cannot exclude a lower value of
$X$(\trecem), by a factor $\sim$2-3, in which case the abundances
derived for the rest of the species should be scaled down in the same
proportion.  The fractional abundances obtained from the
rotational-diagram method are listed in Table\,\ref{tabun}. As shown
in \S\,\ref{nonLTE}, the true values of \xmol\hcomas\ and
\xmol\ndoshmas\ could be slightly larger than those given in the table
if mild non-LTE excitation effects are present (discrepancies
$\ga$15\% are unlikely).

The average fractional abundance of \hcomas\ in \oh\ we derived,
\xmol\hcomas$\approx$10$^{-8}$, is in good agreement with previous
estimates \citep{morr87,san97,san00}. Small differences are expected
due to data calibration uncertainties and the slightly different
$X$(\trecem) abundance and excitation temperature adopted in different
works.  The average abundance of \hcomas\ in \oh\ is also within the
range of values observationally determined on a small sample of O-rich
envelopes, $\sim$0.15-1.3$\times$10$^{-7}$ \citep{pul11}. The
abundance of \htrececomas, and its comparison with that of \hcomas, is
consistent with a low $^{12}$C/$^{13}$C isotope ratio of $\sim$7, in
agreement with previous estimates for this object (see above).

The abundances of the other ions we presented have not been
previously reported in O-rich CSEs. The fractional abundance of
\somas\ in \oh\ implies an abundance relative to SO of
$X$(\somas)/$X$(SO)$\approx$2$\times$10$^{-3}$; this is taking into
account the fractional abundance of SO in \oh\ computed in other works,
$X$(SO)$\sim$[1-3]$\times$10$^{-6}$ \cite[][and this survey, S\'anchez
  Contreras et al., in prep.]{gui86,morr87,san00}.
For \ndoshmas, our abundance estimate is consistent with the upper
limit provided by \cite{morr87} based on the nondetection of the
\mbox{\ndoshmas\,$J$=1--0} line by these authors
($X$(\ndoshmas)$<$1.6$\times$10$^{-8}$). We deduce an average
\hcomas\ abundance relative to \ndoshmas\ of
$X$(\hcomas)/$X$(\ndoshmas)$\sim$6; as we will see in
\S\,\ref{model}, 
the previous ratio may represent an
upper limit to the [CO]/[N$_2$] proportion in \oh.
For \htresomas, we obtain a fractional abundance of
$\approx$7$\times$10$^{-9}$, although we caution that this value is
particularly uncertain given the low S/N of the \htresomas\ spectra.
Our estimate is consistent with the range of values measured in
interstellar clouds, $X$(\htresomas)$\approx$10$^{-10}$-10$^{-9}$
\citep{woo91,phi92,goi01}.
In these interstellar regions, the fractional abundance of
\htresomas\ is expected to relate to that of water by a factor
$\approx$1/1000-1/6000, based on gas-phase model
predictions and observations.
If this proportion also holds in \oh, then one would infer an
order-of-magnitude water content of
$X$(\water)$\approx$10$^{-5}$-10$^{-6}$.
This value is at the low edge of the water abundance range deduced for
a sample of six evolved O-rich stars,
$X$(\water)$\sim$[10$^{-3}$-2$\times$10$^{-6}$], with the majority of
the stars having $X$(\water)$\sim$6$\times$10$^{-4}$ \citep{mae08}. 
\begin{table}
\caption{Results from the rotational diagram analysis
  (\S\,\ref{rotdiagr}) and molecular fractional abundances, $X$,
  i.e.,\,relative to H$_2$, derived (\S\,\ref{s-abund}). We have
  adopted $X$(\trecem)=5$\times$10$^{-5}$ \citep{morr87}. 
}
\label{tabun}      
\centering                          
\begin{tabular}{l c c c c}        
\hline\hline                  
Species  & $\theta_{\rm a} \times \theta_{\rm b}$ & \ntot & \texc & $X$ \\
         &  ('')$\times$('')    &  (cm$^{-2}$)       & (K)             &       \\
\hline                        
\trecem      &  4$\times$12  &  3.0$\times$10$^{17}$ & \hfill  20  & 5.0$\times$10$^{-5}$\\ 
\hdocecomas  &  4$\times$12  &  8.0$\times$10$^{13}$ & \hfill  14  & 1.3$\times$10$^{-8}$ \\ 
\htrececomas &  4$\times$12  &  1.2$\times$10$^{13}$ & \hfill  10  & 2.0$\times$10$^{-9}$ \\ 
\somas       &  4$\times$12  &  2.2$\times$10$^{13}$ & \hfill  20  & 3.7$\times$10$^{-9}$ \\
\ndoshmas    &  4$\times$12  &  1.3$\times$10$^{13}$ &  \hfill  8   & 2.2$\times$10$^{-9}$ \\   
\multirow{2}{*}{\htresomas} &  4$\times$12  & $\sim$4$\times$10$^{13}$ & \hfill $\approx$100 & \multirow{2}{*}{$\sim$7$\times$10$^{-9}$} \\
 &  4$\times$4  &  $\sim$5$\times$10$^{13}$ & \hfill $\sim$55 & \\
\hline                                   
\end{tabular}
\end{table}

We recall that the abundances given in Table\,\ref{tabun} represent
average values within the emitting regions covered by the telescope
beam, that is, the central core and (partially) the fast lobes. The
large wing-to-core intensity ratio of the \hcomas\ profiles is known
to reflect an abundance contrast of a factor $\sim$3-4 between the
fast lobes and the slow central clump I3, as determined by
\cite{san97} from their analysis by individual spectral and spatial
components of the \mbox{\hcomas\,$J$=1--0} emission maps; see their
Table 2.  This result is also corroborated by the interferometric
observations of this transition (Fig.\,\ref{mapa}), which show that,
in contrast to \docem, \trecem, and most molecules, \hcomas\ is found
in abundance in the lobes of \oh,\ but it is scarce at the nebula
center \citep{san00}.
 
\begin{figure}[htbp!]
  \centering
  \includegraphics[width=0.75\hsize]{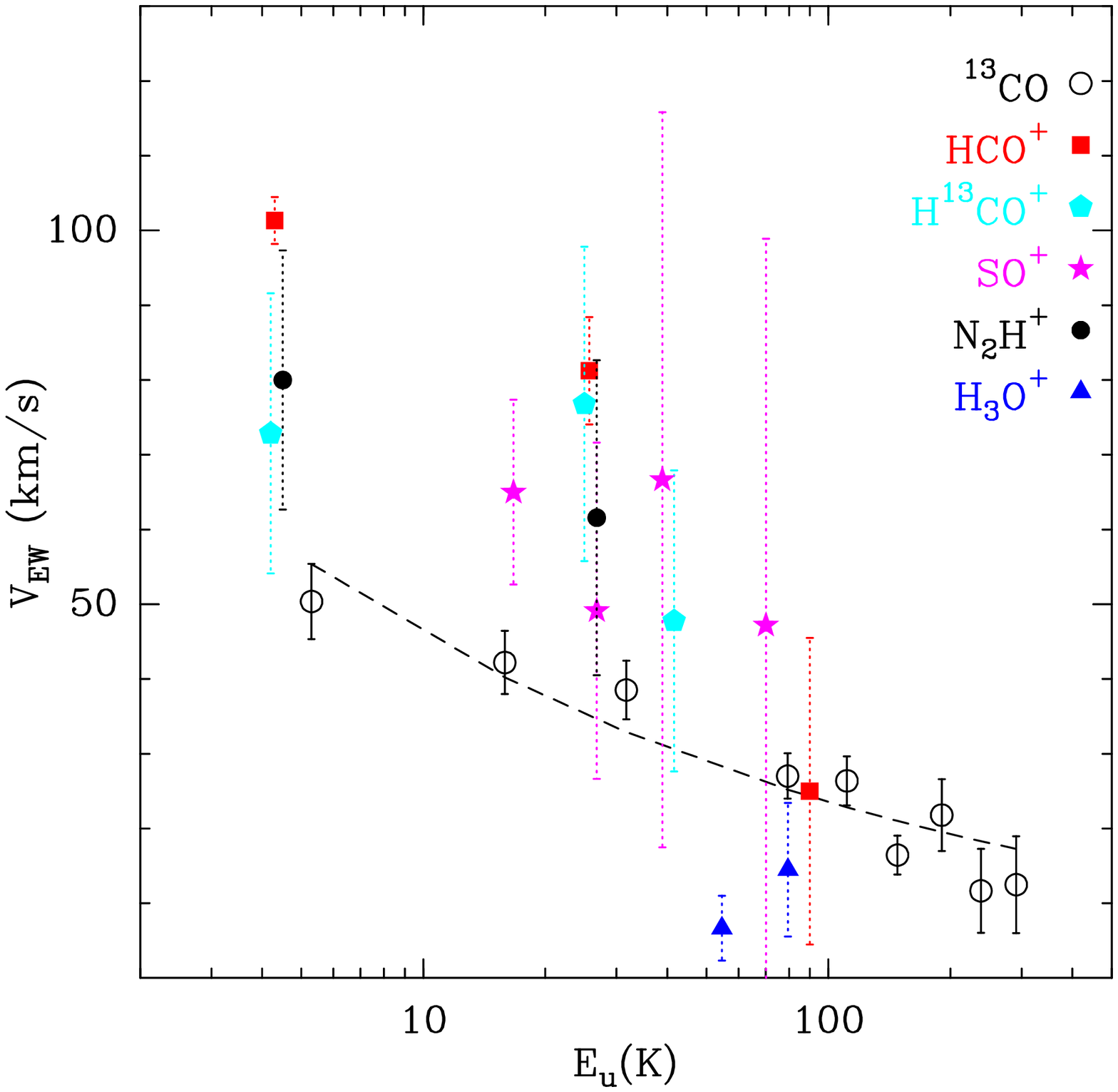}
  \caption{Plot of the equivalent-width velocity ($V_{\rm EW}$) versus
    upper-level energy (\eu) for \mions\ transitions (filled symbols)
    and \trecem, for comparison (empty circles; \trecem\ profiles and
    line parameters are given in the Appendix). The dashed line is a fit to
    the $V_{\rm EW}$ of the \trecem\ lines. For all 
    ions, except for \htresomas, the values of $V_{\rm EW}$ are systematically larger than for
    \trecem,\ indicating a larger wing-to-core emission contribution to
    the observed profile; the \htresomas\ emission, in contrast, is
    dominated by the narrow, core emission component as indicated by
    the small $V_{\rm EW}$.}
  \label{EWs}
  \end{figure}

The broad profiles of \htrececomas, \somas, and \ndoshmas\ suggest
analogous lobe-to-core abundance enhancements for these
ions. Unfortunately, except for the \mbox{\hcomas\,$J$=1--0}
transition, obtaining column densities separately for the different
velocity ranges corresponding to the slow core and the fast lobes is
hampered by the weakness of the molecular lines. However, a
straightforward
argument in support of a lobe-to-core abundance ratio $>$1 for
\htrececomas, \somas, and \ndoshmas\ (as well as, of course, for
\hcomas) is provided by the large equivalent-width velocity ($V_{\rm
  EW}$) of these lines; the equivalent-width velocity is given by the
ratio of the line flux, \intta, to the peak intensity,
\ta\ \citep[see, e.g.,][]{san12}. As can be seen in Fig.\,\ref{EWs},
the $V_{\rm EW}$ of the emission lines by these species are
systematically larger than those of \trecem. The \trecem\ profiles are
dominated by the most massive component, that is, the slow central
core (clump I3). The larger $V_{\rm EW}$ of the \htrececomas, \somas,
and \ndoshmas\ profiles indicates that the contribution to the
emitting profile from the less massive fast lobes is larger for these
ions than for \trecem, implying that their fractional abundances in
the fast lobes are higher than in the slow massive clump I3. This also
indicates that the true abundances of \hcomas, \htrececomas, \somas,
and \ndoshmas\ in the lobes and in the slow central clump are higher
and lower, respectively, than the beam-averaged values given in
Table\,\ref{tabun}.

In the case of \htresomas, the narrow profile of the weak
$J^p_K$=1$^-_1$-2$^+_1$ transition at 307\,GHz suggests that the bulk
of the emission is produced in the slow central clump I3.  The $V_{\rm EW}$ of
this transition is very uncertain but if it is confirmed to be lower
than that of \trecem\ (by higher S/N observations), as suggested by
our data (Fig.\ref{EWs}), then one would infer a larger abundance of
this ion in the slow core than in the fast lobes (i.e.,\,a lobe-to-core
abundance ratio $<$1).  Moreover, since the higher \texc\ suspected
for \htresomas\ compared to that deduced for \trecem\ favors the
origin of the \htresomas\ emission in envelope layers closer to
the center, the average abundance in Table\,\ref{tabun} could represent
a lower limit to the true value in the central, compact emitting regions. 


\subsubsection{Upper limits to nondetections}
We provide upper limits to the column densities and fractional
abundances of other ions that are not detected in \oh\ and that may be
of interest (Table\,\ref{t_UpLim}). We have adopted common values for
the source emitting size, $\theta_{\rm a} \times \theta_{\rm b}$
$\sim$\,4\arcsec$\times$12\arcsec, \texc=15\,K, and FWHM=40\,\kms\ and
LTE conditions. For the ions considered, adopting a smaller size,
larger excitation temperatures, and narrower profiles (as may be the
case of \htresomas) normally results in slightly smaller values of $N$
and $X$ than those given in the Table.  For the different ions, the
upper limits to \ntot\ are derived from the rms noise around the
transition that is expected to have the largest S/N in our data
(adopting \texc=15\,K); we have always checked that the upper limits
obtained are also consistent with the nondetections of the rest of
the transitions in the frequency range of our observations. 

%
\begin{table}
\tiny 
\caption{Upper limits to the column densities and fractional abundances of other ions in \oh\ adopting 
a 4\arcsec$\times$12\arcsec-sized outflow, \texc=15\,K, and FWHM=40\,\kms.}
\label{t_UpLim}      
\centering                          
\begin{tabular}{l l c | l l c}          
\hline\hline                     
Ion & $N$(\cm2) & $X$ & Ion & $N$(\cm2) & $X$ \\
\hline 
OH$^+$       & <3\ex{13} & <5\ex{-9}  & HOC$^+$      & <1\ex{12} & <2\ex{-10} \\ 
H$_2$O$^+$   & <2\ex{13} & <3\ex{-9}  & H$_2$COH$^+$ & <1\ex{13} & <2\ex{-9}  \\ 
CO$^+$       & <2\ex{12} & <3\ex{-10} & H$_2$NCO$^+$ & <5\ex{12} & <8\ex{-10} \\   
HCS$^+$      & <8\ex{12} & <1\ex{-9}  & HSCO$^+$     & <7\ex{13} & <1\ex{-8}  \\ 
SH$^+$       & <2\ex{13} & <3\ex{-9}  & HOCS$^+$     & <4\ex{13} & <7\ex{-9}  \\ 
CS$^+$       & <5\ex{13} & <8\ex{-9}  & HC$_3$NH$^+$ & <4\ex{13} & <7\ex{-9}  \\ 
CH$^+$       & <1\ex{13} & <2\ex{-9}  & HOCO$^+$     & <7\ex{12} & <1\ex{-9} \\  
HCNH$^+$     & <6\ex{13} & <1\ex{-8}  & H$_2$Cl$^+$  & <4\ex{13} & <8\ex{-9}  \\
NO$^+$       & <4\ex{13} & <7\ex{-9}  & NeH$^+$      & <1\ex{13} & <2\ex{-9}  \\ 
CF$^+$       & <8\ex{12} & <1\ex{-9}  & H$_2$F$^+$   & <2\ex{13} & <3\ex{-9}  \\ 
C$_3$H$^+$   & <2\ex{13} & <3\ex{-9}  & ArH$^+$      & <5\ex{12} & <8\ex{-10} \\  
& & & & & \\ 
OH$^-$       & <3\ex{13} & <5\ex{-9}  & NCO$^-$      & <2\ex{13} & <3\ex{-9}  \\ 
C$_2$H$^-$   & <1\ex{12} & <2\ex{-10} & C$_4$H$^-$   & <2\ex{12} & <3\ex{-10} \\   
CN$^-$       & <1\ex{13} & <2\ex{-9}  & C$_3$N$^-$   & <1\ex{13} & <2\ex{-9}  \\ 
SH$^-$       & <3\ex{14} & <5\ex{-8}  & & & \\  
\hline                                  
\end{tabular}
\end{table}




\section{Formation of \hcomas, \somas, \ndoshmas, and \htresomas}
\label{formation}

\subsection{Chemical model}
\label{model}
We  performed chemical kinetics models to investigate the
formation of the observed ions in \oh.
Our model is based on that by \cite{agu06} and we have used it recently  to study the new N-bearing species detected in \oh\ by
\cite{vel14}. The code has also been  employed to model the chemistry
in different astrophysical environments, including the prototypical 
C-rich star IRC+10216 \citep[see,
  e.g.,][]{agun07,agun08,agu10,agu12,cer10,cer13}, and the O-rich
yellow hypergiant IRC+10420 \citep{qui13}. The chemical network in our
code includes gas-phase reactions, cosmic rays, and photoreactions
with interstellar UV photons; it does not incorporate reactions
involving dust grains, X-rays or shocks.  Our code does not explicitly consider
 the chemistry of isotopologues, that is, does not include
isotopic fractionation, such as selective photodissociation or
isotope-exchange reactions; as a consequence, predictions for the
\htrececomas\ abundance distribution are not made.


As input we adopt the same physical model for the envelope used in
\cite{vel14}, which consists of two main components: $a$) a dense,
spherically symmetric wind expanding at low velocity, which represents
the slow, central nebular component of \oh\ (clump I3); and $b$) a
rectangular slab of gas (plane-parallel geometry) with characteristics
similar to those of the walls of the hollow bipolar lobes.  We modeled separately these two
components, which are externally illuminated by the interstellar UV
field.

The main parameters of the simple physical model for \oh\ used as
input in our chemical code are summarized in Table\,\ref{tab:oh231};
additional details justifying the temperature, density, and velocity
laws adopted are given in \S\,6.1 of \cite{vel14}.  The region of
the envelope modeled in case $a$ goes from \rin$\sim$10$^{15}$\,cm
($\sim$20\,\rs) to its end, at \rout$\sim$7$\times$10$^{16}$\,cm (we
refer to this region as the \ioe). The outer radius is observationally
determined \citep{san97,alc01} and the inner radius has been chosen to
be well beyond the dust condensation radius
so that the full expansion velocity of the gas has been reached. In
these intermediate/outer regions of the envelope, the chemistry is
driven by chemical kinetics. 
Another major input of our chemistry model is the set of initial
abundances of the 'parent' species. These are formed in deeper layers
and are injected to the intermediate/outer envelope. The parent
species, typical of O-rich environments, and the initial abundances
adopted in our model are as in \cite{vel14} and are
reproduced again in Table\,\ref{t_parent}.


As in \cite{vel14}, the sources of ionization and dissociation adopted
in our model are cosmic-rays and the interstellar ultraviolet
radiation field. The cosmic ray ionization rate adopted is
1.2$\times$10$^{-17}$\,s$^{-1}$ \citep{dal06}.  The intensity of the
UV field assumed is the Draine field or G$_0$=1.7 in units of the Habing
field \citep[G$_0$=1.6\ex{-3}\,erg\,s$^{-1}$\,\cm2;][]{hab68,dra78}.
We computed the dust optical extinction, $A_V$, across the envelope
layers adopting the standard conversion from H$_2$ column density
$N_{H_2}$=9.3$\times$10$^{20}$$\times$$A_V$\,\cm2, derived by
\cite{boh78}. The gas-to-dust mass ratio implicit in this conversion,
$\sim$100, is comparable within errors to the average value measured
in nearby O-rich AGB stars and, in particular, in
\oh\ \citep[e.g.,][]{kna85,kas95,san98}.


\begin{table} 
\caption{Parameters of the central Mira star and molecular envelope of
  \oh\, used for the chemical kinetics models (\S\,\ref{model}.)}
\label{tab:oh231}
\small 
\centering    
\begin{tabular}{l l l}
\hline\hline
Parameter & Value & Reference      \\ 
\hline   
Distance ($d$)                             &  1500\,pc                                   & b \\
Stellar radius (R$_{*}$)                   &  4.4x10$^{13}$\,cm                           & g  \\ 
Stellar effective temperature (T$_{*}$)    &  2300\,K                                     & g  \\ 
Stellar luminosity (L$_{*}$)               &  10$^{4}$\,\ls                               & g  \\
Stellar mass (M$_{*}$)                     &  1\,\msun                                   & g  \\ 
\vspace{0.1cm} \\
\multicolumn{2}{c}{\sl model $a$: slow, core} & \\
AGB CSE expansion velocity (\vexp)       &  20\,\kms                            &  a,e,f,g,i \\
AGB mass loss rate (\mloss)                    &  10$^{-4}$\,\my                              &  e,a,f \\
Kinetic temperature (\tkin) &  T$_*(r/R_{*})^{-0.70}$\,K &  i \\
\vspace{0.1cm} \\
\multicolumn{2}{c}{\sl model $b$: lobe walls} & \\
Wall thickness  & 1\arcsec\ & a \\
H$_2$ number density (\dens) & 10$^5$\,\cm3 & a \\  
Kinetic temperature (\tkin) & 20\,K & a \\ 
\hline                        
\hline                                  
\end{tabular}
\tablefoot{a: \cite{alc01}, b: \cite{choi12}, c: \cite{coh81}, d: \cite{kas92}, e: \cite{morr87},
f: \cite{san97}, g: \cite{san02}, h: \cite{san04}, i: \cite{vel14}.  }
\end{table}
\begin{table}[ht!] 
\caption{Initial abundances relative to H$_2$ for representative elements 
and parent molecules used as input for the chemical kinetic models.}
\label{t_parent}
\centering    
\begin{tabular}{c c c}
\hline\hline
Species & Abundance & Reference \\
\hline                        
He      &  0.17                & a   \\
H$_2$O  &  3.0$\times$10$^{-4}$ & b,TE  \\
CO      &  3.0$\times$10$^{-4}$ & c,TE  \\
CO$_2$  &  3.0$\times$10$^{-7}$ & d  \\
NH$_3$  &  4.0$\times$10$^{-6}$ & e  \\
N$_2$   &  4.0$\times$10$^{-5}$ & TE \\
HCN     &  2.0$\times$10$^{-7}$ & f,g\\ 
H$_2$S  &  7.0$\times$10$^{-8}$ & h  \\
SO      &  9.3$\times$10$^{-7}$ & f  \\
SiO     &  1.0$\times$10$^{-6}$ & i  \\
SiS     &  2.7$\times$10$^{-7}$ & j \\                                                  
\hline                                  
\end{tabular}
\tablefoot{a:\,\cite{asp09}, b:\,\cite{mae08}, c:\,\cite{tey06}, d:\,\cite{tsu97}, e:\,\cite{men95}, f:\,\cite{buj94}, g:\,\cite{sch13}, h:\,\cite{ziu07}, i:\,\cite{gon03}, j:\,\cite{sch07} 
TE: From LTE chemical calculations by \cite{vel14}). 
}
\end{table}

\begin{figure*}[]
  \centering
\includegraphics*[bb=20 20 400 540, width=0.45\hsize]{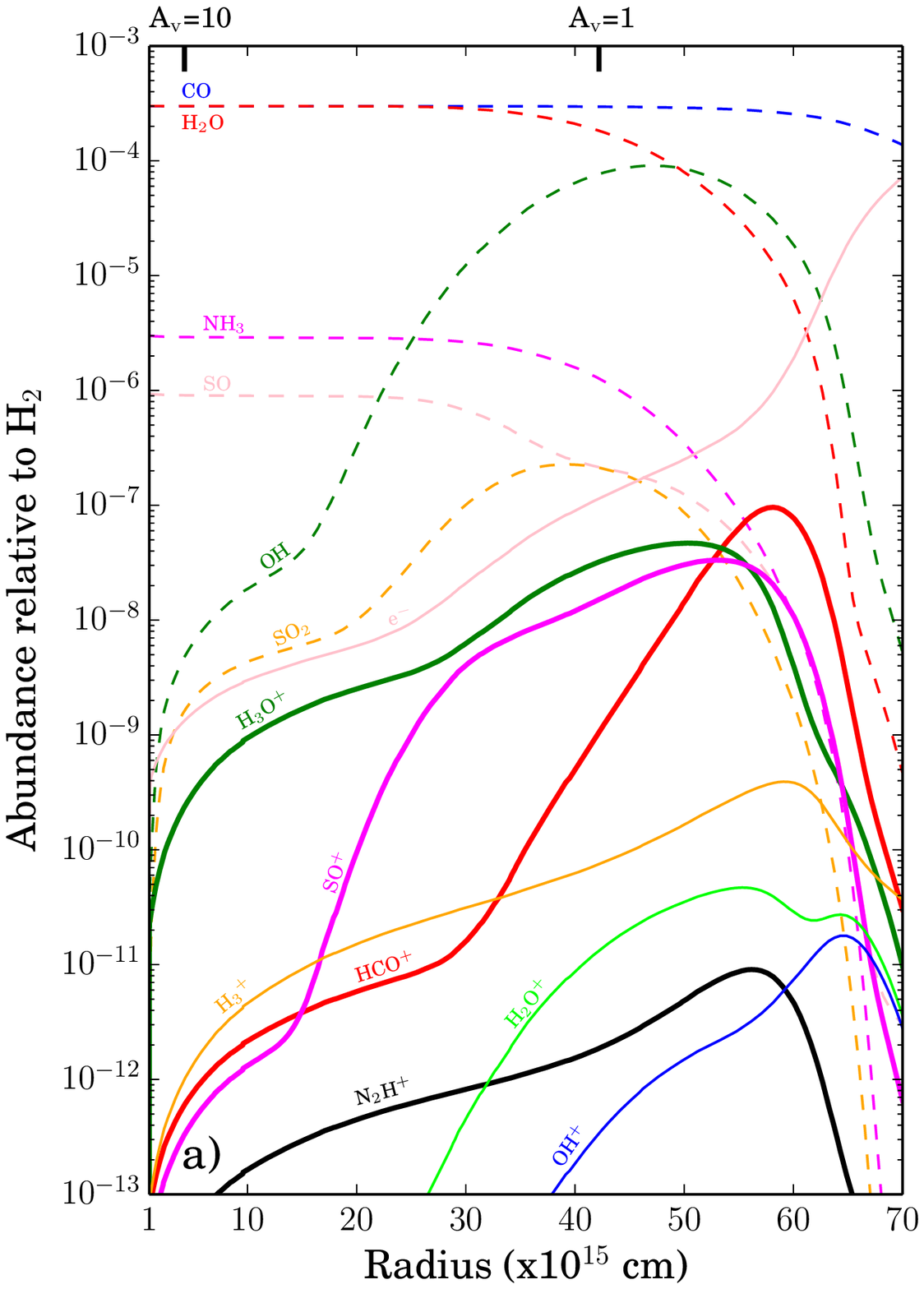} 
\includegraphics*[bb=20 20 400 540, width=0.45\hsize]{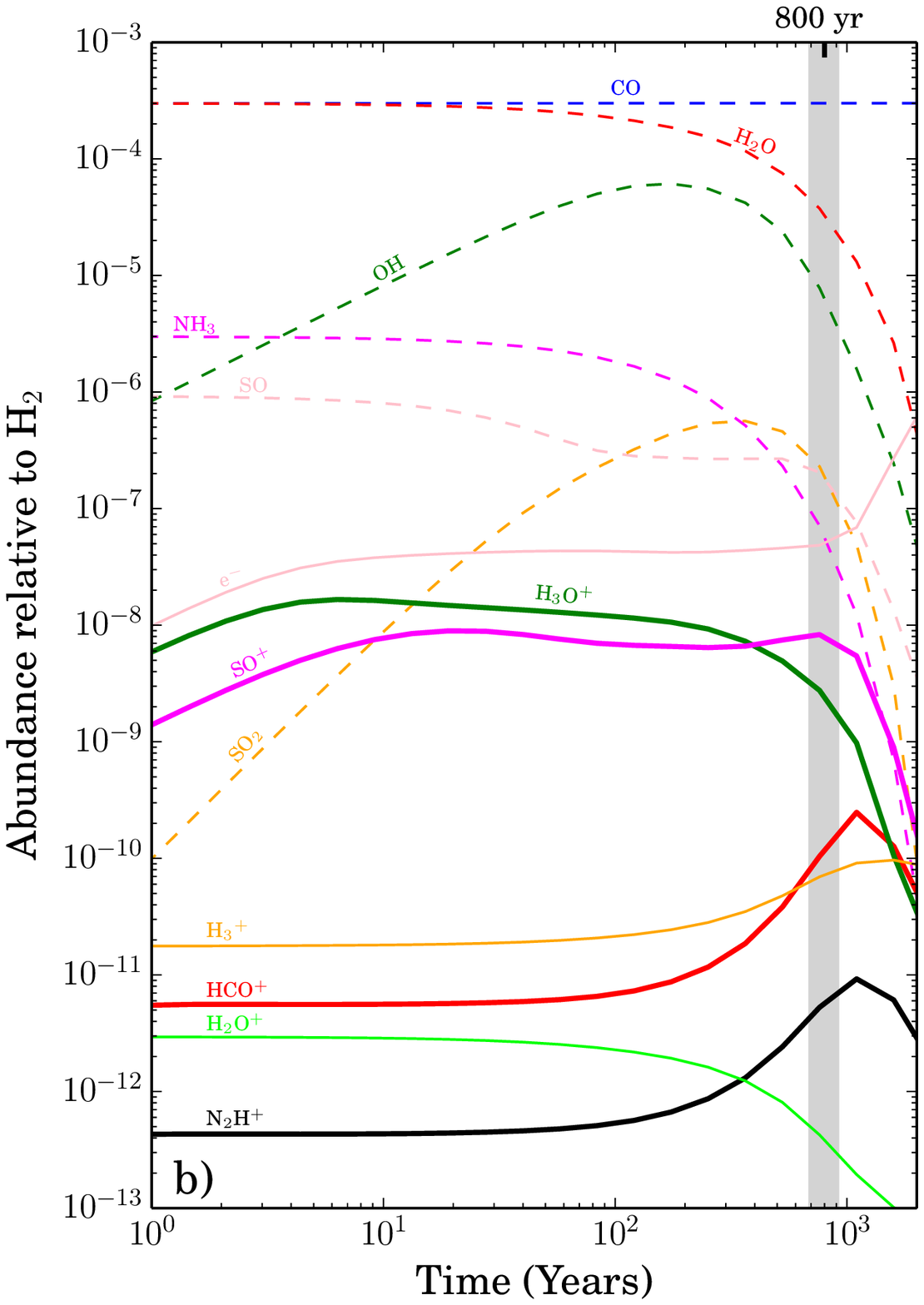}

\includegraphics*[angle=270, width=0.65\hsize]{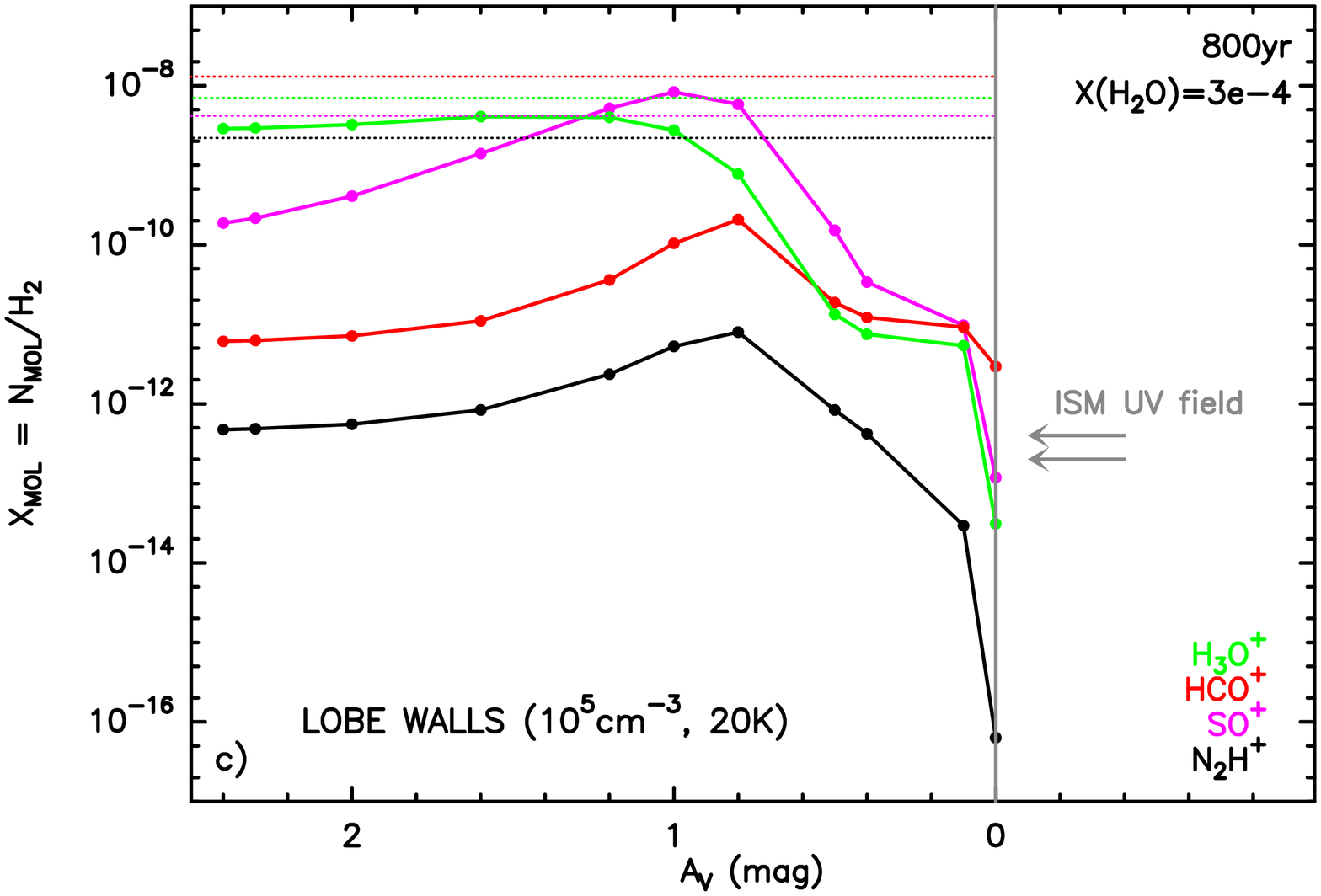}
     \caption{Results from our chemical kinetics model adopting a
       fractional water abundance of $X$(\water)=3$\times$10$^{-4}$
       (\S\,\ref{model}).  {\bf a)} Spatial distribution of the model
       fractional abundances in the slow central core of the envelope
       (fossil remnant of the AGB CSE); {\bf b)} evolution with time
       of the model fractional abundances in a
       representative gas slab or cell within the lobe walls
       (approximately in the midpoint between the outer and inner end
       of the wall) where the extinction is A$_V$ $\sim$1\,mag; {\bf
         c)} model fractional abundances as a function of the depth
       (expressed as A$_V$) into the lobe walls, at
       $\sim$800\,yr. Dotted horizontal lines in panel c) represent
       the observed beam-averaged values of our estimated fractional abundances. }
  \label{f-mod1}
  \end{figure*}

{\bf The slow, central component. }  We used our chemical kinetics
model  first to investigate the formation of \hcomas,
\somas, \ndoshmas, and \htresomas\ in an O-rich AGB CSE similar to the
slow, core component of \oh\ (model $a$, Table\,\ref{tab:oh231}).
The fractional abundances predicted by the model as a function of the
distance to the center are shown in Fig.\,\ref{f-mod1}.  For \hcomas,
\ndoshmas, and \htresomas, we  checked that there is good 
agreement between our predictions and those obtained from analogous
photochemistry simulations by \cite{mam87}; these authors modeled the
chemistry of these three ions for O-rich CSEs with standard properties and a
set of different mass-loss rates, including \mloss=10$^{-4}$\,\my.  We found no
previous modeling attempts for \somas\ in circumstellar envelopes
 in the literature.

As the gas in the envelope expands, parent molecules start to be
exposed to the interstellar UV radiation and photochemistry drives the
formation of new species in the outer layers. Penetration of photons
through the envelope is gradually blocked by dust and also by
self-shielding of abundant gas species (mainly H$_2$, \water\ and
CO). At the deepest layers of the \ioe\ inaccessible
to the UV radiation, cosmic rays are major drivers of the
chemistry. Cosmic rays ionize molecular hydrogen (H$_2^+$) , which
combines with neutral H$_2$ to form \htresmas. The latter is a
fundamental ion for breaking down parent molecules leading to the
formation of new ions in the inner layers of the envelope.

It is well known that the radical OH is an important product of the 
photodissociation of \water\ in the external 
regions of O-rich CSEs. Our model predicts that OH reaches maximum
abundance at $\sim$5$\times$10$^{16}$\,cm, in good agreement
with the characteristic radius of the OH maser toroidal shell observed at the
center of \oh\ \citep{morr82,zij01}. Our
model indicates that, as the radical OH, molecular ions should be most
efficiently formed in the outermost layers of the envelope, 
showing a shell-like distribution with maximum abundances at typical distances
from the star of $\sim$[5-7]$\times$10$^{16}$\,cm. Farther out, there
is an abrupt fall of the ions' abundances due to their destruction by
dissociative recombination with electrons ($e^-$) and to photodissociation
of their main progenitors.


According to our model, \hcomas\ forms in the outer layers of O-rich
envelopes mainly as a photodissociation product of CO and \water\ (in
order of importance):

\begin{equation}
\begin{array}{l l}
\mathrm{HOC^+ + H_2 \raw HCO^+ + H_2} & \\ 
\mathrm{CO^+ + H_2 \raw  HCO^+ + H} & \\ 
\mathrm{C^+ + H_2O \raw  HCO^+ + H} & \\
\label{eq-hco+}
\end{array}
,\end{equation}

\noindent
where the C$^+$ ion is predominantly produced in the CO photoionization
chain, and the molecular ions HOC$^+$ and CO$^+$ are
formed through

\begin{equation}
\begin{array}{l l}
\mathrm{C^+ + OH \raw  CO^+ + H} & \\
\mathrm{CO^+ + H_2 \raw  HOC^+ + H} & \\ 
\end{array}
.\end{equation}

An additional, but nondominant pathway to \hcomas\ production is by
proton transfer from \ndoshmas\ to CO (\ndoshmas\ + CO $\raw$
\hcomas\ + N$_2$). The model predicts that \hcomas\ reaches highest
abundances in a very thin outer shell with a characteristic radius of 
$\sim$6$\times$10$^{16}$\,cm. \hcomas\ has, together with
\ndoshmas, the most sharply peaked distribution amongst the ions
modeled. The model peak abundance in the shell is
$X$(\hcomas)$\approx$10$^{-7}$, that is 
larger than the average value derived in \oh\ from these
observations.
Deeper into the envelope, i.e.,\,in the $\sim$[10$^{15}$-3$\times$10$^{16}$]\,cm
region before the inner rim
of the peak abundance shell,
\hcomas\ is formed in much lower amounts ($X$(\hcomas)$<$10$^{-11}$)
directly from CO and H$_3^+$, which is a product of cosmic-ray
ionization (H$_3^+$ + CO \raw\ \hcomas\ + H$_2$).


The \somas\ shell-like abundance distribution predicted by the model
is broader than that of \hcomas: for \somas\ the inner rim of the
shell is located deeper in the envelope than for \hcomas.
Within the \somas\ peak abundance shell, the fractional abundance 
varies between
$X$(\somas)$\sim$2$\times$10$^{-12}$ 
and $\sim$[3-4]$\times$10$^{-8}$, the maximum value being reached at a
distance of $\sim$5$\times$10$^{16}$\,cm.  The \somas\ 
abundance distribution shows a ``hump'' or secondary peak at
$\sim$3$\times$10$^{16}$\,cm where the abundance is
$X$(\somas)$\sim$7$\times$10$^{-9}$. The values of the \somas\ model
abundance inside the shell seem, in principle, compatible with
the beam-averaged value deduced in \oh\ from our observations.
The main formation paths of \somas\ in the outer shell are
\begin{equation}
\begin{array}{l l}
\mathrm{S^+ + OH \raw SO^+ + H} & \\
\mathrm{SO + h\nu \raw SO^+ +} e^- & \\
\mathrm{C^+  + SO_2 \raw SO^+ + CO} & \\
\label{eq-so+}
\end{array}
.\end{equation}

The reaction of S$^+$ with OH is the dominant formation route
near the peak abundance radius; the hump or secondary peak of the
\somas\ abundance distribution results from direct photoionization of
SO by UV photons and the reaction of C$^+$ with SO$_2$
(Eq.\ref{eq-so+}).
In regions deeper into the envelope, 
cosmic rays are the fundamental source of direct
ionization of SO leading to the production of \somas\ (although in 
much lower amounts, $<$10$^{-12}$, than within the shell)
and free electrons. 

The spatial distribution of \ndoshmas\ predicted by the model is
sharply peaked, similar to that of \hcomas, but with a very low peak
abundance of $X$(\ndoshmas)$\sim$[0.9-1.0]$\times$10$^{-11}$ that is reached at
$\sim$[5-6]$\times$10$^{16}$\,cm.
This maximum abundance is more than two orders of magnitude lower than the mean
value measured in \oh.
Based on our model, \ndoshmas\ is formed mainly via

\begin{equation}
\begin{array}{l l}
\mathrm{H_3^+ + N_2 \raw N_2H^+ + H_2} & \\
\mathrm{N_2^+ + H_2 \raw N_2H^+ + H} & \\
\mathrm{HCO^+ + N_2 \raw N_2H^+ + CO} & \\
\label{eq-n2h+}
\end{array}
,\end{equation}

\noindent
where the first reaction above plays a major role all the way
throughout the different layers of the \ioe, whereas
the second and third reactions become an important factor in
\ndoshmas\ production only in the outermost parts near the peak abundance
radius.
In these outer envelope layers, the ion N$^+_2$ is mainly produced by the
proton transfer reaction of He$^+$, which is produced by direct ionization
of He by UV photons, with N$_2$.

Regarding \htresomas, our model predicts efficient formation of this
ion in the outer envelope mainly as a product of the photodissociation
chain of \water\ and CO. The model peak abundance, 
$X$(\htresomas)$\sim$5$\times$10$^{-8}$, which is larger than the
observed value, is reached at $\sim$5$\times$10$^{16}$\,cm. Near the
peak abundance radius, the distribution of \htresomas\ is relatively
broad and similar in shape and values to that of \somas. In these
outer regions, the main reactions involved in the production of
\htresomas\ are

\begin{equation}
\begin{array}{l l}
\mathrm{H_2O^+ + H_2 \raw H_3O^+ + H} & \\
\mathrm{HCO^+ + H_2O \raw CO + H_3O^+} & \\
\label{eq-h3o+}
\end{array}
.\end{equation}

In contrast to the other ions, \htresomas\ is rather abundant not only
within the peak abundance shell but also deeper into the envelope, with a
fractional abundance that varies from
$X$(\htresomas)$\sim$2$\times$10$^{-11}$ at 10$^{15}$\,cm to
$X$(\htresomas)$\sim$4$\times$10$^{-9}$ at the inner rim of the shell 
(at $\sim$2$\times$10$^{16}$\,cm). In these inner envelope regions,
\htresomas\ is by far the most abundant of the ions modeled and
is produced by protonation of water (\water + H$_3^+$
\raw\ \htresomas\ + H$_2$) after cosmic-ray ionization of H$_2$
leading to H$_3^+$.
In the outer layers of the envelope, dissociative recombination with
electrons is the major mode of destruction of all the ions, except for
\ndoshmas, which instead is most rapidly destroyed by reactions with
\water\ and CO.
Dissociative recombination with electrons ceases to be the main
destruction mechanism for all ions deeper into the envelope, where
the electron abundance is low ($<$few$\times$10$^{-9}$). In these inner
regions, molecular ions are quickly disassembled primarily by
reactions with \water\ and NH$_3$ (and, in the case of \ndoshmas, also
with CO), i.e., 

\begin{equation}
\begin{array}{l l}
\mathrm{HCO^+ + H_2O \raw CO + H_3O^+} & \\   
\mathrm{N_2H^+ + H_2O \raw N_2 + H_3O^+} & \\  
\mathrm{SO^+ + NH_3 \raw SO + NH_3^+} & \\   
\mathrm{H_3O^+ + NH_3 \raw NH_4^+ + H_2O} & \\ 
\end{array}
\label{eq-destruction}
.\end{equation}

{\bf The lobes.}  We have also modeled the chemistry in the walls of
the lobes (model $b$), which are on average more tenuous than the
central regions and, thus, enable deep penetration of the ionizing
interstellar UV radiation. The main input parameters of the simple
physical model adopted for the lobe walls (a rectangular gas slab
externally illuminated by one side by the interstellar UV field) are
given in Table\,\ref{tab:oh231}.
Our chemical kinetics code computes the evolution with time of the
molecular fractional abundances inside a series of individual gas
cells at different depths into the lobe walls. An example is given in
Fig.\,\ref{f-mod1}b for a representative cell midway 
between the inner and outer edge of the walls (i.e.,\,at
$\sim$0\farcs5 through the lobe wall).  The depth of the cell into the
wall is expressed as a function of the dust optical extinction, $A_V$. 
For the cell whose abundances
are represented in Fig.\,\ref{f-mod1}b, the optical
extinction is $A_V$$\sim$1\,mag.

The spatial distribution of the model fractional abundances across the
lobe walls attained at $\sim$800\,yr, which is the dynamical age of
the outflow (\S\,\ref{intro}), is shown in Fig.\,\ref{f-mod1}c. As a
result of photodissociation, the abundances of all molecular ions
rapidly increase inward across the lobe walls, reaching maximum
values near $A_V$$\sim$1\,mag; at larger depths,
the ions' abundances gently decrease as destruction processes by
reactions with \water, NH$_3$, and CO start prevailing over formation
routes. The peak abundances predicted by the model are comparable to
the observed beam-averaged values  in the case of \somas\ , and
possibly in the case of \htresomas\ . In the latter case, the model value is lower than the observational value
but the model-data discrepancy, of a factor $\sim$2, is probably
within the total uncertainties in the abundance
calculation. According to the model, these two ions are the only ones
that would form in detectable amounts\footnote{For the sensitivity
  limit reached by our observations.} in the fast lobes of \oh.  In
fact, in clear disagreement with the observations, our
model predicts \htresomas\  to be the second most abundant ion (after
\somas) in the fast lobes of \oh. For \hcomas\ and \ndoshmas,\ the
model predicts fractional abundances two orders of magnitude lower
than observed. We have checked that the model abundances for these
ions never reach values comparable to the observed values for a
reasonable range of outflow ages, $\sim$500-1000\,yr.

For all the ions discussed in this research, the maximum abundances
reached within the lobes (model $b$) are smaller than the
maximum abundances reached in the slow, central component
(model $a$). The largest core-to-lobe peak abundance ratio ($\sim$500)
is found for \hcomas. The ion \ndoshmas\ is predicted to be
almost equally deficient
in the lobes and in the central, peak abundance shell. \somas\ and
\htresomas\ are intermediate cases, with core-to-lobe peak abundance
ratios of $\sim$4 and 13, respectively.

It may appear a priori somewhat surprising that \hcomas, which is
predicted by the model to be efficiently produced in the outer
regions of the slow, central component (with a peak abundance of
$X$(\hcomas)$\approx$10$^{-7}$), is so scarce across the lobe walls,
where the maximum abundance reached is only
$X$(\hcomas)$\la$10$^{-10}$. This is because 
the \hcomas\ formation and destruction processes in the lobes are
similar to those prevailing in the deepest regions of the central
component, where \hcomas\ is indeed not very abundant because of the 
rapid destruction of this ion by reactions with the abundant
\water\ molecule.
In the peak abundance shell of the central component, however, the
balance between photochemical formation processes and dissociative
recombination with $e^-$ results in an efficient net production of
\hcomas.

\subsubsection{Initial abundance of \water.}
\label{water}

\begin{figure*}[]
  \centering
\includegraphics*[bb=20 20 400 540, width=0.45\hsize]{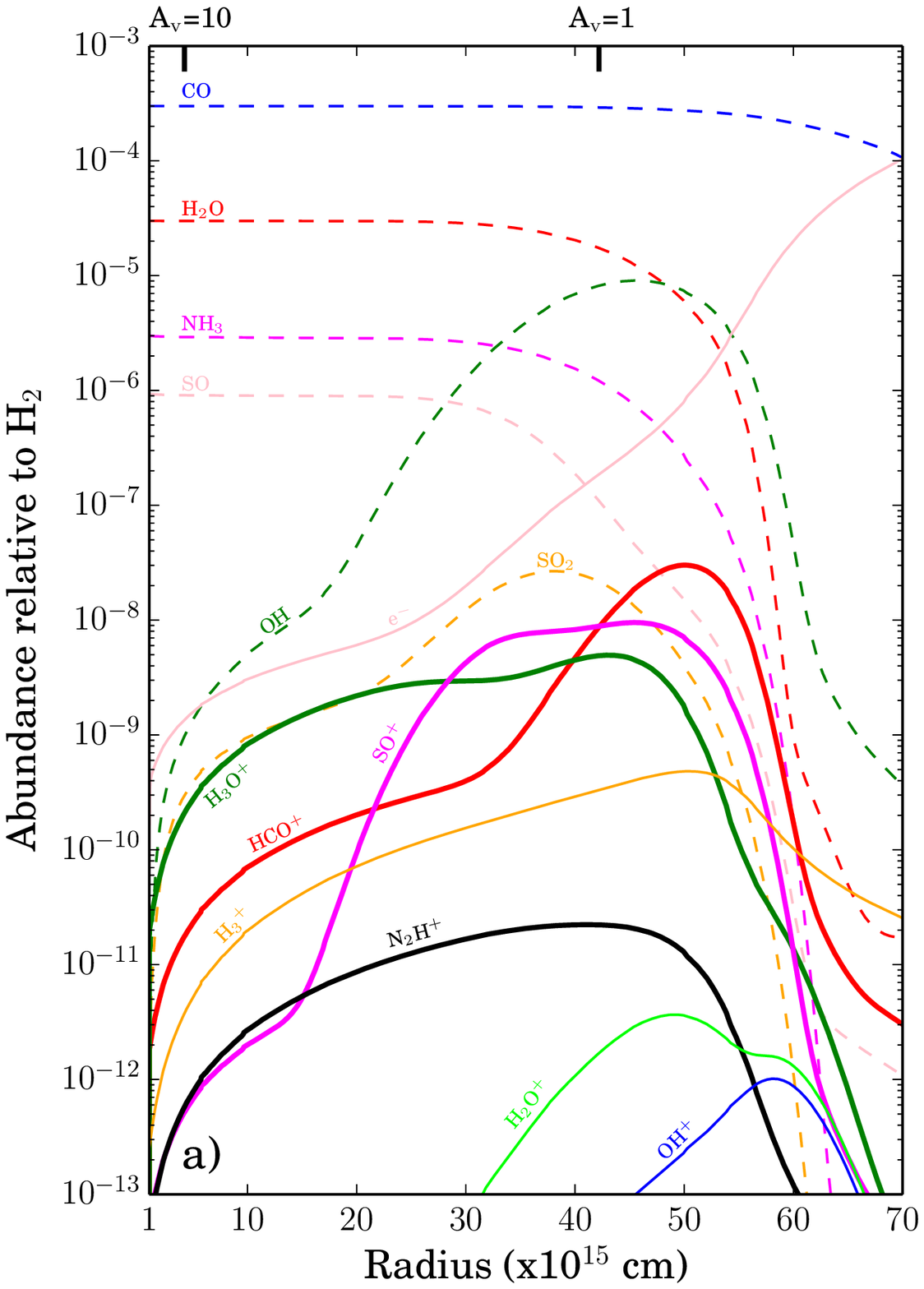} 
\includegraphics*[bb=20 20 400 540, width=0.45\hsize]{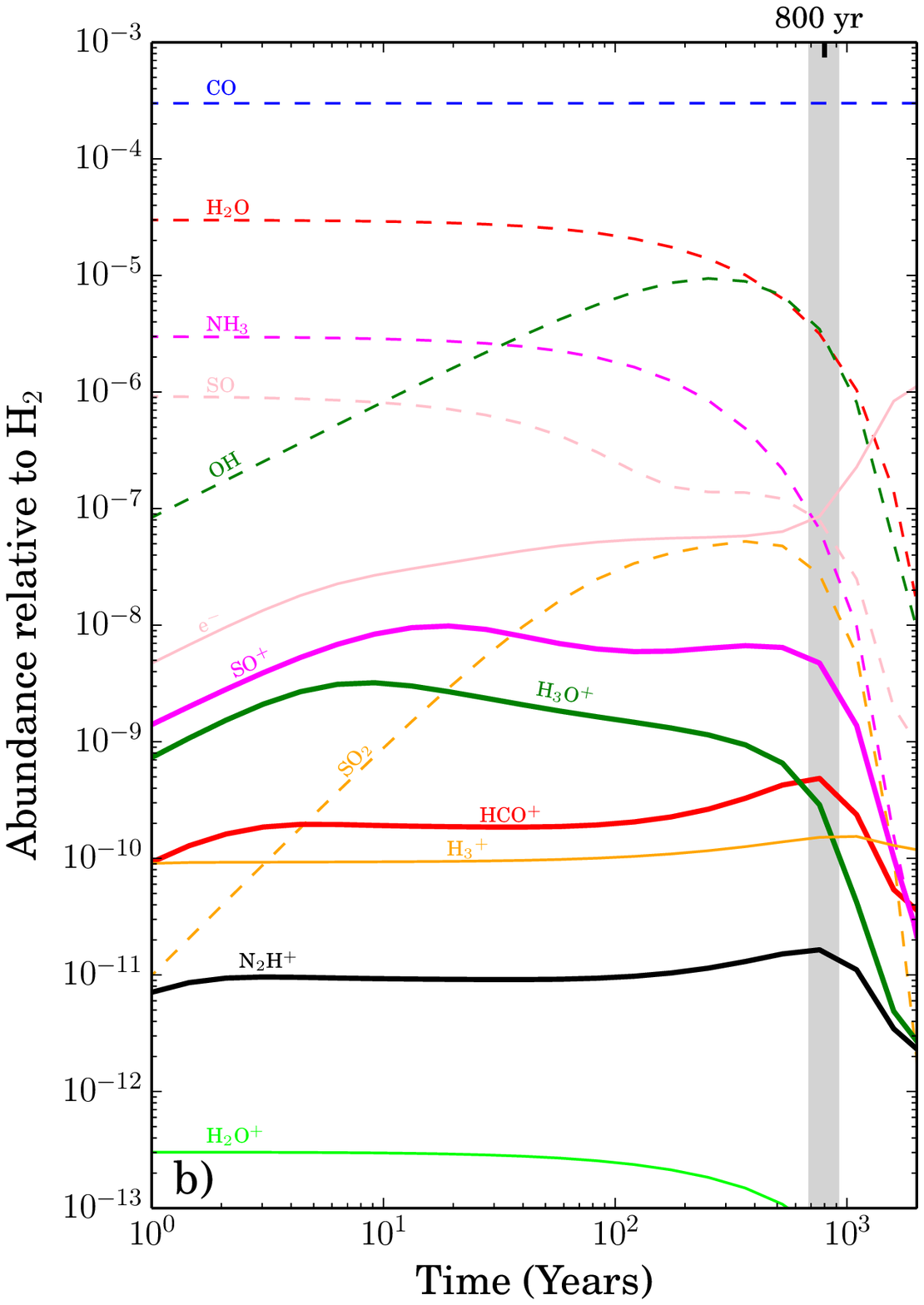}
\includegraphics*[angle=270, width=0.65\hsize]{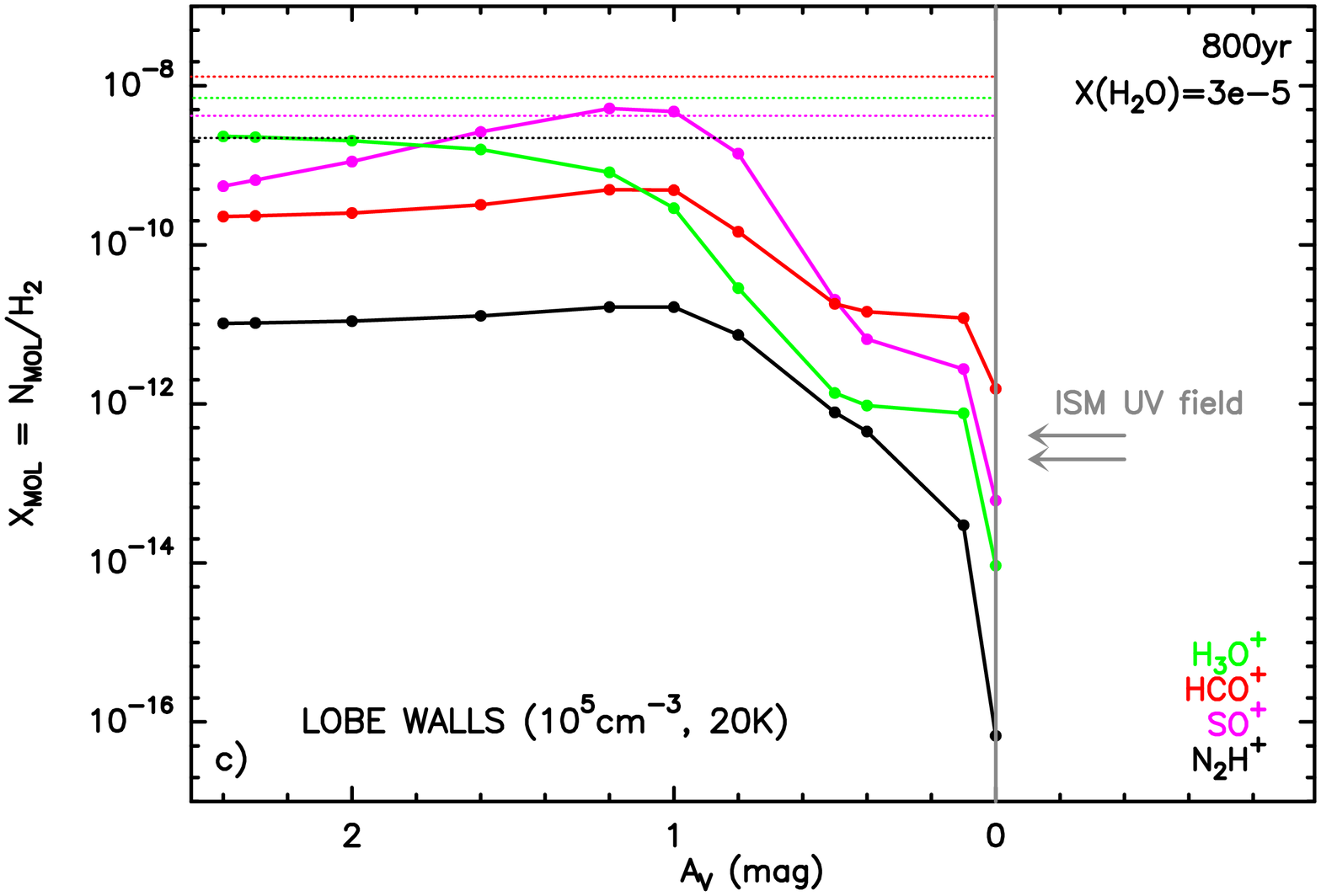}
     \caption{Same as in Fig.\,\ref{f-mod1} except for a lower abundance of
       water, $X$(\water)=3\ex{-5} -- see \S\,\ref{water}.}
  \label{f-mod2}
  \end{figure*}

We have investigated how the abundances of the molecular ions under
discussion depend on the initial abundance of \water, which is a 
prime 
parent molecule tied up
to the chemistry of these and other species.
We ran our code using the same input parameters as in the previous
section but with a relative abundance of water of
$X$(\water)=3$\times$10$^{-5}$ that is ten times lower
(Fig.\,\ref{f-mod2}).  For a given ion, the final role of
\water\ either as a net builder or a net destroyer, after the balance
of formation and destruction reactions with water, varies throughout
the envelope and, of course, is different for the different ions
considered.

%

We find that, within the peak abundance shell in model $a$, a decrease
of the initial abundance of \water\ results in a decrease of the peak
abundances for all ions except for \ndoshmas. The peak abundances of
\hcomas\ and \somas\ are lowered by a factor $\sim$1/3-1/4 and the
abundance of \htresomas\ (and also that of the radical OH) by a factor
$\sim$1/10. In contrast, the peak abundance of \ndoshmas\ increases by
a factor $\sim$2-3. Moreover, in the low \water\ abundance case,
\ndoshmas\ no longer presents  a sharply peaked, shell-like
distribution, but rather its abundance progressively increases from the
inner layers
up to a radius of $\sim$[4-5]$\times$10$^{16}$\,cm; beyond this point,
the abundance falls abruptly as for the rest of the species.  The
markedly different reaction 
 of \ndoshmas\ to a decline in the \water\ abundance with respect to
 the other ions is because, in the case of \ndoshmas, \water\ provides
 the main mode of destruction of this ion but is not a required
 formation ingredient in any of the regions of the envelope (see
 \S\,\ref{model}).
In contrast, in the outer shell, for \hcomas, \htresomas, and \somas,
water acts only as a main production agent (directly or indirectly via H$_2$O$^+$
and OH). 

Deeper into the envelope, \hcomas\ and \ndoshmas\ are the two ions
whose abundances are most sensitive to $X$(\water): our model predicts
\hcomas\ and \ndoshmas\ enhancements of one order of magnitude for a
similar decrease of the water abundance. In contrast to
\ndoshmas, the response of $X$(\hcomas) to a change in the water
abundance is  different in the outer peak abundance shell and
in regions deeper into the envelope. This is because, as explained
earlier, in the inner envelope regions the formation of \hcomas\ is
principally the result of protonation of CO from H$_3^+$, with
\water\ acting almost exclusively as a main destroyer. We also find
that the model abundances of \somas\ and \htresomas\ vary only
marginally in the inner regions when the \water\ abundance is
lowered. For \somas\ this result is readily expected since neither the
main formation nor destruction processes in the inner envelope involve
water in this case (as discussed in \S\,\ref{model}). The marginal
response of \htresomas\ to a decrease of $X$(\water), which a priori
seems less predictable than for \somas, is a direct consequence of the
enhanced production of \hcomas\ in the inner envelope when $X$(\water)
is lowered. In the low-water case, because of the $X$(\hcomas) increase,
the dominant formation route of \htresomas\ in the inner envelope is
by \hcomas\ + \water\ \raw\ CO + \htresomas\ (instead of water
protonation by H$^+_3$). Therefore,
the lower 
\water\ abundance is compensated by the higher abundance of \hcomas, which translates to  
an unaffected \htresomas\ abundance. 


In the lobes (model $b$), lowering the initial \water\ abundance to
$X$(\water)=3$\times$10$^{-5}$ produces a different effect at different
depths into the lobe walls for \hcomas, \somas, and \ndoshmas. In the
outer edge of the walls, fully exposed to the ISM UV radiation field
($A_V$=0\,mag), the fractional abundances of these ions remain almost
equal or slightly smaller than in the high water abundance case
($X$(\water)=3\ex{-4}), whereas in the shadowed ($A_V$$\sim$2.3\,mag) inner
rim of the lobe walls the predicted abundances are higher, with the
largest enhancements (of a factor $\sim$20-40) being for \hcomas\ and
\ndoshmas. In the case of \htresomas, a lower water abundance results
in a lower abundance of this ion in all regions throughout the lobe
walls;  at the shadowed inner edge of the lobe walls only, both the high
and low water abundance models predict a similar value of the
\htresomas\ abundance of $X$(\htresomas)$\sim$2$\times$10$^{-9}$. %

For all ions modeled, it appears that lowering the water abundance
results in the maximum abundance zone 
moving deeper into the lobe walls. As seen above, another effect is an
increase of $X$(\hcomas) and $X$(\ndoshmas) in the lobes, however, the
model predictions are still significantly 
lower ($\sim$2 orders of
magnitude) 
than the abundances observationally deduced.
Moreover, the core-to-lobe abundance contrast for \hcomas\ and
\somas\ in the low water abundance case ($\sim$60 and 20,
respectively) is still irreconcilable
with the broad profiles observed
for these species. As in the high water abundance case, \ndoshmas\ is
equally deficient in the core and in the lobes. In the case of
\htresomas, the core-to-lobe abundance ratio diminishes to $\sim$2-3
for a low \water\ abundance. 

Finally, by pushing the water abundance to the low end of the range of
observed values in O-rich stars, $X$(\water)$\approx$10$^{-6}$, the
abundances predicted by our chemistry model in all cases, except maybe
for \somas, become much smaller than the observed values both in the
slow core component and in the fast lobes.

\subsubsection{Enhanced elemental abundance of nitrogen}

In this section we examine the behavior of the ions discussed here against an overabundance of the elemental nitrogen (up to a factor
$\times$40)\footnote{Elemental N-enrichment can result from hot bottom
  burning (HBB) processes in the nucleus of $\ga$3\,\msun\ stars.} and
whether such an N-enhancement can explain the large fractional
abundance of \ndoshmas\ observed in \oh. As in \cite{vel14}, we ran
 our chemistry code again modifying the initial abundances of those
parent species that are most sensitive to the elemental abundance of
nitrogen, namely, N$_2$, NH$_3$, and HCN: N$_2$ is increased by a
factor 40 and NH$_3$ and HCN by a factor 7 relative to the values in
Table\,\ref{t_parent}. We find that the \ndoshmas\ peak abundance
increases
to $\sim$2$\times$10$^{-10}$ 
in the slowing expanding shell 
and to $\sim$10$^{-10}$ in the fast lobes. Therefore, despite the
substantial N-enhancement, the model still predicts \ndoshmas\ to be
underabundant (by one order of magnitude).  Moreover, as in the case
  of a standard N abundance, the model fails to reproduce the
  enrichment of \ndoshmas\ in the fast lobes relative to the slow core
  deduced from the observations. This was, in fact, expected since
  changing the elemental abundance of nitrogen  comparably affects  both the fast and slow outflow components.

The elemental nitrogen enhancement has only a minor
effect on the other ions discussed in this research, except for
\htresomas\ and \somas\ but
only in the deepest layers of the slow
central component. In these inner regions, the abundances of
\htresomas\ and \somas\ decrease (by a factor $\sim$6 and $\sim$10,
respectively) as these ions are destroyed at larger rates through their
interaction with more NH$_3$ molecules (eq.\,\ref{eq-destruction}). 

Given the set of chemical reactions involved in the production and
destruction of \hcomas\ and \ndoshmas, the abundance ratio between
these two ions provides an upper limit to the CO to N$_2$ abundance
ratio. In \oh, we deduce $X$(\hcomas)/$X$(\ndoshmas)$\sim$6 from the
observations and, therefore, one could conclude that
[CO]/[N$_2$]$\la$6. This value, which is slightly smaller than, but
comparable to, the [CO]/[N$_2$]=7.5 ratio in O-rich sources (adopted
in our model, Table\,\ref{t_parent}), does not rule out some N enrichment
but it also  does not clearly point to a significant nitrogen excess.
Based on observations and modeling of the
[\ion{N}{ii}]$\lambda\lambda$6548,6584\AA\ emission (and other optical
lines) from the shock-excited regions at the tips of the lobes of \oh,
\cite{coh85} suggest that N may be overabundant in these regions by a
factor $\sim$5 relative to the solar value.
This factor is much smaller than that used here to test this effect
($\times$40), and that has been proven to be unable to bring the model
\ndoshmas\ abundance close enough to the observed value.

We also recall that an enhancement of the elemental nitrogen is not
able to reproduce satisfactorily the observed abundances of HNCO,
HNCS, HC$_3$N, and NO recently discovered in \oh\ \citep{vel14}: on
the one hand, HNCO, HNCS, and HC$_3$N would be underestimated and, on the
other hand, NO (and maybe others, such as NH$_3$) would be
significantly overestimated. Moreover,  there are other
N-bearing molecules that, in contrast to \ndoshmas, are not
particularly abundant in \oh.  One example is HCN: while the average
abundance of this molecule in O-rich CSEs is relatively large
$X$(HCN)$\approx$10$^{-7}$ \citep{buj94,sch13}, and actually larger
than predicted by standard chemistry models \citep[e.g.,][]{ner89}, the
value found in \oh\ is $X$(HCN)$\sim$few$\times$10$^{-8}$
\citep{san97,san14}.

In any case, an enhancement of the elemental N abundance would at most
be able to reduce the model-data discrepancies for 
N-bearing species, but it is unlikely to help improve
the model predictions for all the other molecules that are not chemical
daughter products of N$_2$, such as the ions \hcomas, \somas, and
\htresomas\ reported in this work. 




\section{Discussion and conclusions}
\label{conc}

Based on single-dish observations with the \iram\ and
\hso\ telescopes, we have detected \hcomas, \htrececomas, \somas,
\ndoshmas, and (tentatively) \htresomas\ in \oh.
The broad line profiles (with FWHM of up to $\sim$90\,\kms) and low-excitation temperature ($\sim$10-20\,K) deduced for \hcomas,
\htrececomas, \somas, and \ndoshmas, support the location of these
ions preferentially in the base of the cold high-velocity lobes of \oh.
The fact that the narrow emission component from the massive central
parts of the nebula (clump I3) is minimal in the profiles of these
ions, clearly points to
an abundance enhancement in the fast lobes relative to the slow
core. Although uncertain, the much narrower profile
(FWHM$\sim$14\,\kms) and higher excitation conditions
($\approx$100\,K) suspected for \htresomas\ suggest a different
spatial distribution for this ion, which may be more concentrated
toward the central parts of the slow, central core.


Our chemical kinetics model predicts that \hcomas, \somas, and
\htresomas\ can form profusely (with peak abundances of
$\approx$10$^{-8}$) in the external layers
of the slow central regions of \oh\ and thus, most generally, in similar
\mbox{O-rich} CSEs. 
A moderate production of \htresomas\ (with fractional abundances of up
to $\sim$10$^{-9}$) is also expected  in regions deeper into the
envelope.
The production of \ndoshmas\ is, however, rather inefficient, 
leading to very low fractional abundances of $X$(\ndoshmas)$\la$10$^{-11}$. 


In the fast lobes of \oh, our standard
chemistry model predicts relatively high abundances for \somas\ and
\htresomas\ (of up to $\approx$10$^{-9}$); however, we predict that \hcomas\ and
\ndoshmas\  are very scarce, with abundances of
$\la$10$^{-10}$ and $\la$10$^{-12}$, respectively.


There are various observational results that our chemistry model fails
to reproduce satisfactorily. In particular, the high \hcomas\ and \ndoshmas\ abundances in the fast lobes deduced from the observed
profiles. If these ions were present in the lobes in the low amounts
predicted by the model, their emission would have remained undetected
in our survey. Another model prediction that is inconsistent with the
observations is that
all ions should be more 
abundant in the slow core than in the fast lobes, however, the 
broad profiles of \hcomas, \ndoshmas, and \somas,
reflect exactly the opposite. One of the most extreme cases is
\hcomas, for which the model predicts a core-to-lobe abundance
ratio of $\sim$500 versus the measured value of $\sim$1/3-1/4; the latter 
value is deduced  from the broad, flattened profiles we report,  and from accurate characterization of the
\hcomas\ spatio-kinematic distribution obtained from previous
single-dish and interferometric mapping (\S\,\ref{intro}).

Regarding \htresomas, the model predicts a spatial distribution
analogous to those of the other ions, i.e., peaking at the external
layers of the slow central core due to photodissociation
processes. This distribution is not easily reconcilable with the
suspected origin for the \htresomas\ emission in the deep (slow and
warm) regions of the central core. The model
predicts a detectable amount of \htresomas, of $\approx$10$^{-9}$, in
the fast lobes that is not consistent with our observations. The
tentative \htresomas\ abundance distribution in \oh\ deduced from our
observations needs, however, to be confirmed by higher quality data.
 
The large core-to-lobe abundance ratio of \hcomas, \somas, and
\ndoshmas\ predicted by the model can be reduced (but not suppressed)
by adopting a lower initial \water\ abundance of 3$\times$10$^{-5}$
(instead of 3$\times$10$^{-4}$ used in our standard model). Also,
lowering the water content injected to the \ioe\ would produce
somewhat lower values of the peak abundances of \htresomas, which is
probably overabundant in the external envelope layers in our standard
model.

We also find that an elemental nitrogen overabundance in \oh\ (by a
factor of up to $\sim$40) is not able to produce sufficient \ndoshmas,
which remains one order of magnitude less abundant than observed. We
believe that the model's inability to explain the large abundance of
\ndoshmas\ is not mainly a problem caused by  an assumed nitrogen abundance
that is too low. A similar problem also resides in NH$_3$ in the majority of
O-rich CSEs, which contain ammonia in amounts that exceed predictions
from conventional chemical models by many orders of magnitude
\citep[e.g.,][]{men10,men95}.

\subsection{Molecule reformation after the passage of fast dissociative shocks?}
Taking into consideration the various important model-data
discrepancies deduced from our analysis, we must conclude that the
standard molecular formation scenario adopted (triggered by ionization
by UV photons and cosmic rays of gas whose initial state is
predominantly molecular) is too simple and/or does not correctly depict
 the recent molecular formation history of \oh. 
The adopted scenario, however, may represent  the
chemistry of most molecules in the intermediate/outer layers of normal
AGB CSEs reasonably well \citep[see, e.g.,][]{nej88,gla96,mill00,agun08}.

The notable chemical differences between \oh\ and most O-rich AGB CSEs
are widely recognized and are not restricted to the ions discussed
here but affect most of the species identified in this object
(\S\,\ref{intro}).  Since the earliest studies on the molecular
composition of \oh, the remarkable molecular richness of this source
has been related to shocks. Indeed, the main difference between
\oh\ and 'normal' O-rich AGB CSEs is the presence of a fast
($\sim$400\,\kms) outflow in \oh.  If the acceleration of the bipolar
lobes resulted from the violent collision between underlying fast jets
on the pre-existing slow AGB wind, as proposed by many authors,
then shocks {\sl must} have necessarily played a major role in
defining the current chemical composition of the outflow.

In particular, given the high velocities observed in the fast outflow
of \oh, it is unlikely that molecules survived this kind of acceleration
process but rather they were probably dissociated. Molecules are
indeed expected to be destroyed by shocks with speeds larger than
$\ga$50\,\kms\ \citep[e.g.,][]{holl80}.  After the shocked gas has
cooled below $\sim$10$^4$\,K, atoms begin to associate and form
molecules again; if the post-shock gas is moderately dense
(e.g., $\approx$10$^5$-10$^6$\,\cm3), molecule reformation can happen
indeed rather quickly, within a few years from the shock
\cite[e.g.,][]{gla89,neu89}.  We believe that most molecules observed
in the fast outflow of \oh, if shock accelerated, must have
dissociated and reformed in the post-shock gas. A similar scenario has
been proposed to explain the chemistry in the high-velocity molecular
outflow of the C-rich PPN CRL\,618 \citep{neri92}.



At present, the shocks that accelerated the outflow $\sim$800\,yr ago
are not active, therefore, the bulk of the molecular outflow in
\oh\ is probably swept-up AGB wind material that has cooled down
sufficiently to allow molecule reformation. Note that the lobe axial acceleration took place in
less than $\approx$150\,yr, which may have been the duration of, e.g.,
a pulsed fast wind.


The physical properties (density and temperature) of the fast
post-shock gas, which could have been  fully dissociated and
fully ionized $\sim$800\,yr ago, must have evolved as a function of
time. In particular, the temperature in the lobes has certainly
decreased  to its current value of $\sim$10-20\,K.
%
The assembly of new molecules would have happened progressively while
the shock-heated material was cooling down. The ionized, atomic, and
molecular fractions in the post-shock gas must have also evolved with
time in a way that is not easy to predict, over the course of the cooling
and molecular regeneration. [In particular, a transition from an
H$^+$-dominated regime to an H$_2$-dominated regime, passing through
an intermediate H-dominated phase, has probably happened.]  Since
these are the basic ingredients for putting together new molecules,
their evolution with time must have dictated the set of chemical
reactions that were dominant at each
precise moment over the course of molecular regeneration leading to the final
molecular content observed at present. 

We increase the number of observational studies and
chemistry modeling of molecular ions, which is needed to make
progress in determining the role of ion-molecule reactions in the
chemistry scheme of O-rich CSEs. Taking  the presence of
fast outflows and the probable shock-acceleration history of the
envelope of \oh\ into account, the observations reported here may contribute to the
emergence of an observational picture of the effects of shocks on the
molecular chemistry of circumstellar envelopes.
In the future, it would be desirable to attempt new chemical kinetics
models adopting the molecule reformation scenario proposed here for
\oh. This is a difficult task since
determining the time dependence of the relevant physical parameters of
the cooling gas, where second-generation species have 
formed, is a challenging theoretical problem. Undoubtedly, it would
also be advantageous to enlarge the reaction network used in most
standard chemistry models by including additional chemical reactions
that may affect the predicted abundances of certain
species (e.g.,\,grain-surface reactions, X-rays, etc).

\begin{acknowledgements}
{\it Herschel} is an ESA space observatory with science instruments
provided by European-led Principal Investigator consortia and with
important participation from NASA. HIFI is the Herschel Heterodyne
Instrument for the Far Infrared. HIFI has been designed and built by a
consortium of institutes and university departments from across
Europe, Canada and the United States under the leadership of SRON
Netherlands Institute for Space Research, Groningen, The Netherlands
and with major contributions from Germany, France and the
US. Consortium members are: Canada: CSA, U. Waterloo; France: CESR,
LAB, LERMA, IRAM; Germany: KOSMA, MPIfR, MPS; Ireland, NUI Maynooth;
Italy: ASI, IFSI-INAF, Osservatorio Astrofisico di Arcetri- INAF;
Netherlands: SRON, TUD; Poland: CAMK, CBK; Spain: Observatorio
Astronómico Nacional (IGN), Centro de Astrobiología (CSIC-INTA);
Sweden: Chalmers University of Technology – MC2, RSS \& GARD; Onsala
Space Observatory; Swedish National Space Board, Stockholm University
– Stockholm Observatory; Switzerland: ETH Zurich, FHNW; USA: Caltech,
JPL, NHSC. HCSS / HSpot / HIPE is a joint development by the Herschel
Science Ground Segment Consortium, consisting of ESA, the NASA
Herschel Science Center, and the HIFI, PACS and SPIRE consortia. This
work has been partially supported by the Spanish MINECO through grants
CSD2009-00038, AYA2009-07304, and AYA2012-32032. This research has
made use of the SIMBAD database, operated at CDS, Strasbourg, France,
the NASA's Astrophysics Data System, and Aladin.
\end{acknowledgements}

%

%

%

%

\clearpage 
\begin{appendix}

\section{The HIFI beam pattern}
\label{app-neff}

\begin{table}
\caption{\label{t-neff} Adopted values for HPBW, $\eta_{\rm mb}$, and
  Ruze-like scaling factors $\eta_{\rm mb,0}$ and $\sigma_{\rm mb}$
  (see Eq.\,\ref{eq-neff} in \S\,\ref{obs-hso}) for one spot frequency
  per mixer and per polarizations (H and V) of the HIFI
  instrument. Adapted from tables 2 and 3 of Mueller et al.\,(2014).}
\small
\begin{tabular}{r r r r c c}      
\hline\hline         
Mixer & $f$ & HPBW & $\eta_{\rm mb}$ & $\eta_{\rm mb,0}$ &  $\sigma_{\rm mb}$ \\ 
      & (GHz) & (\arcsec) & & & ($\mu$m) \\ 
\hline 
1H   & 480  &    43.1  &   0.62  & 0.624 $\pm$ 0.007 & 2.278 $\pm$ 0.004 \\ 
1V   & 480  &    43.5  &   0.62  & 0.618 $\pm$ 0.007 & 2.248 $\pm$ 0.004 \\ 
2H   & 640  &    32.9  &   0.64  & 0.643 $\pm$ 0.009 & 2.219 $\pm$ 0.003 \\ 
2V   & 640  &    32.8  &   0.66  & 0.662 $\pm$ 0.009 & 2.224 $\pm$ 0.002 \\ 
3H   & 800  &    26.3  &   0.62  & 0.626 $\pm$ 0.008 & 2.227 $\pm$ 0.004 \\ 
3V   & 800  &    25.8  &   0.63  & 0.639 $\pm$ 0.008 & 2.287 $\pm$ 0.010 \\ 
4H   & 960  &    21.9  &   0.63  & 0.639 $\pm$ 0.008 & 2.227 $\pm$ 0.004 \\ 
4V   & 960  &    21.7  &   0.64  & 0.643 $\pm$ 0.008 & 2.252 $\pm$ 0.006 \\ 
5H   & 1120 &    19.6  &   0.59  & 0.595 $\pm$ 0.006 & 2.068 $\pm$ 0.005 \\ 
5V   & 1120 &    19.4  &   0.59  & 0.597 $\pm$ 0.006 & 2.114 $\pm$ 0.005 \\ 
\hline 
\end{tabular}
\end{table}

The HIFI beam pattern (size, efficiency, etc) is described by
\cite{roe12}. In October 2014, a release note was issued by the HIFI
calibration team (Mueller et al.\,2014\footnote{{\sl The HIFI Beam: Release
  \#1. Release Note for Astronomers} -- {\tt
  http://herschel.esac.esa.int/twiki/bin/view/Public/Hifi\-CalibrationWeb\#HIFI\_performance\_and\_calibration}.}) 
reporting improved measures  
of the half-power beam width (HPBW) and main-beam efficiency
($\eta_{\rm mb}$) that supersede previous estimates in \cite{roe12}.
The new efficiency estimates are systematically lower than previous
values by typically 15–20\% for $\eta_{\rm mb}$.

For convenience, we reproduce again in this appendix the updated
values of the relevant beam-model parameters provided by Mueller et
al.\,(2014) in Tables 2 and 3 of their release note.  These values,
given in Table\,\ref{t-neff}, are used to obtain the formulae
(Eq.\,\ref{eq-neff} in \S\,\ref{obs-hso}) that describe the
frequency/wavelength dependence of HPBW and $\eta_{\rm eff}$ for the
HIFI bands (1 through 5) used in this work. Since we average the H and
V polarization data to obtain final spectra, we use the average of
$\eta_{\rm mb}$ for H and V at a given frequency in order to convert
antenna temperatures (\ta) to a main-beam scale (\tmb) at that
frequency.  We recall that, as stated in \S\,\ref{obs}:

\[
\tmb=\ta\,/\eta_{\rm eff}=\ta\,\frac{\eta_{\rm l}}{\eta_{\rm mb}}  
. \]



%

\section{Additional material}
\label{app-1}
In this Appendix, we include additional material for 
the ion \somas\ and the molecule \trecem\ in \oh. 

\begin{figure}[!htbp]  
   \centering
   \includegraphics*[bb=1 90 756 460,width=0.975\hsize]{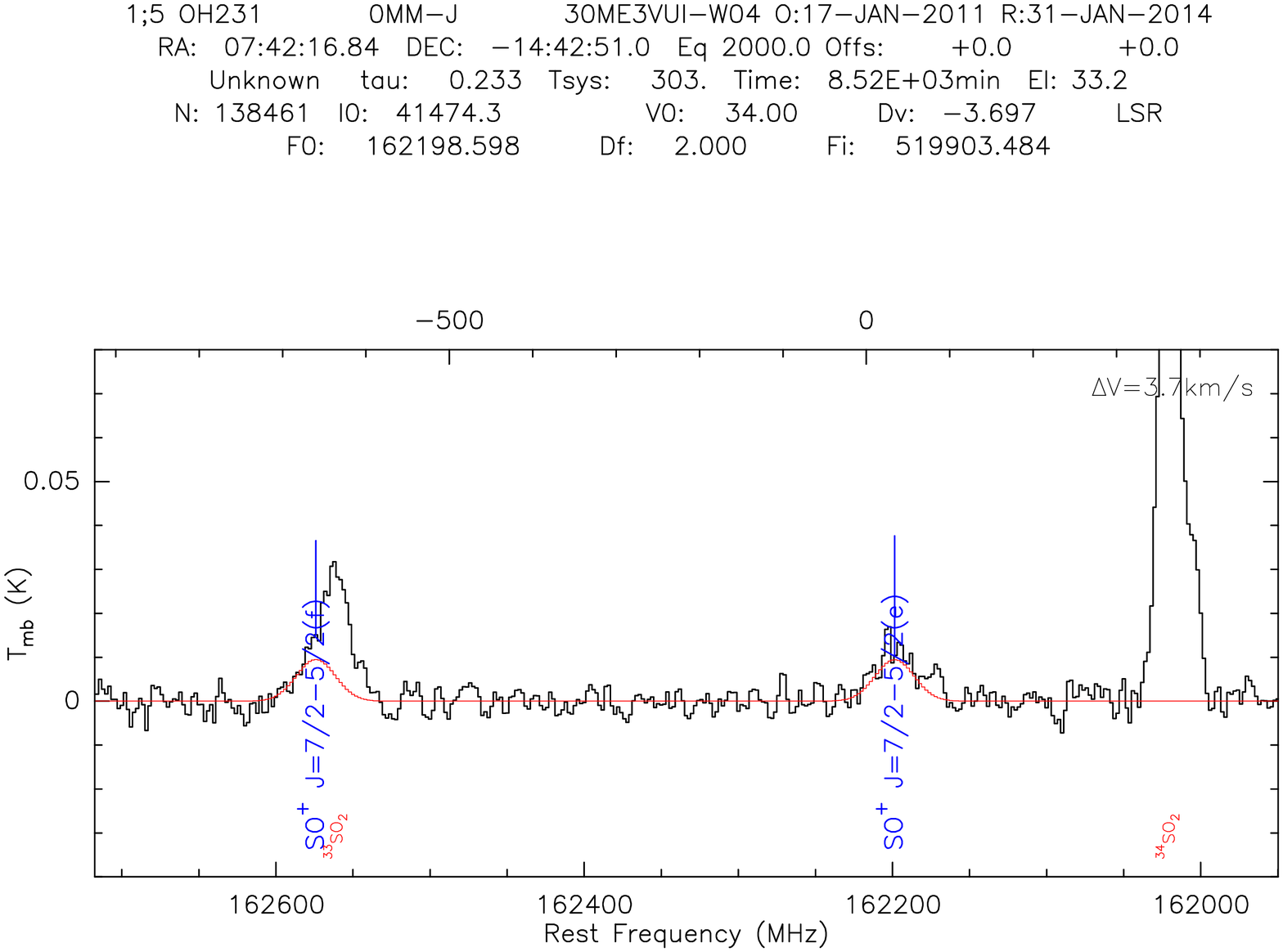}
   \includegraphics*[bb=1 90 756 460,width=0.975\hsize]{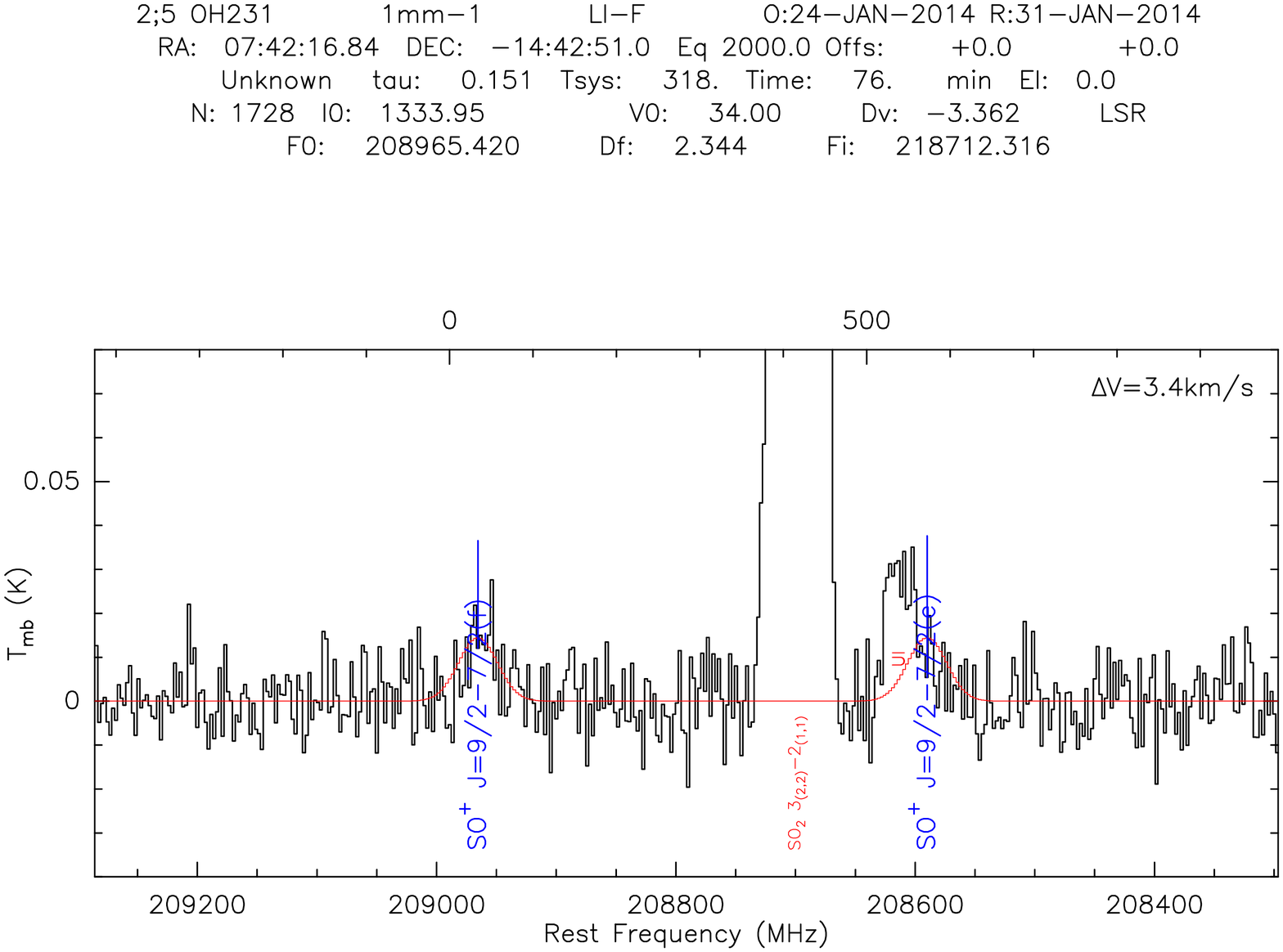}
   \includegraphics*[bb=1 90 756 460,width=0.975\hsize]{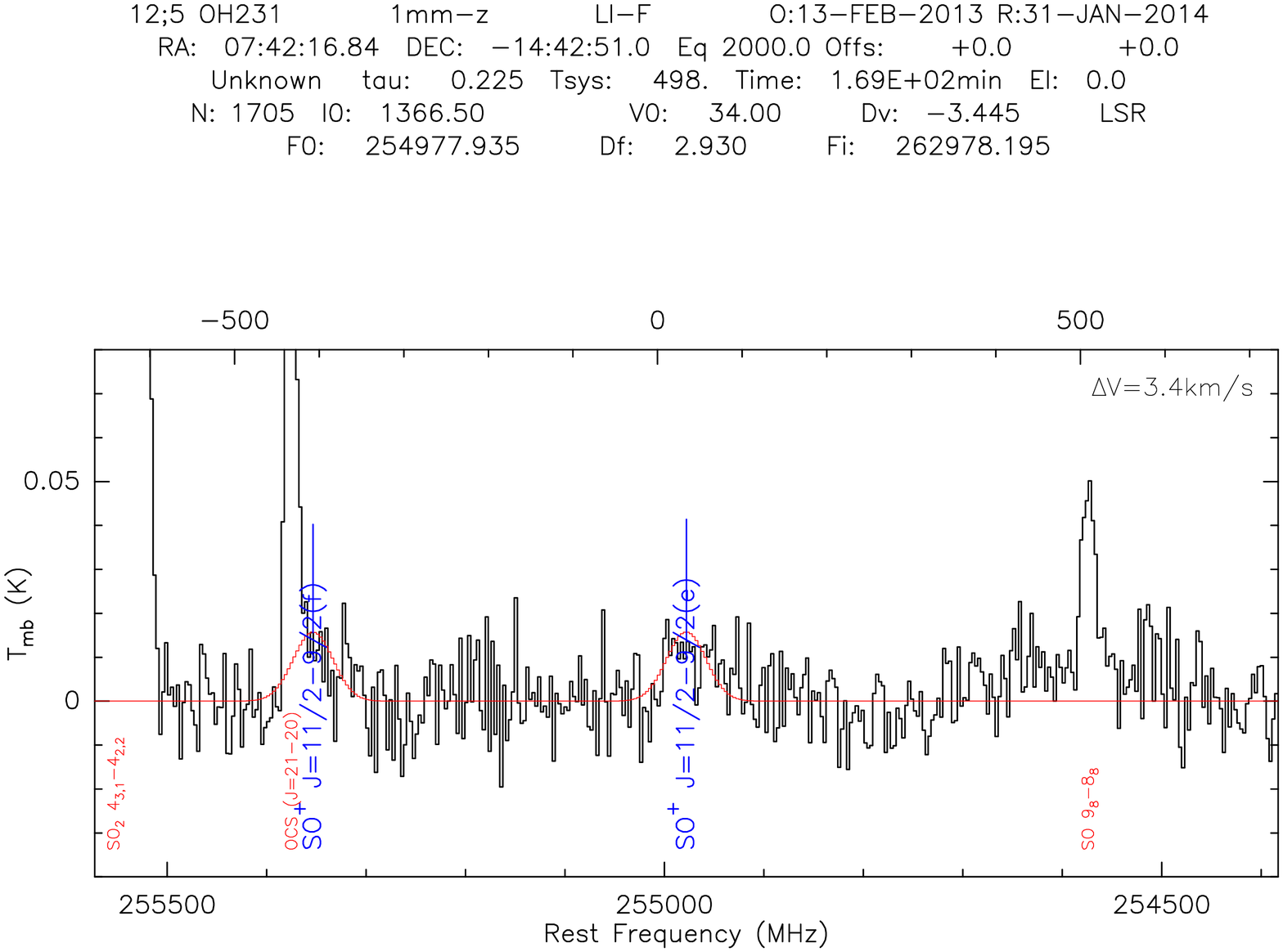}
   \includegraphics*[bb=1 90 756 460,width=0.975\hsize]{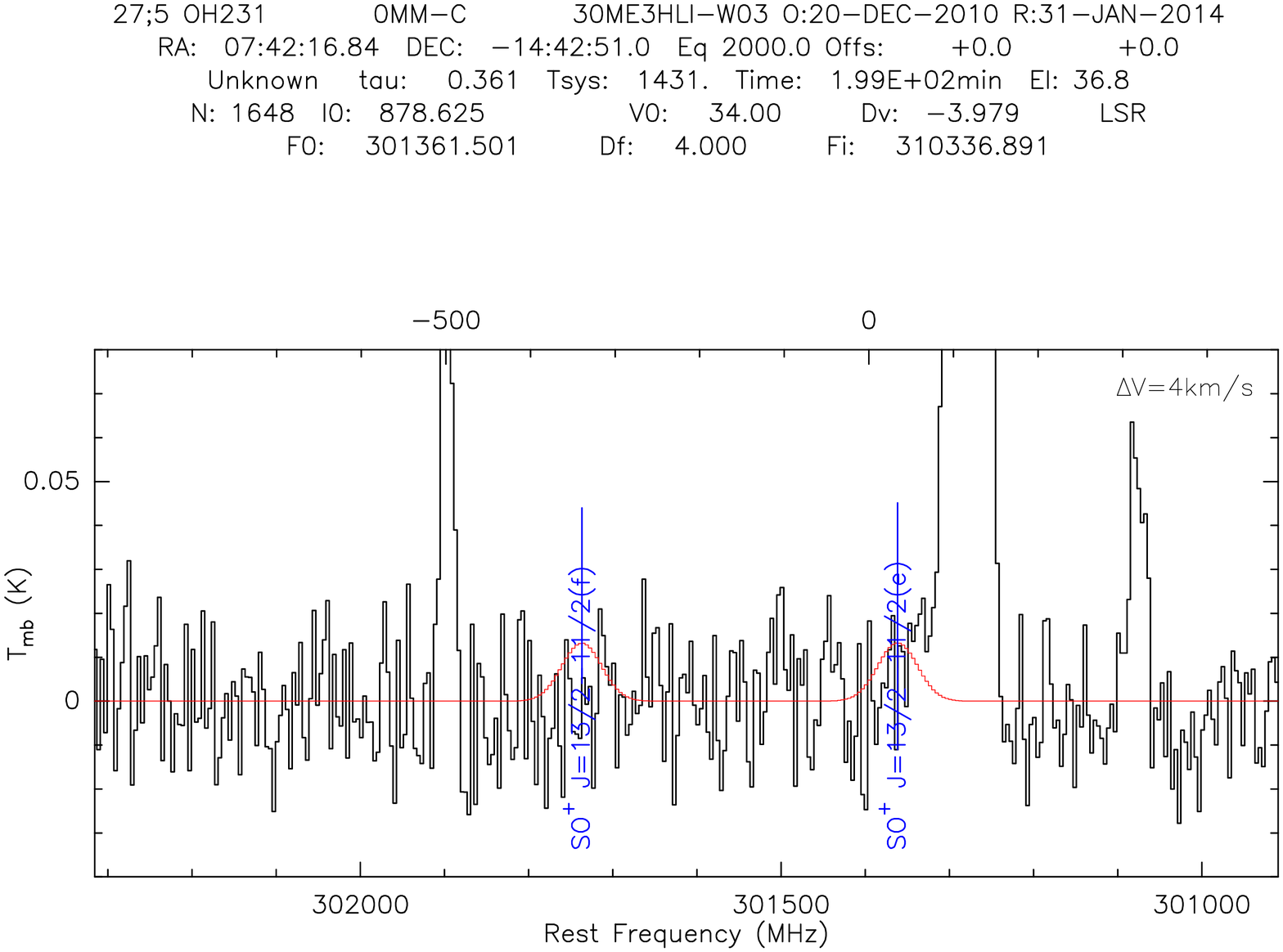}
   \includegraphics*[bb=1 90 756 460,width=0.975\hsize]{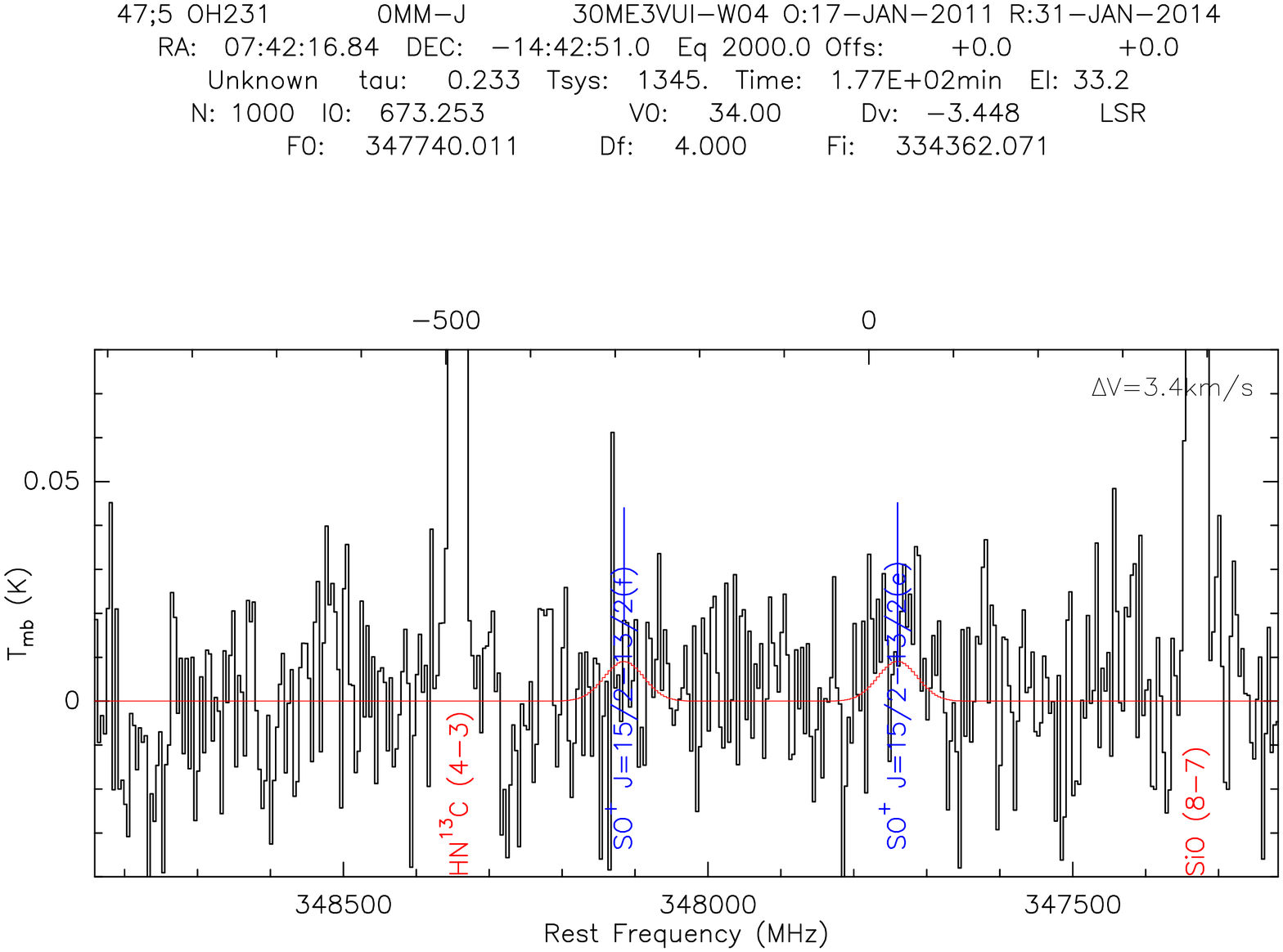}
      \caption{\iram\,spectra toward \oh\ near the five mm-wave
        doublets of \somas\ (see Table\,\ref{t_res}). An LTE model compatible with values in Table\,\ref{tabun} 
and adopting an average FWHM=55\kms\ for the lines is overplot (red line).}
         \label{II-so+}
   \end{figure}

In Fig.\,\ref{II-so+}, we show all the \somas\ transitions (detected
and nondetected) observed within the frequency range covered in our
survey with the \iram\ telescope.  Some of the \somas\ lines are
blended with transitions of other molecules and/or are in particularly noisy regions and have not been considered for the rotational
diagram analysis presented in Section \ref{rotdiagr}.

\begin{figure*}[!htbp]  
   \centering
 \includegraphics*[bb=0 97 721 510,width=0.45\hsize]{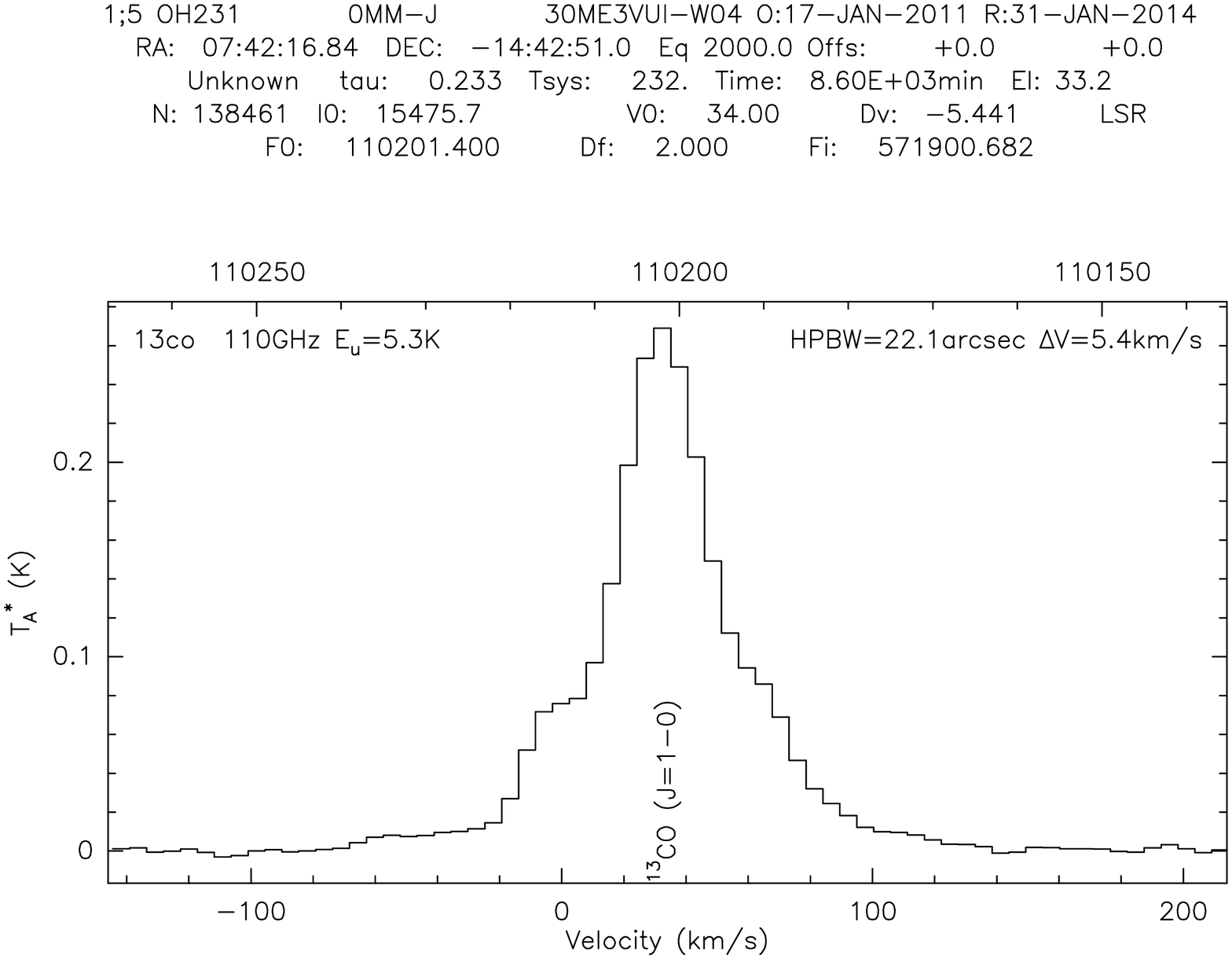}
 \includegraphics*[bb=0 97 721 510,width=0.45\hsize]{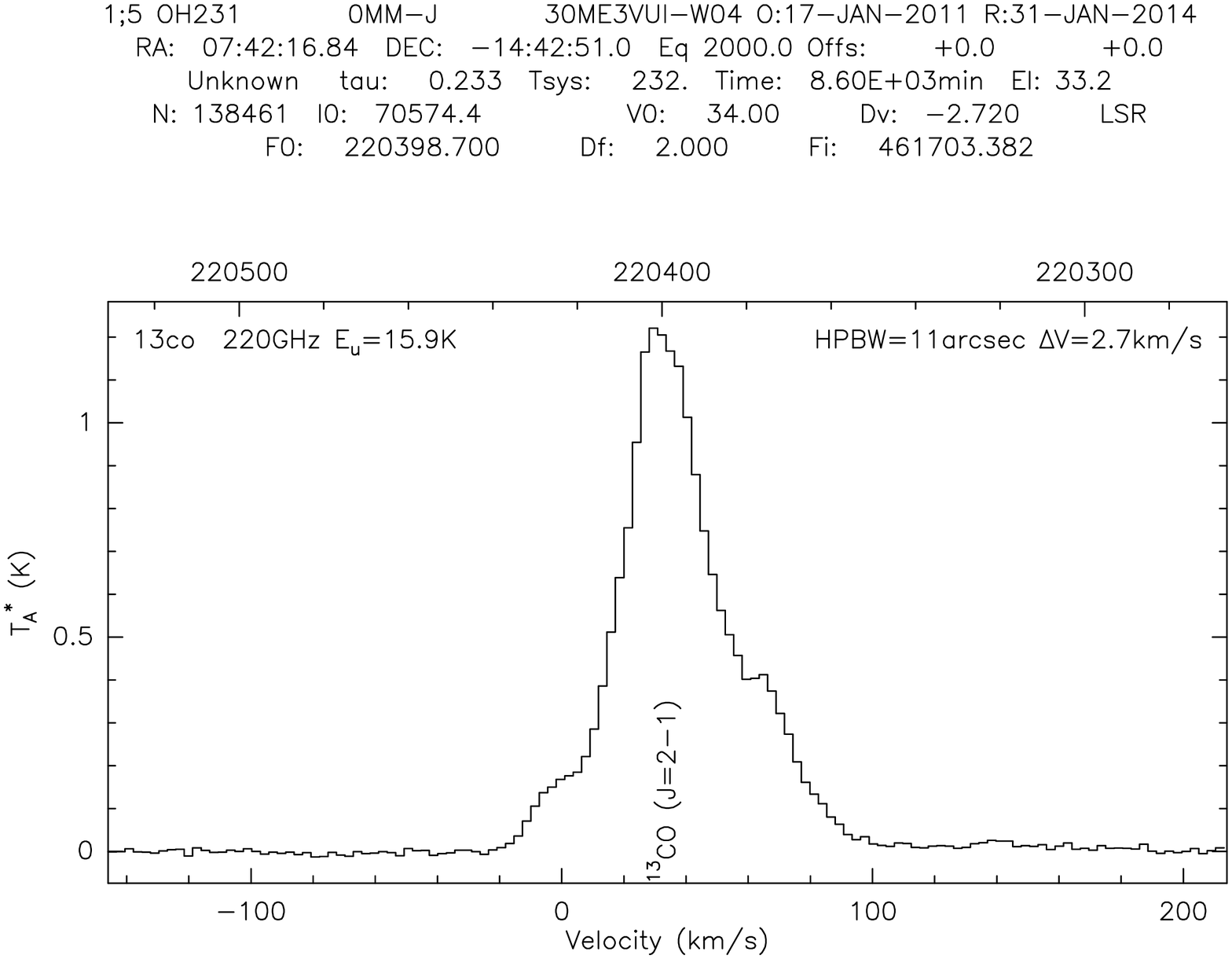}
 \includegraphics*[bb=0 97 721 510,width=0.45\hsize]{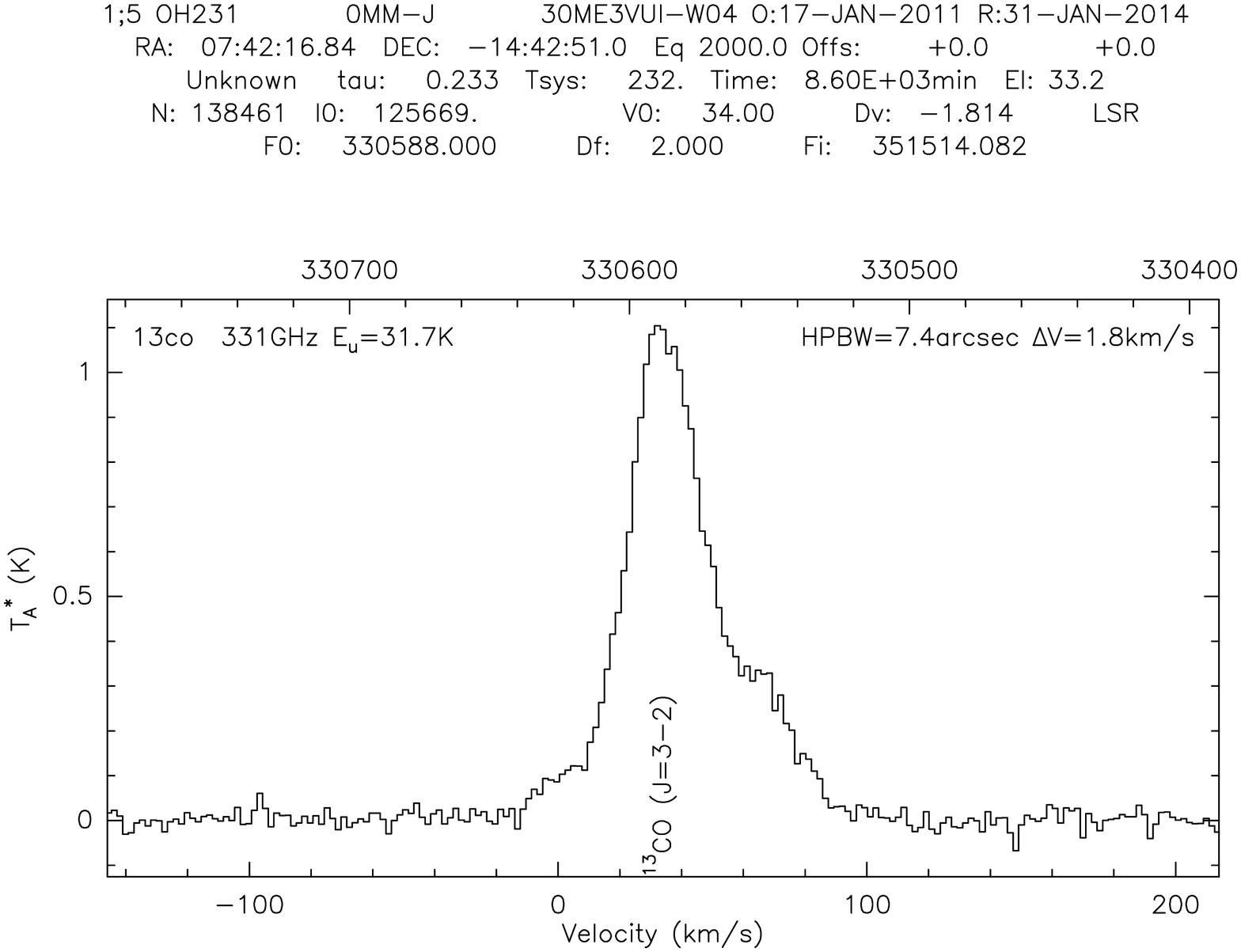}
 \includegraphics*[bb=0 97 721 510,width=0.45\hsize]{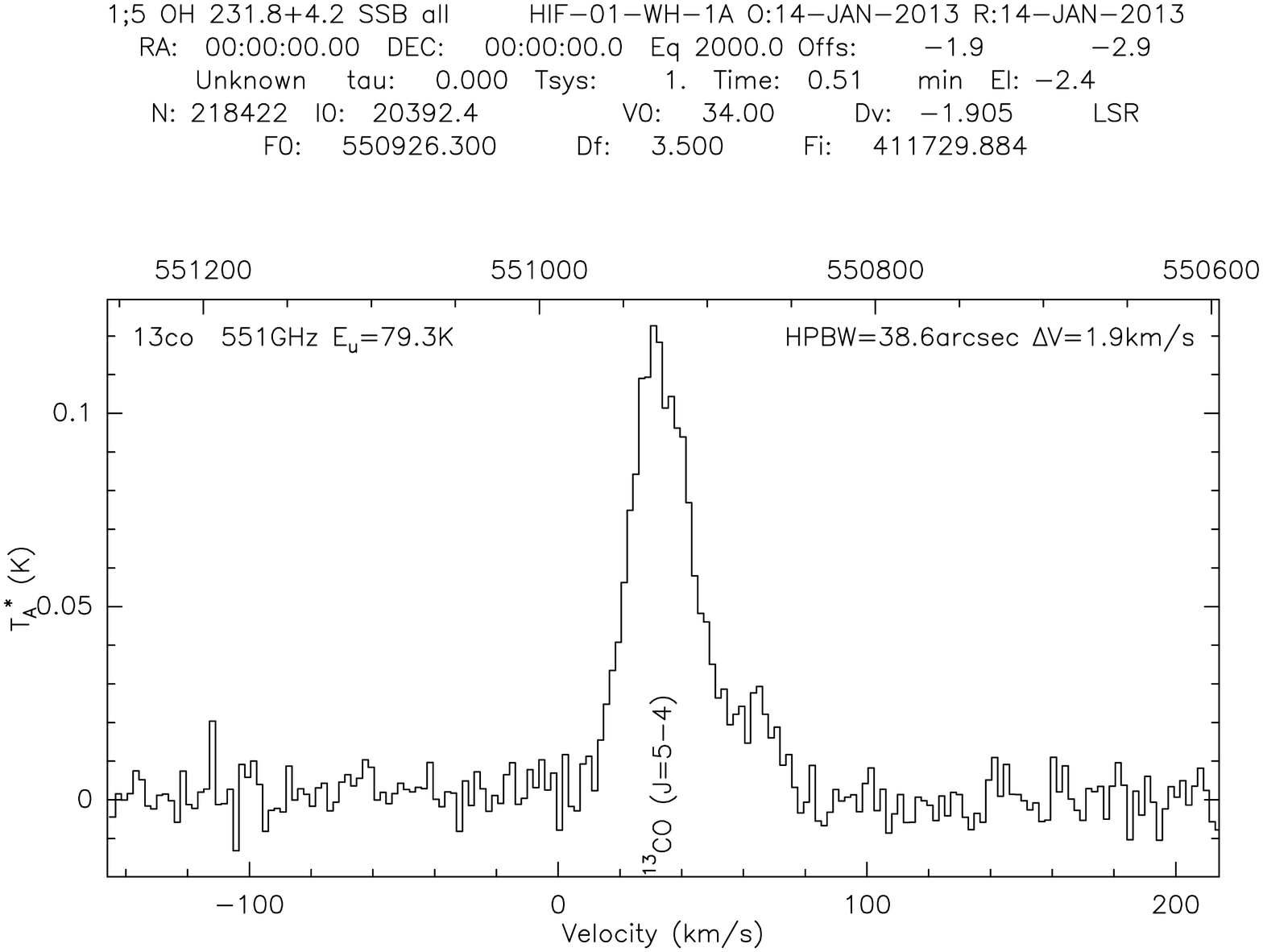}
 \includegraphics*[bb=0 97 721 510,width=0.45\hsize]{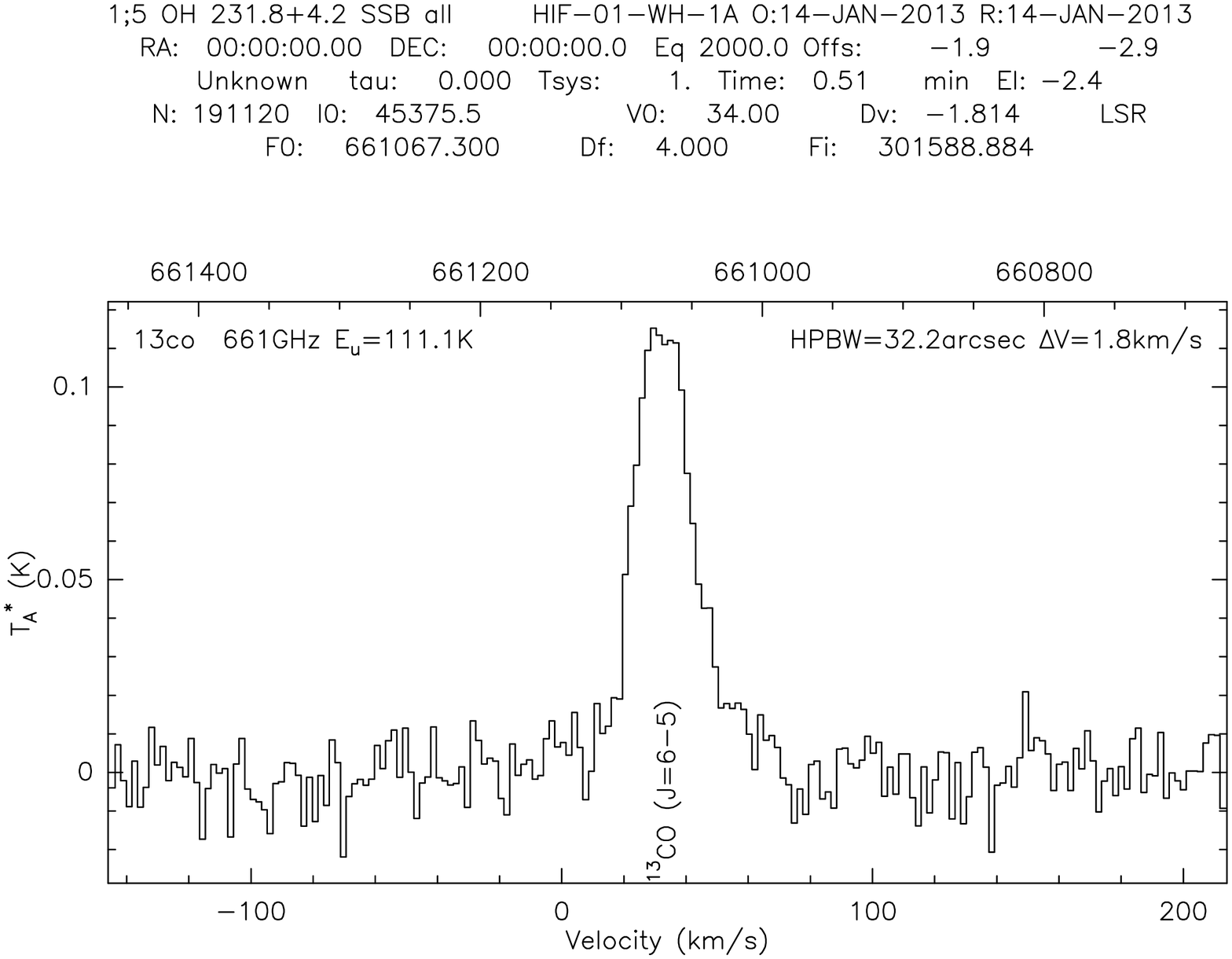}
 \includegraphics*[bb=0 97 721 510,width=0.45\hsize]{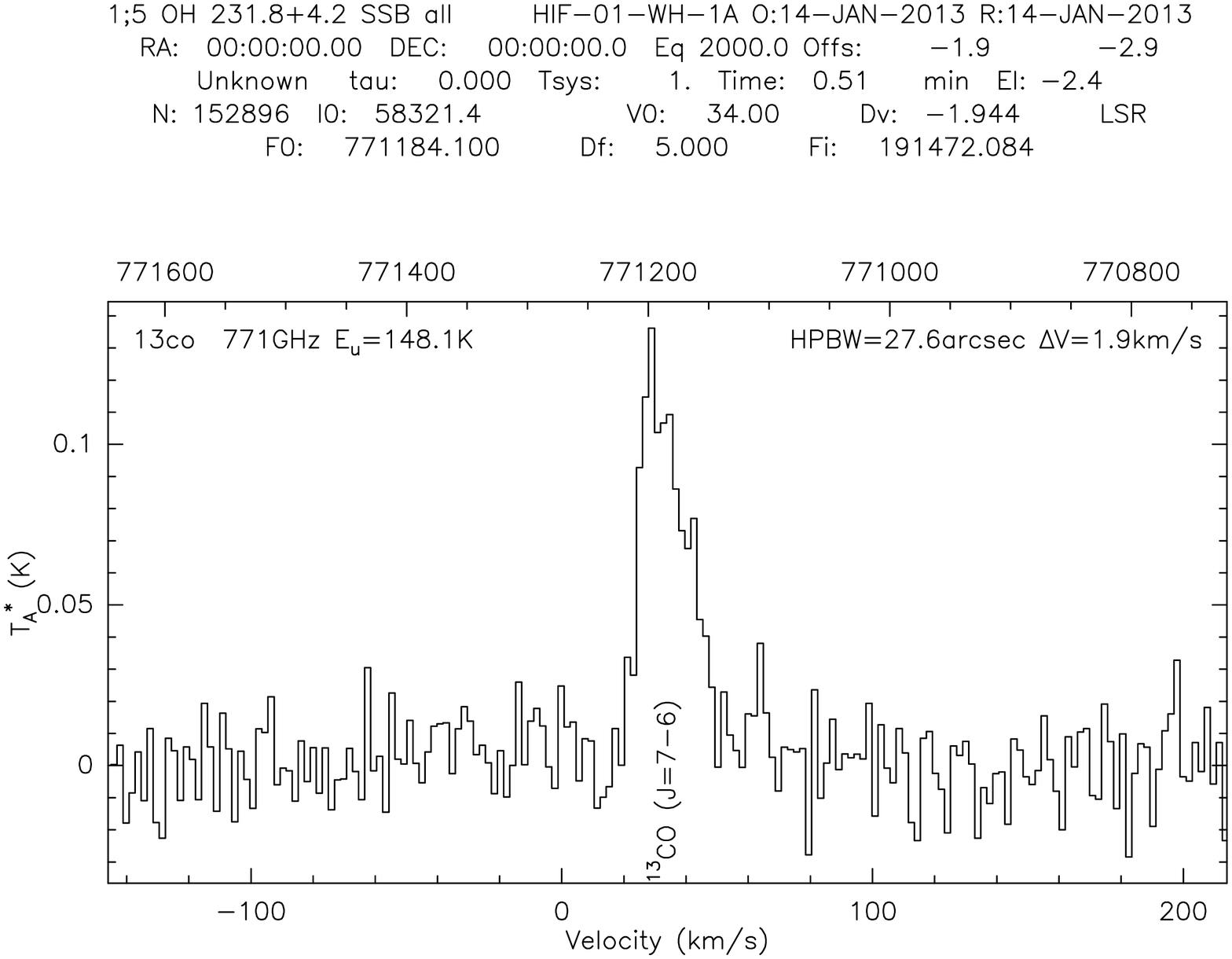}
 \includegraphics*[bb=0 97 721 510,width=0.45\hsize]{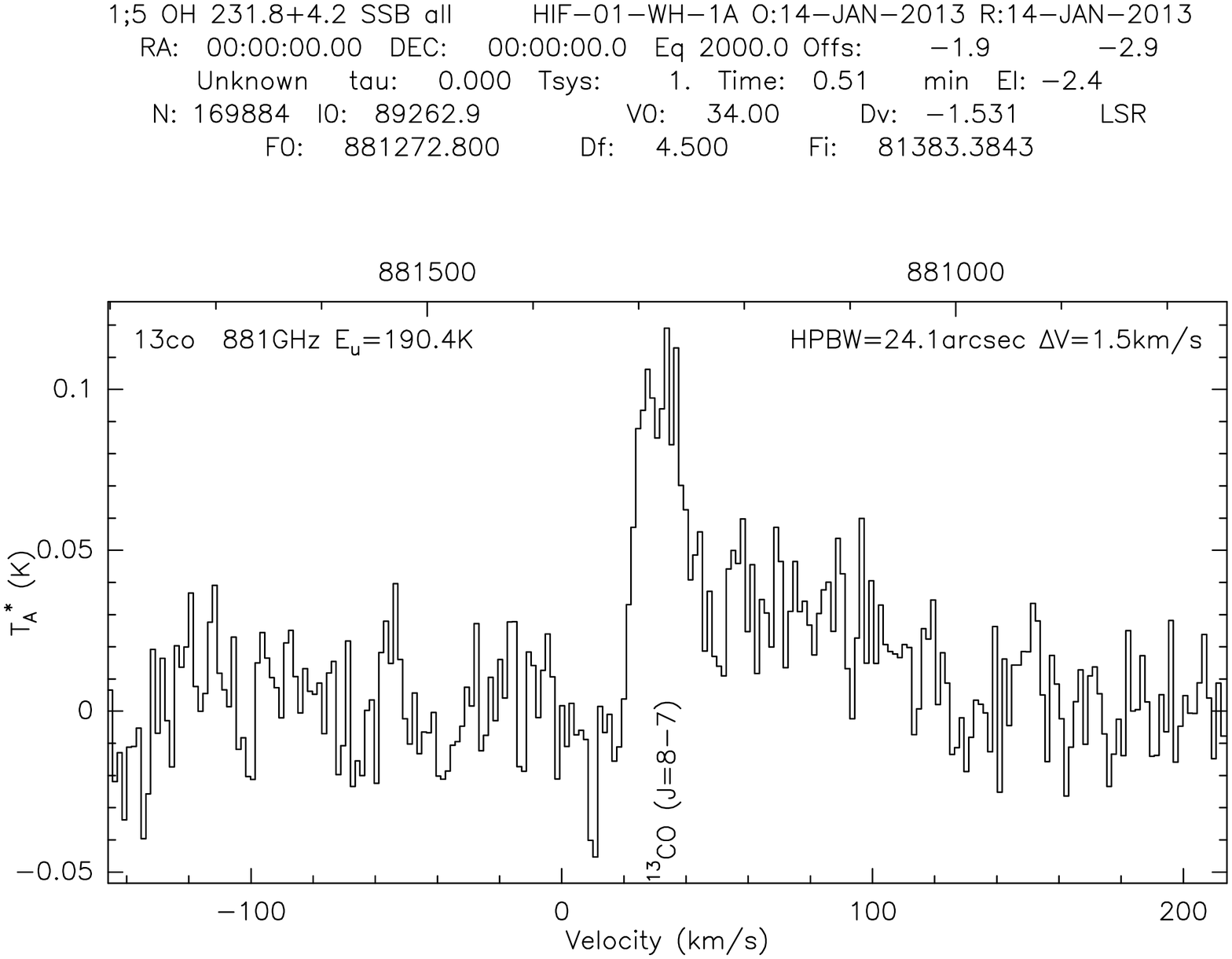}
 \includegraphics*[bb=0 97 721 510,width=0.45\hsize]{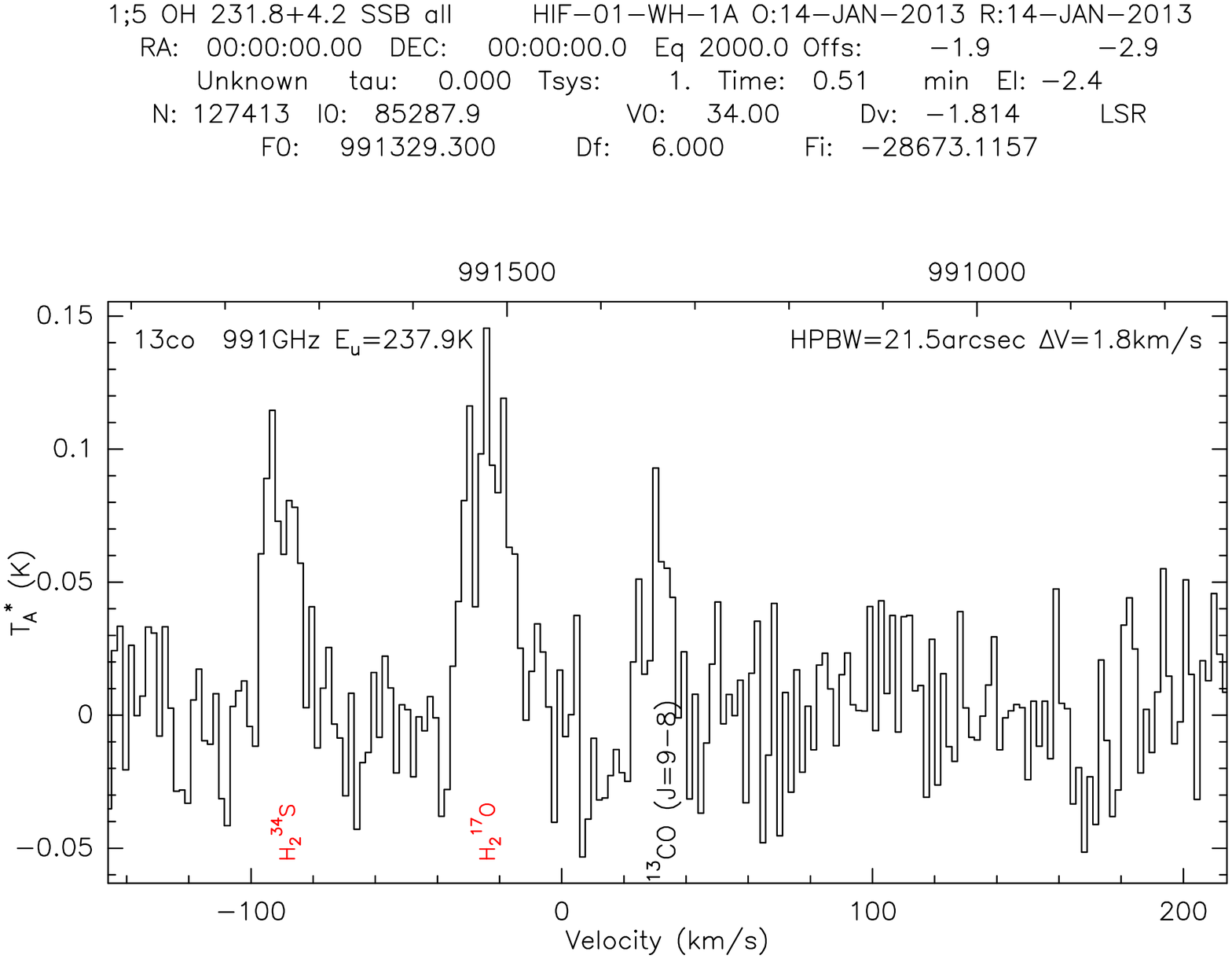}
 \includegraphics*[bb=0 97 721 510,width=0.45\hsize]{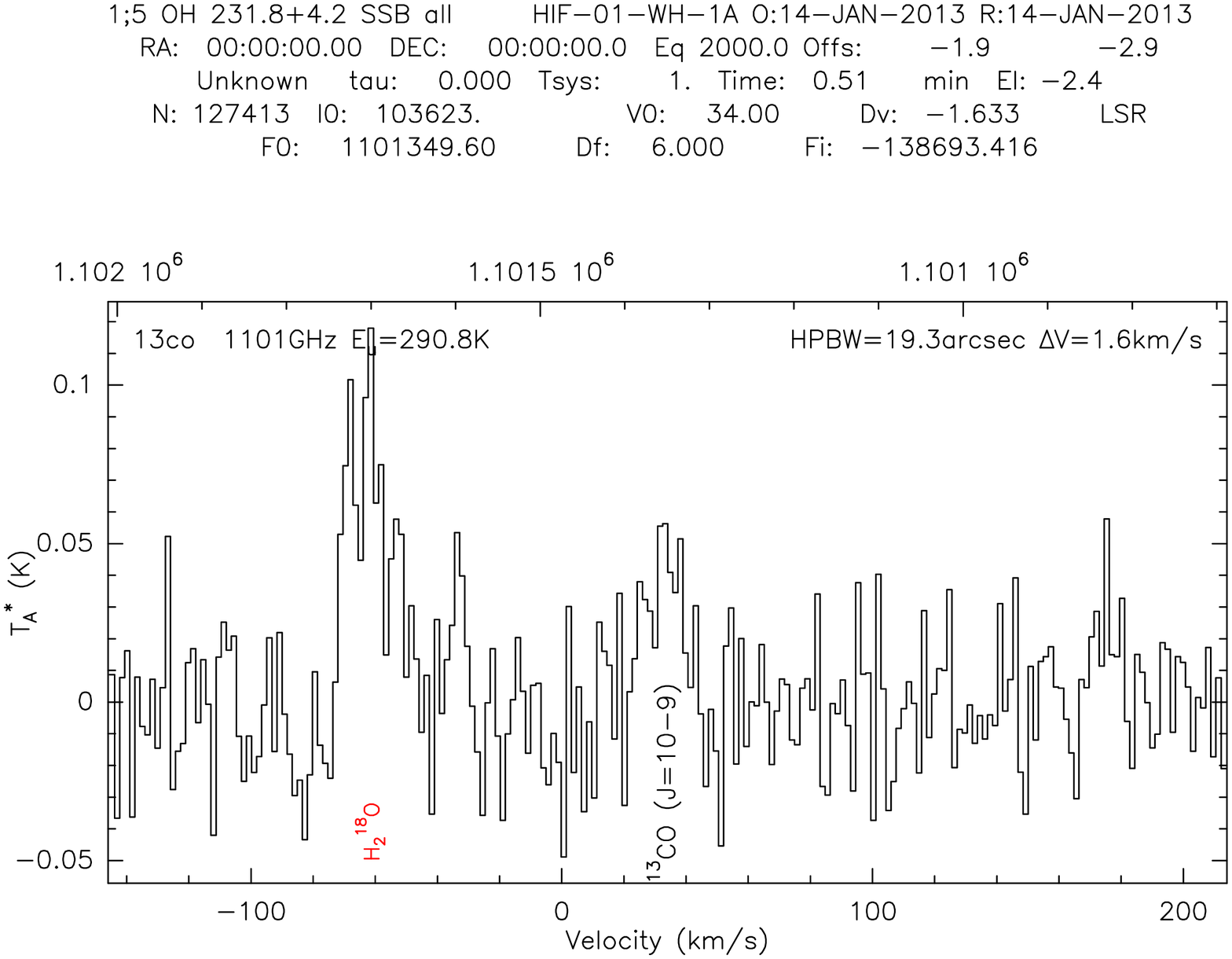}

      \caption{$^{13}$CO transitions detected in OH\,231.8+4.2 with
        \iram\ and \hso. The red broad shoulder observed in
        the \trecem\,($J$=8--7) line is probably an artificial
        feature due to residual baseline distorsion. }  
         \label{13co}
   \end{figure*}

A total of nine \trecem\ transitions have been detected in \oh\ from
our surveys with \iram\ and \hso\ (Fig.\,\ref{13co} and
Table\,\ref{t_13co}). The \trecem\ lines, which are good tracers of
the density and kinematics of the nebula, show structured profiles
with two main components: (1) a prominent, relatively narrow core
(centered around the systemic velocity of the source,
\vsys$\sim$33-34\,\kms) plus weaker, broad wings. The line core arises
at the slow, dense central (low-latitude) parts of the nebula, which
expand at relatively low velocities of $\sim$10-35\kms\ (clump I3),
whereas the wing emission mainly arises at the fast, bipolar lobes
(see \S\,\ref{intro} and Fig.\,\ref{mapa}).  (As explained in the
Introduction, the spatio-kinematic structure of the molecular outflow
of \oh\ is well known from previous single-dish and interferemtric
maps of \docem). The large core-to-wing intensity ratio of the
\trecem\ profiles is consistent with most ($\sim$70\%) of the mass in
the molecular outflow being in the slow central regions
\citep{san97,alc01}. 

The \trecem\ wing emission progressively weakens, relative to the
core, at higher frequencies. For transitions observed with the same
telescope (i.e.,\,for the three $J_{\rm u}$$\leq$3 lines observed with
\iram\ or for the six $J_{\rm u}$$\geq$5 lines observed with \hso),
this trend is partially explained by the smaller beam and, thus, the
smaller fraction of the fast outflow sampled by the observations at
higher frequencies. However, this is not the only reason, since at all
frequencies the \hso\ beam is comparable to, or larger than, the
\iram\ beam, with a maximum size of HPBW=22\farcs1 at 110\,GHz for the
observations reported here.

Both the full width of the wings and the full width at half maximum of
the \trecem\ profiles decrease as the upper energy level
increases. The full width of the wings is largest for the
\trece\ transition, which is observed over a velocity range of
\vlsr=[$-$80:$+$140]\,\kms\ with a beam HPBW=22.1\arcsec, and
decreases for higher-$J$ transitions down to
\vlsr=[$+$15:$+$55]\,\kms\, for the $J$=7--6 line, observed with a
comparable beam of HPBW=27.6\arcsec; for transitions with $J$$>$7, we
do not detect wing emission.  The FWHM of the \trecem\ lines ranges
between $\sim$36 to $\sim$14\,\kms\ for the $J$=1--0 and the 10--9
transitions, respectively. An analogous behavior is observed in most
molecules observed in our surveys with \iram\ and \hso, including
\docem\ \citep[][S\'anchez Conteras et al., in prep.]{san14}. The
observed trend suggests that the envelope layers with higher
excitation conditions (i.e.,\,warmer and, thus, presumably closer to
the central star) are characterized by lower expansion velocities.

\begin{table}

\caption{\label{t_13co} Same as Table\,\ref{t_res} except for \trecem\ lines 
observed with \iram\ and \hso.
} 
\small
\begin{tabular}{r c r c r}      
\hline\hline                  
Rest Freq.\ & Transition  & E$_{\rm u}$     &   A$_{\rm ul}$  &  \intta\\    
(MHz)       & QNs         & (K)            &     (s$^{-1}$)  &  (K\,\kms) \\     
\hline                        
\multicolumn{4}{c}{$^{13}$CO  \hspace{0.5cm} $\mu$= 0.11\,Debyes} &\\ 
110201.4 & 1   - 0   & 5.3 & 6.336E-08 & 13.45 (0.06) \\
220398.7 & 2   - 1   & 15.9 & 6.082E-07 & 51.50 (0.13) \\
330588.0 & 3   - 2   & 31.7 & 2.199E-06 & 42.00 (0.30) \\
550926.3 & 5   - 4   & 79.3 & 1.079E-05 & 3.33 (0.07) \\
661067.3 & 6   - 5   & 111.1 & 1.894E-05 & 3.03 (0.09) \\
771184.1 & 7   - 6   & 148.1 & 3.040E-05 & 2.30 (0.13) \\
881272.8 & 8   - 7   & 190.4 & 4.574E-05 & 2.00 (0.15) \\
991329.3 & 9   - 8   & 237.9 & 6.554E-05 & 0.65 (0.15) \\
1101349.6 & 10   - 9   & 290.8 & 9.034E-05 & 0.70 (0.15) \\
\hline 
\end{tabular}
\end{table}

\end{appendix}


\end{document}